\documentclass[12pt]{article}

\usepackage{adjustbox}
\usepackage{amsmath, amssymb, mathtools}
\usepackage{appendix}
\usepackage{bm, bbm}
\usepackage{booktabs} 
\usepackage{caption} 
\usepackage{subcaption}
\usepackage{etoolbox} 
\usepackage[margin = 1in]{geometry}
\usepackage{graphicx}
\usepackage{longtable}
\usepackage{multirow}
\usepackage[comma]{natbib}
\usepackage[parfill]{parskip} 
\usepackage{rotating} 
\usepackage{setspace} 
\usepackage{textcomp} 
\usepackage{threeparttablex} 
\usepackage{xurl}  
\usepackage{tikz} 
\usetikzlibrary{arrows, intersections, shapes}
\usepackage[colorlinks = true, citecolor = blue]{hyperref} 

\usepackage{lipsum}

\DeclareMathOperator{\Cat}{Categorical}
\DeclareMathOperator{\Dir}{Dirichlet}

\DeclareMathOperator{\vecv}{vec}
\DeclareMathOperator{\vech}{vech}
\newcommand{\N}{\mathcal{N}}
\newcommand{\IW}{\mathcal{IW}}
\newcommand{\T}{^{\top}}
\newcommand{\uBar}[1]{\underline{#1}}

\AtBeginEnvironment{footnote}{\onehalfspacing} 
\allowdisplaybreaks 
\providecommand{\JEL}[1]{ 
  \small	
  \textbf{\textit{JEL:}} #1
}
\providecommand{\Keywords}[1]{ 
  \small	
  \textbf{\textit{Keywords:}} #1
}

\title{EASI Drugs in the Streets of Colombia: Heterogeneous Drug Preferences and Marijuana Legalization}
\author{Santiago Montoya-Blandón\thanks{Adam Smith Business School, University of Glasgow. Room 620, ASBS Building, 2 Discovery Place, Glasgow, United Kingdom, G11 6EY. Email: \href{mailto:santiago.montoya-blandon@glasgow.ac.uk}{santiago.montoya-blandon@glasgow.ac.uk}} \and Andr\'es Ramírez--Hassan\thanks{School of Finance, Economics and Government. Valor Público and OMEGA research group, Universidad EAFIT, Medellín, Colombia. E-mail: \href{aramir21@eafit.edu.co}{aramir21@eafit.edu.co}. Carrera 49 7 Sur-50, phone: + 57 2619500}}

\date{\today}

\begin{document}

\maketitle
\vspace{-5mm}

\begin{abstract}
The response of illicit drug consumers to policy changes like legalization is mediated by demand behavior. Since individual drug use is driven by many unobservable factors, accounting for unobserved heterogeneity becomes crucial for designing targeted policies. This paper introduces a finite Gaussian mixture of EASI demand systems to estimate joint demand for marijuana, cocaine, and basuco (a low-purity cocaine paste) in Colombia, accounting for corner solutions and endogenous prices. Our method classifies users into two groups with distinct preferences over consumption: ``soft'' and ``hard'' users. Nationally representative survey estimates find drugs are unit-elastic, with marijuana and cocaine complementary. International marijuana legalization episodes along with Colombia's low marijuana production cost suggest legalization is likely to drive prices down significantly. Legalization counterfactuals under the most likely scenario of a 50\% marijuana price decrease reveal \$363/year welfare gains for consumers, \$120M in governement revenue, and \$127M dealer losses.
\end{abstract}
\JEL{D12, C11, C35.}


\section{Introduction}
\label{Sec:Intro}
Drug consumption is a growing global market involving an increasing number of users. According to the 2023 report by the United Nations Office on Drugs and Crime \citep{UNODC2023}, an estimated 219 million people used marijuana in 2021, representing about 4.3\% of the adult population. This was followed by opioids (64 million), amphetamines (36 million), cocaine (22 million), and ecstasy (20 million). In order to cope with the public health outcomes from increasing drug consumption, several nations across the world have enacted laws legalizing recreational drug use. Specifically for cannabis, the most-widely consumed substance, recreational use has been legalized in Canada, Georgia, Germany, Luxembourg, Malta, Mexico, South Africa, Thailand, Uruguay, along with 24 US states, three territories, and the District of Columbia, with many other nations currently considering similar policies. Assessing the impacts of these policies \textit{ex-ante} requires understanding how consumers might react to variations in the prices of drugs associated with their legalization \citep{Becker2006, Hall2016}. Additionally, joint policy interventions are usually employed to support the legalization or decriminalization of substances. These interventions are more effective if they target specific segments of the population that face particular necessities. Characterizing population segments that are similar in their preferences for drugs can directly translate to more effective targeting of interventions aimed at these groups.


This paper aims to evaluate the potential effects of marijuana legalization among regular consumers on their welfare, government collection, and drug dealers' revenues, using Colombia as a study case. To study the effect of such a policy \textit{ex-ante}, we must understand how consumers' drug demand behavior responds to price changes, such as those associated with the possible legalization of marijuana in the country. To this end, we introduce a method to estimate the joint demand for illicit drugs in Colombia, taking preference heterogeneity into account to identify potentially diverse responses to price changes. Given the large rates of violent crime associated with Colombian drug activity, policymakers advocate legalization as a strategy to reduce drug profitability given its causal link to violent crime \citep{Fajnzylber2002, Pinotti2015, gavrilova2019legal, queirolo2019uruguay}. Evidence supports the claim that marijuana legalization introduces legal competition in the drug market, which decreases profitability and can help reduce crime  \citep{huber2016cannabis,dragone2019crime,brinkman2019not,burkhardt2019short,wu2020spillover,anderson2023public}, particularly violent crimes associated with drug trafficking and resolving disputes \citep{burkhardt2019short,chu2019joint,gavrilova2019legal,wu2022effects,anderson2023public}. The issue of violent drug crime is of great relevance in Colombia, where homicide rates and violence linked to drug trafficking are pronounced \citep{puerta2024spatial}.

Our paper also contributes to the literature by studying the demand for illicit drugs using a novel data set in a particularly relevant developing country such as Colombia. The Colombian case is of special interest as it is one of the world's leading producers of marijuana and cocaine \citep{UNODC2021}, resulting in relatively low prices and widespread access to these substances. The research question is also timely, as the Colombian National Congress recently debated the legalization of recreational marijuana in 2023 but failed to secure sufficient votes in the final round of discussions.\footnote{\url{https://www.reuters.com/world/americas/colombia-senate-votes-down-recreational-marijuana-bill-2023-06-21/}.} According to our main dataset of interest, in a 2019 nationally representative survey of Colombians aged 12 to 65 regarding their consumption of psychoactive substances, marijuana is found to be the most-consumed drug, followed by cocaine, and then basuco (a cheap, highly addictive, and toxic by-product of cocaine production), with other drugs showing only trace amounts of measurable national consumption. Therefore, in this research we focus on studying the joint demand for these three most-consumed drugs in Colombia.\footnote{Currently, possession for personal use is decriminalized in Colombia, allowing individuals to carry up to 20 grams of marijuana and 1 gram of cocaine or basuco.}

To address the technical challenges associated with modeling the demand for illicit drugs and to understand the mechanisms driving the potential effects of marijuana legalization, we propose a new estimation framework based on a microeconomically founded model that takes unobserved heterogeneity, corner (zero) solutions, and endogeneity into account. Specifically, we propose a Bayesian inferential framework that allows for clusters of unobserved preference heterogeneity through a finite mixture of Exact Affine Stone Index (EASI) demand systems. The Bayesian framework allows us some additional advantages that are relevant for our application. First, it allows us to account for corner outcomes where individuals can decide not to consume a given set of goods, which is relevant as many consumers in our sample only report consuming one illicit drug out of the three we consider. Second, it becomes straightforward to include membership to unobserved heterogeneity clusters and associated probabilities as parameters to perform data-driven consumer segmentation and recover heterogeneous drug responses. Third, it allows us to obtain inference on the structural quantities of interest, such as price-demand elasticities or predicted revenue under counterfactual scenarios as a by-product of estimation, all of which are highly non-linear functions of data and model parameters. Finally, we can easily impose and test relevant microeconomic restrictions such as symmetry, strict cost monotonicity, and concavity of the cost function, which ensure the recovered demand functions satisfy standard theoretical conditions \citep{RamirezHassan2021, RamirezHassan2024}. For further details on Bayesian estimation of demand systems, see \cite{Tiffin2010, Kasteridis2011, Kehlbacher2020, Jacobi2021, RamirezHassan2021}.

Applying our methodology to the analysis of demand for marijuana, cocaine, and basuco in Colombia, the results suggest that unobserved heterogeneity is crucial to obtaining precise and economically relevant demand behavior when taking into account censoring and price endogeneity. Our method automatically partitions our sample into two sub-populations: ``soft'' and ``hard'' drug consumers. The former group represents the largest user segment, spending a majority of their expenditure on marijuana leading to sensible and economically rational demand patterns. In contrast, the latter group has large consumption of the cocaine-based substances and scores highly in the survey questions meant to screen for substance addiction, providing a rationale behind our labels. In addition, we find for the ``soft'' user group that all three own-price elasticities for each drug are statistically indistinguishable from unity, with a complementary relationship between marijuana and cocaine, and a substitution between higher-quality cocaine and lower-quality residual.

In the most likely scenario that the legalization of marijuana results in a 50\% decrease in the price of marijuana (given the high illegal price markup for this drug in Colombia), our estimates suggest that commercial sale of marijuana would imply a considerable increase in marijuana consumption among regular consumers, with relatively much smaller effects for cocaine and basuco. This increase is valued by consumers at approximately \$363 USD of annual utility-equivalent expenditure as measured by the Equivalent Variation of the price change. In addition, we find a high probability that the government will capture a large amount of the legal market resulting in considerable revenue gains even in the face of the reduced post-legalization price. Specifically, assuming that post-legalization legal sales of marijuana accounted for two thirds of all sales (with cocaine and basuco remaining illegal) the government accrues \$120 million annual USD at the expense of dealers who experience a loss of \$127 million annual USD in revenue. These figures imply that drug dealers would need to reach around 130\% new domestic drugs users compared with pre-legalization to offset such a revenue decline, which is not likely to be realized based on international legalization experiences. Taken together, these findings suggest that a marijuana legalization policy in Colombia is likely to succeed at disincentivizing drug-related criminal activity; the current largest source of homicides and other violence in the country.

This paper connects with a wealth of previous literature on demand systems, demand for addictive substances, unobserved preference heterogeneity, among others. Demand systems serve as essential tools for learning about consumer behavior and its responses to diverse market conditions \citep{Deaton1980}. There is extensive previous literature using demand systems to understand consumption patterns for some addictive substances, particularly alcohol and tobacco \citep{Thies1993, Deza2023}. However, there is a conspicuous gap in the literature regarding the study of joint drug consumption patterns, likely due to issues of data availability and general attitudes towards illicit substances. A key contribution of our analysis is to examine not just a single drug in isolation, but the joint demand for cannabis, cocaine, and basuco. This broader perspective is critical as it enables us to explore whether these substances exhibit substitution or complementary relationships and whether these patterns differ across population groups. 

An important advantage of demand systems is their ability to consider joint consumption across categories of goods as well as providing insight into determinants of these behaviors. For instance, \cite{duffy2003} uses the Quadratic Almost Ideal Demand system \citep[QAID,][]{Banks1997} to investigate consumer spending patterns in the United Kingdom, challenging the prevailing assumption that advertising significantly influences preferences, particularly for products like tobacco. \cite{aristei2010} conduct an innovative exploration of alcohol and tobacco consumption patterns in Italy, effectively addressing criticisms related to unmeasured preferences and correlated unobserved heterogeneity, shedding some light on the joint determinants of these behaviors.

An additional feature of demand systems is that they allow us to study demand under counterfactual scenarios. That is, we can construct conterfactual price scenarios associated with the legalization of marijuana, to account for mixed effects of legalization on marijuana price (due to shifts away from the black market structure). Evidence highlights varied outcomes depending on the region, market conditions, and regulatory frameworks. For example, the Uruguayan government deliberately kept marijuana prices low to undercut the black market, while in Canada, the price of marijuana increased by 32\% after legalization, primarily due to regulatory factors.\footnote{See the official website from Statistics Canada on the price of cannabis post-legalization: \url{https://www150.statcan.gc.ca/n1/daily-quotidien/190710/dq190710c-eng.htm}.} In the specific context of Colombia, the production cost of one gram is approximately US\textcent 1 \citep{velez2021medicinal}. In contrast, the current illegal market price is around US\textcent 83 (see descriptive statistics from our dataset in Table \ref{Tab:Summary_Statistics}). This significant disparity indicates a considerable potential for lowering marijuana prices in a legalized market. Thus, the price elasticities of drug dealers' product offerings are critically important for assessing their potential revenue impacts in a post-legalization scenario.

Other recent research adds further structure to the models in order to dive deeper into the determinants of addictive substance consumption. For example, \cite{Lacruz2009} give an in-depth examination of alcohol demand among youth in Spain, highlighting the impact of factors such as income, prices, and the addictive nature of alcohol. Using German survey data, \cite{Tauchmann2013} challenge the notion of tobacco and alcohol as substitutes by employing a structural model to directly estimate their interdependence, with profound implications for anti-smoking policies. \cite{bokhari2018} considers pharmaceuticals, as they are legal but still potentially addictive substances. This paper analyzes the demand for drugs used to treat attention deficit hyperactivity disorder (ADHD) through the lens of demand systems, focusing on the impact of pharmaceutical mergers. Their study emphasizes the importance of selecting the demand model when evaluating the effects of large-scale market changes, particularly to capture drug substitution patterns. Collectively, these findings highlight the value of demand system analysis in understanding drug consumption patterns and informing effective policy strategies across a wide range of substances.

Previous studies additionally and heavily emphasize the relevance of heterogeneity, in both observable and unobservable factors, to understanding complex patterns found in demand data \citep{duffy2003, Lacruz2009, aristei2010, Tauchmann2013, bokhari2018}. In estimation of demand systems, researchers often rely on observable features to approximate unobserved preference heterogeneity. This has been implemented in the literature through the incorporation of interaction terms \citep{Blundell1993, Moro2000} or by categorizing individuals \textit{ex-ante}  using observable features \citep{Bonnet2013}. Other studies address unobserved heterogeneity using random parameter models \citep{Haefen2004, Lewbel2009} or segmentation algorithms \citep{Bertail2008, Kehlbacher2020}. In the former, heterogeneity emerges from parameter values that vary across units according to some distribution function, while the latter assumes that units are clustered into sub-populations, with similar preferences within clusters and distinct preferences across clusters.

Unobserved heterogeneity in the demand for illicit drugs is particularly relevant when examining the \textit{gateway hypothesis} \citep{Kandel1975, Lynskey2018}, which posits that individuals often transition from using softer drugs, such as marijuana, to harder ones like cocaine or basuco. This, in turn, may lead to complementary consumption patterns between drugs. \cite{DeSimone1998} found evidence that marijuana consumption can potentially lead to the consumption of other drugs, where they allow for structural estimates of unobserved heterogeneity to affect both marijuana and cocaine. \cite{Ours2003, Ours2006} provide evidence of a causal relationship between cannabis and cocaine use, where the use of multiple drugs is generally driven by unobserved individual characteristics. \cite{BrettevilleJensen2008} argues that the empirical association between cannabis and heroin use could be spurious. The authors interpret unobserved heterogeneity as ``antisocial behavior" and explain that this could simultaneously affect both cannabis and heroin use. 

\cite{BrettevilleJensen2008} agree that the gateway effect diminishes greatly when unobserved factors are considered, and \cite{Melberg2010} examine the gateway hypothesis, explaining that factors like traumatic childhood experiences could be associated with both cannabis and heroin use. If this is the case, the causal impact of cannabis use on the use of harder drugs would be confounded by the presence of the unobserved trauma factor, mistakenly leading to the conclusion that cannabis serves as a gateway to other drugs. \cite{Deza2015} explores the dynamic patterns of drug use, distinguishing between the effects of prior drug consumption and unobserved heterogeneity. The former implies that past experiences influence future choices, while the latter suggests inherent tendencies toward drug consumption in certain individuals. This study finds evidence supporting the hypothesis that consumption of hard drugs complements the consumption of alcohol and marijuana. \cite{Jorgensen2022} found that marijuana is not a reliable gateway cause of illicit drug use when taking unobserved effects into account, meaning prohibition policies are unlikely to reduce illicit drug use. In summary, the literature finds that heterogeneous effects are relevant when analyzing demand of illicit drugs, and that there is not conclusive evidence about the relationship between marijuana use and the use of ``harder'' drugs like cocaine.

After this introduction, we present in Section \ref{Sec:Data} our data set, descriptive statistics, and some preliminary descriptive patterns. Section \ref{Sec:Methodology} shows the microeconomic and econometric framework, and Section \ref{Sec:Results} shows the results applied to our consumption dataset. We present in Section \ref{Sec:Legalization} counterfactual exercises from implied price changes after a simulated legalization policy on consumer welfare, government revenue, and drug dealers' illegal market size. Section \ref{Sec:Conclusions} presents our policy recommendations based on results and concluding remarks. 

\section{Descriptive Statistics of Colombian Drug Market}
\label{Sec:Data}
We use the \emph{National Survey of the Consumption of Psychoactive Substances} performed in 2019 (\textit{ENCSPA} from its name in Spanish) by the Colombian National Administrative Department of Statistics (DANE). This is a nationally representative survey aiming to measure the consumption of legal and illegal psychoactive substances. Individuals between 12 and 65 years old from several municipalities were randomly selected, and the enumerators privately performed the survey. If the chosen individual was absent during the survey, the enumerator should return later. The survey resulted in a total sample of 49,439 individuals, representative of approximately 23.6 million individuals, which is equivalent to roughly the total urban Colombian population in 2019 within the considered age range.

Our primary objective is to examine the potential effects of significant drug price variations in Colombia on the consumption of marijuana and harder drugs (specifically cocaine and basuco); for example, as a result of a marijuana legalization policy. These three drugs were identified as the most relevant illicit drugs in Colombia in terms of consumption and expenditure. Therefore, we restrict our sample to current regular consumers, defined in the survey as individuals who report using at least one of the three illicit substances at least once per month during the interview year. Limiting our sample to regular consumers results in a total of 1,236 users that are representative of 633,490 users nationwide.

A natural concern with self-reported use of psychoactive substances collected via survey is the potential for under-reporting and reliability of consumption amounts disclosed. While this issue is recognized in the survey design by implementing repeated visits, we separately assess the reliability of the survey in Appendix \ref{Sec:Appendix_Survey_Reliability}. By conducting similar descriptive statistic analysis for consumers in single-dweller households, we provide evidence that those with the lowest incentive to misrepresent their information behave similarly to the full sample. This assuages potential concerns of data reliability, and is consistent with previous literature that finds generally low instances of under-reporting with larger cases occurring mostly in younger individuals or specialized groups \citep{Needle1983, Harrison1993, Darke1998}.

\subsection{Drug market}

\begin{figure}[htbp]
    \centering
    \caption{Shares of total expenditure on drugs and users in Colombia by age groups and drugs}
    \label{Fig:Drug_Expenditure_Users}
    \begin{subfigure}{0.49\linewidth}
        \includegraphics[width = \linewidth, trim={18pt 20pt 20pt 38pt}, clip]{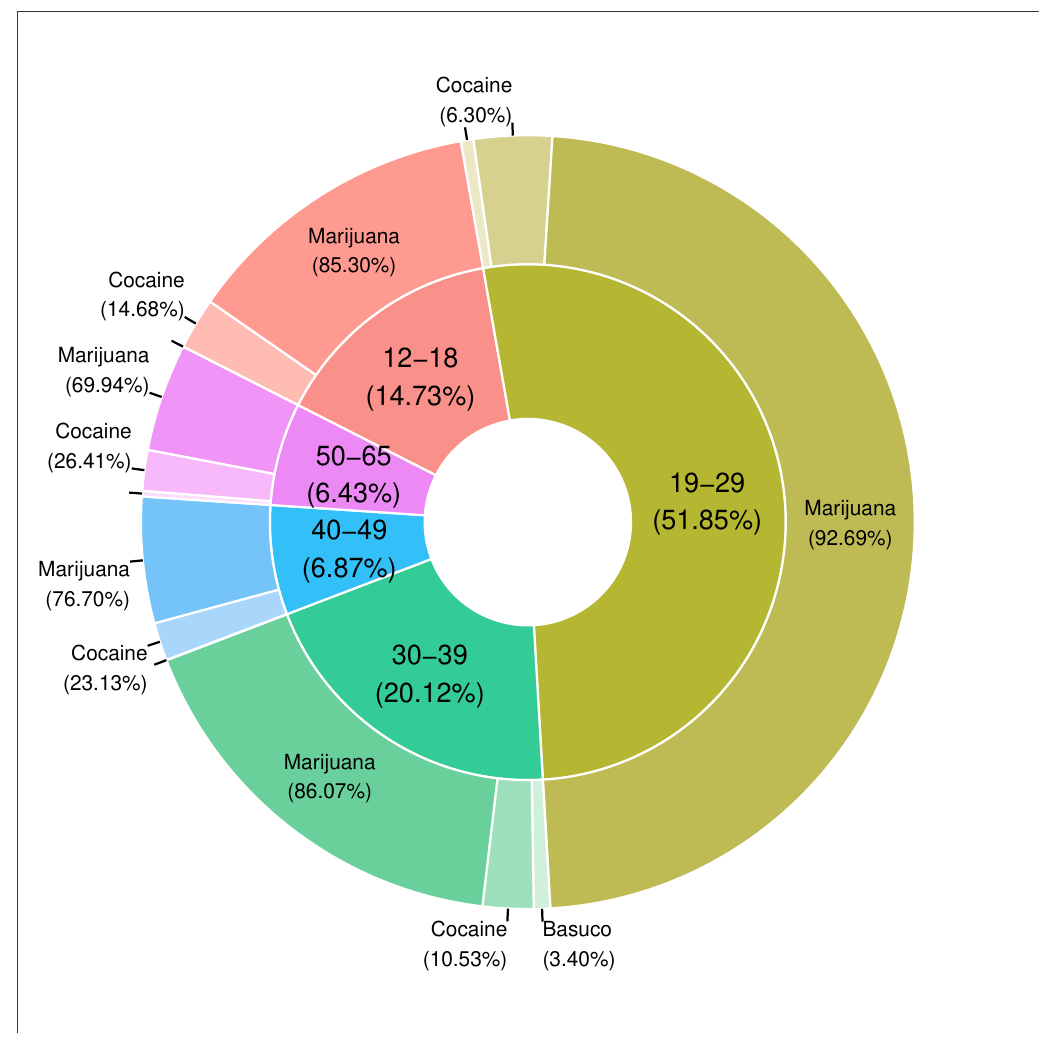}
        \caption{Expenditures}
        \label{Fig:Drug_Expenditure}
    \end{subfigure}
    \begin{subfigure}{0.49\linewidth}
        \includegraphics[width = \linewidth, trim={18pt 20pt 20pt 38pt}, clip]{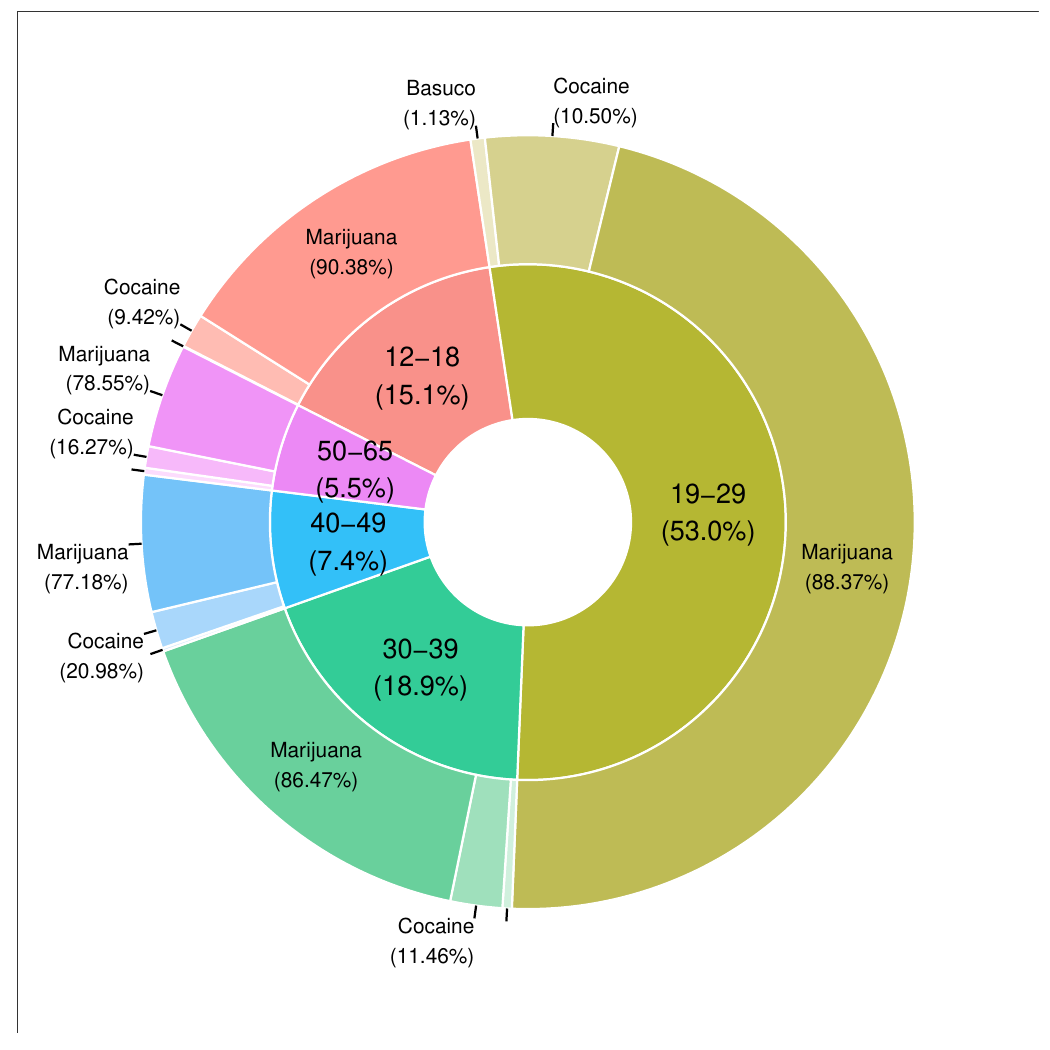}
        \caption{Users}
        \label{Fig:Drug_Users}
    \end{subfigure}
    \begin{tablenotes}
        \item \small{\textbf{Notes:} In 2019, total drug market expenditure in Colombia was approximately 225.73 million with users amounting to 633,490. We classify individuals according to their most consumed drug. Individuals in their twenties represent the largest share of total expenditure and users, followed by individuals in their thirties and then teenagers. Marijuana has the largest share of expenditure and users in all age groups.}
    \end{tablenotes}
\end{figure}

According to the \textit{ENCSPA} survey, the total national expenditure on marijuana, cocaine, and basuco in 2019 was USD 226.3 million among regular users.\footnote{We use the average exchange rate in 2019 (COP/USD 3,282.39) to convert all values in Colombian pesos to US dollars throughout the paper \citep{banrep_trm_2019}.} Figure \ref{Fig:Drug_Expenditure} shows the share of total expenditure by age groups and drug of choice (defined as the most-consumed drug). We see that individuals in their twenties spend the most, approximately 51.9\%, followed by individuals in their thirties (20.1\%), teenagers (14.7\%), forties (6.9\%), and fifties (6.4\%). It is concerning that such a large share is spent by younger individuals. We also see in this figure that marijuana represents by far the most relevant expenditure. Conditional on age group, marijuana shares range between 70\% (fifties) and 92.7\% (twenties). We see that basuco represents the lowest share, ranging between 0.0\% (teenagers) and 3.7\% (fifties). Cocaine remains as in-between these two substances, with the largest expenditure share coming from individuals in their fifties spending 26.4\% of their drug budget on this substance, with a concerningly large share remaining for both teenagers and young adults (14.7\% and 6.3\%, respectively).

Expenditure is composed of three elements: quantities, prices, and total size of the market. Turning our attention first to market size, we classify the total number of regular users according to their drug of choice, resulting in 550,786 total marijuana users, with 73,578 for cocaine and 9,126 for basuco. Figure \ref{Fig:Drug_Users} shows the distribution of these users by age group. We observe again that individuals in their twenties represent the largest share (53.0\%), followed by individuals in their thirties (18.9\%), teenagers (15.1\%), forties (7.4\%), and fifties (5.5\%). We also see again that marijuana comprises the largest share, ranging between 88.4\% (twenties) and 77.2\% (forties). The second largest is associated with cocaine, where its share ranges between 21.0\% (forties) and 9.4\% (teenagers). Finally, there is basuco, whose share ranges between 5.2\% (fifties) and 0.2\% (teenagers). It is concerning that most of drug users are individuals less than 30 years old, and that older individuals have higher concentrations of hard drug use.

Table \ref{Tab:Summary_Statistics} shows descriptive statistics for individual-level expenditure shares, quantities consumed and prices faced in the market for drugs. Marijuana's share is the largest (86.9\%), followed by cocaine (10.7\%). We also report the proportion of zeros in the sample, where marijuana has the lowest figure (6.5\%), while basuco has the highest (95.5\%). This highlights the importance of taking into account the censoring issue in the econometric framework, as standard models cannot account for the exceeding share of consumers that do not demand some of the drugs.

\begin{table}[!htbp]
    \begin{adjustbox}{max totalsize={\textwidth}{\textheight}}
    \begin{threeparttable}
        \caption{Summary statistics for shares of drug consumption and prices}
        \label{Tab:Summary_Statistics}
            \begin{tabular}{lcccc}
            \toprule
            Shares (percentage points, pp.) & Mean & Std. Dev. & \multicolumn{2}{c}{Proportion with no expenditure} \\
            \midrule
            Marijuana & 86.86 & 29.95 & \multicolumn{2}{c}{6.47} \\ 
            Cocaine & 10.75 & 27.15 & \multicolumn{2}{c}{80.74} \\ 
            Basuco & 2.40 & 13.86 & \multicolumn{2}{c}{95.55} \\ 
            \midrule
            Quantity (grams per month) & Mean & Std. Dev. & Min. & Max. \\
            \midrule
            Marijuana & 39.80 & 74.78 & 0.00 & 600.00 \\ 
            Cocaine & 1.28 & 5.04 & 0.00 & 50.00 \\ 
            Basuco & 0.95 & 7.97 & 0.00 & 150.00 \\ 
            \midrule
            Prices (cents per gram, USD \textcent/gr.) & Mean & Std. Dev. & Min. & Max. \\
            \midrule
            Marijuana & 83.31 & 53.10 & 15.27 & 305.44 \\ 
            Cocaine & 309.32 & 162.06 & 61.09 & 1,221.75 \\ 
            Basuco & 73.32 & 37.44 & 30.54 & 305.44 \\ 
            \bottomrule
        \end{tabular}
        \begin{tablenotes}
            \item \small{\textbf{Notes:} Total sample size includes 1,236 consumers from ENCSPA survey in Colombia. Monthly quantity is conditional on consumption and prices are given in 2019 USD. Marijuana is the most relevant drug expenditure, followed by cocaine. There is a high level of censoring, particularly in basuco, and also a high variability regarding drug prices faced by individuals.}
        \end{tablenotes}
    \end{threeparttable}
    \end{adjustbox}
\end{table}

Conditional on the consumption of each drug, the average monthly consumption of marijuana, cocaine, and basuco is 43.1, 6.6, and 21.1 grams, respectively. However, there is substantial variability. For instance, one individual reports consuming 20 grams (joints) of marijuana per day. The unconditional average monthly consumption of marijuana, cocaine, and basuco is 39.8, 1.3, and 0.95 grams, respectively. This is because most individuals consume only marijuana, as illustrated in Figure \ref{Fig:Drug_Consumption_Bundles}. The average marijuana price is 83.3 USD \textcent/gr. with a range from 15.3 USD \textcent/gr. to 305.4 USD \textcent/gr. This high variability in prices is also present in the other drugs: cocaine prices range from 61.1 USD \textcent/gr. to 1,221.8 USD \textcent/gr., and basuco from 30.5 USD \textcent/gr. to 305.4 USD \textcent/gr. This heterogeneity is largely attributed to differences in drug quality and suppliers' location. 

\begin{figure}[!htbp]
    \centering
    \caption{Average consumptions and prices for individuals with positive consumption across age groups and illicit drugs}
    \label{Fig:Consumption_Prices_Radar}
    \begin{subfigure}{0.49\textwidth}
        \centering
        \includegraphics[width = \linewidth, trim={60pt 100pt 50pt 60pt}, clip]{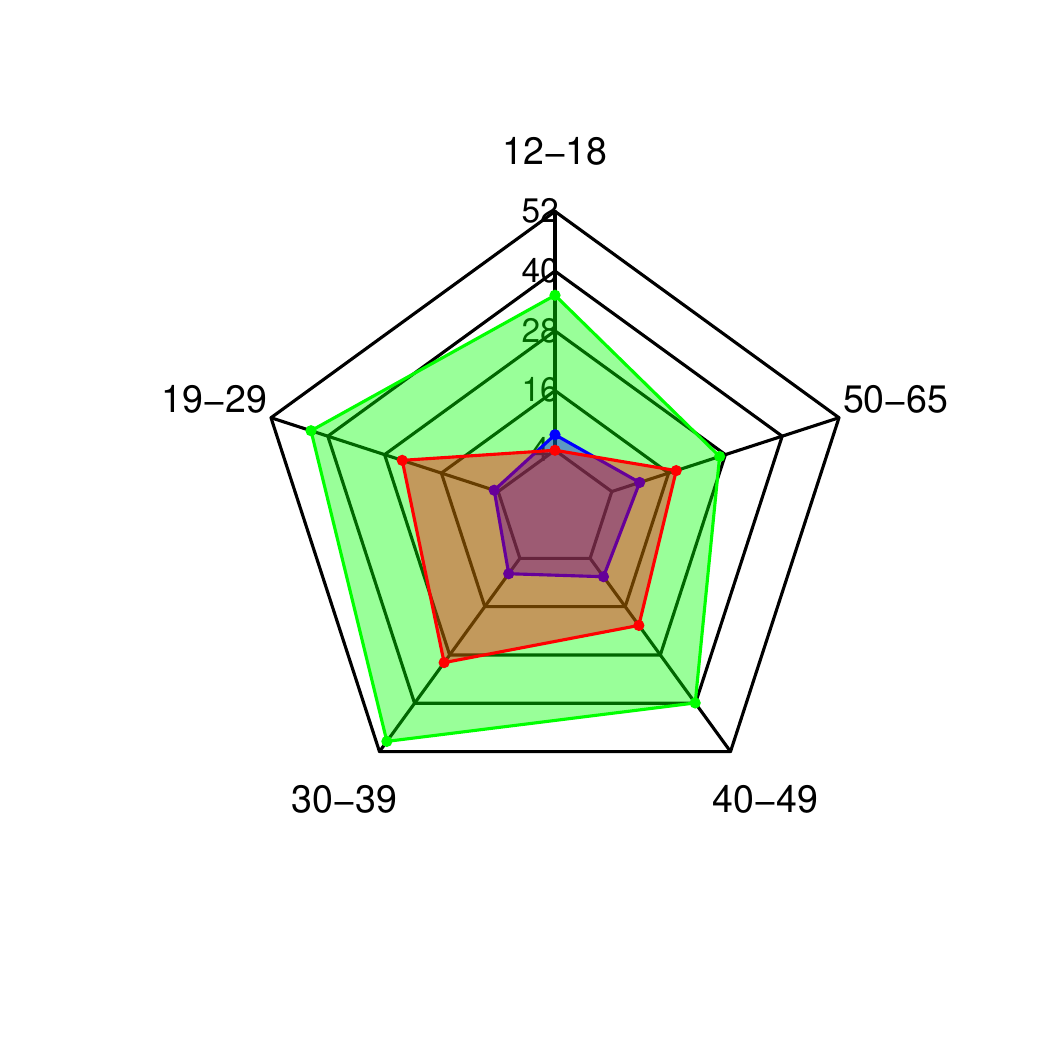}
        \caption{Average consumption}
        \label{Fig:Consumption_Radar}
    \end{subfigure}
    \begin{subfigure}{0.49\textwidth}
        \centering
        \includegraphics[width = \linewidth, trim={60pt 100pt 50pt 60pt}, clip]{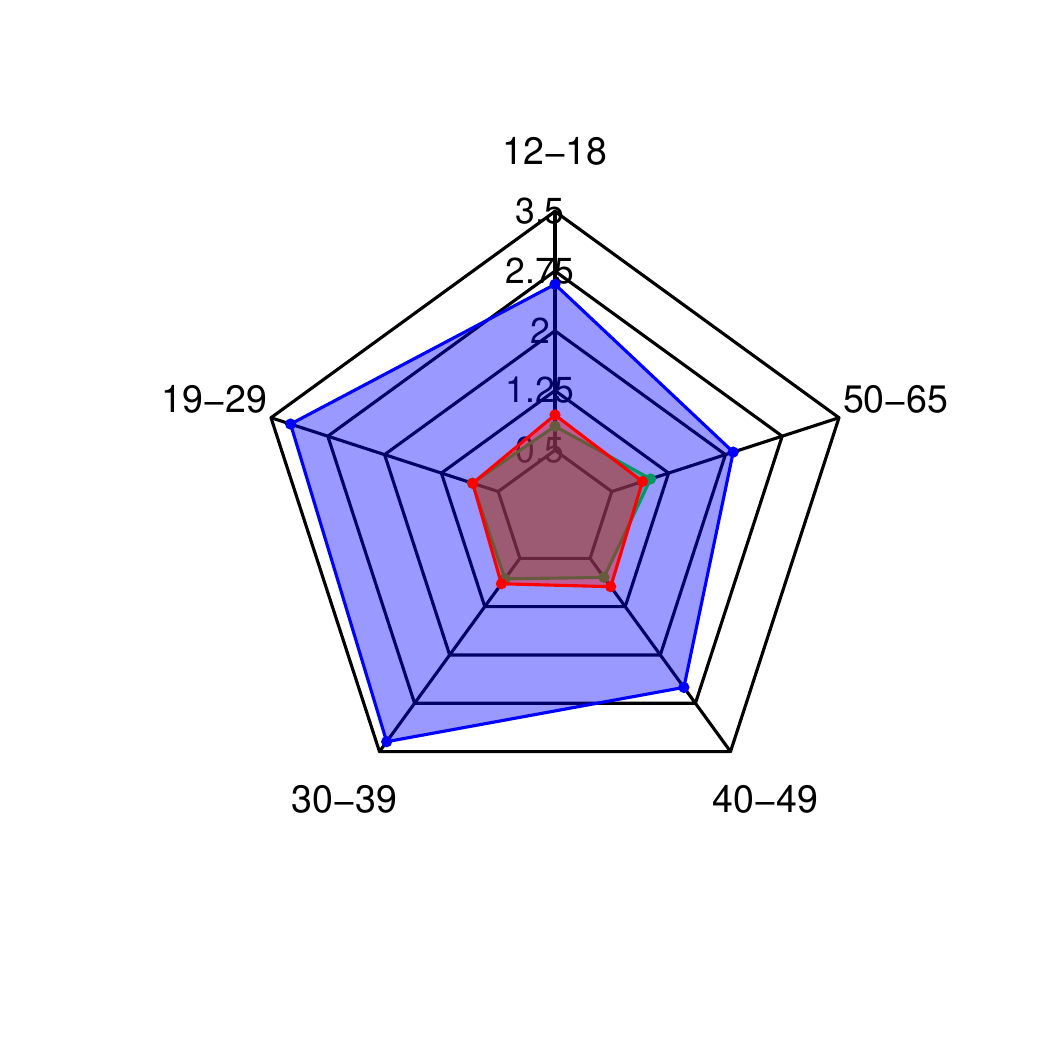}
        \caption{Average price}
        \label{Fig:Price_Radar}
    \end{subfigure}

    \definecolor{Rred}{HTML}{FF0000}
    \definecolor{Rgreen}{HTML}{00FF00}
    \definecolor{Rblue}{HTML}{0000FF}
    \vspace{-2mm} 
    \begin{minipage}{\textwidth}
        \centering
        \begin{tikzpicture}[line width=1pt]
            \draw[Rgreen] (0,0) -- (1,0); \node[anchor=west] at (1.1,0) {\small{Marijuana}};
            \draw[Rblue]  (4,0) -- (5,0); \node[anchor=west] at (5.1,0) {\small{Cocaine}};
            \draw[Rred] (8,0) -- (9,0); \node[anchor=west] at (9.1,0) {\small{Basuco}};
        \end{tikzpicture}
    \end{minipage}
    
    \begin{tablenotes}
        \item \small{\textbf{Notes:} Left panel: Average consumption per month in grams for those with positive consumption of each drug in the last 12 months. Right panel: Average price per gram in 2019 USD paid by those with positive consumption. There is some degree of heterogeneity regarding drug consumption by age group, with relatively homogeneous pricing among age groups. The average consumption amongst consumers of marijuana is the largest followed by basuco, except among teenagers. Cocaine is the most expensive drug, while marijuana and basuco have similar prices per gram.}
    \end{tablenotes}
\end{figure}
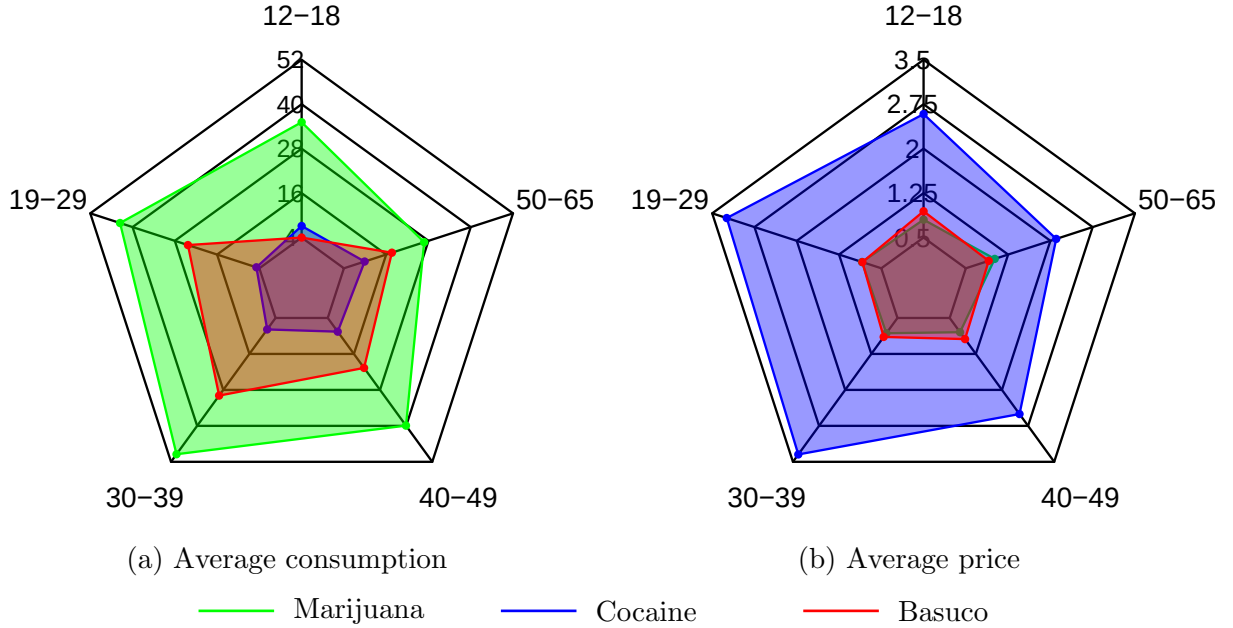

Due to the clear heterogeneity exhibited in both users and expenditure due to age, we consider the distribution of quantities and price similarly disaggregated by age groups for each drug in Figure \ref{Fig:Consumption_Prices_Radar}. In these figures we focus on those individuals who present only positive consumption to rule out the effects of the large outlier at zero shown in Table \ref{Tab:Summary_Statistics}. We observe that there is some degree of heterogeneity regarding consumption of marijuana and basuco according to Figure \ref{Fig:Consumption_Radar}. The average consumption of marijuana ranges between 25.1 (fifties) and 51.2 (thirties) joints per month, and the average consumption of basuco ranges between 2.0 (teenagers) and 29.9 (thirties) grams per month. On the other hand, the average consumption of cocaine is fairly homogeneous, around 6.7 grams per month, except for individuals in their twenties (4.5 grams per month). We note that individuals in their thirties have the largest consumption of all the three drugs, and individuals in their twenties have the second largest consumption of marijuana (43.2 joints) and basuco (20.1 grams).

Figure \ref{Fig:Price_Radar} shows the distribution of average prices by age group among regular users. There is still some small heterogeneity remaining for the prices per dose among age groups. The most expensive drug is cocaine, where its average price ranges between USD/gram 2.1 (fifties) and USD/gram 3.4 (thirties), with marijuana and basuco having similar prices per doses. Marijuana prices faced by consumers ranged between USD/gram 0.61 (teenagers) and USD/gram 1.02 (fifties), and basuco prices ranged between USD/gram 0.77 (teenagers) and USD/gram 0.94 (forties). It is concerning that teenagers get the lowest average prices of marijuana and basuco. Price variation remains at the individual level due to both observed and unobserved factors, which we exploit in our modeling strategy to construct accurate demand estimates.

\subsection{Consumer characteristics}

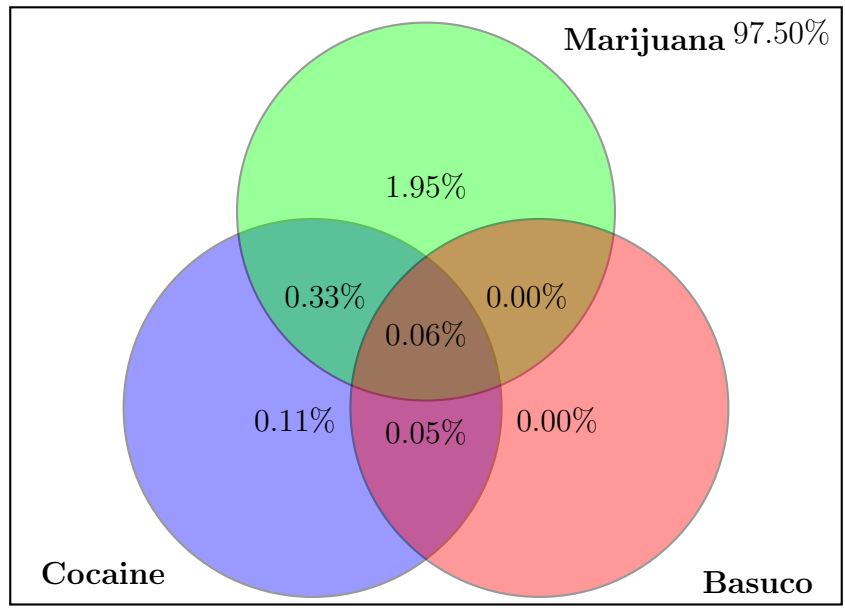
\begin{figure}[htbp]
    \centering
    \caption{Distribution of consumption bundles implicit in the \textit{ENCSPA} survey}
    \label{Fig:Drug_Consumption_Bundles}
    \begin{adjustbox}{max totalsize = {\textwidth}{\textheight}}
        \begin{tikzpicture}[thick]
        \tikzset{venn circle/.style = {draw, circle, minimum width = 5cm, fill = #1, opacity = 0.4}}
        \draw (-4, -2.6) rectangle (7, 5.3) node[below left] {97.50\%};
        \node [venn circle = blue, label = {[label distance = -3cm]200:0.11\%}] (A) at (0,0) {};
        \node [venn circle = green, label = {[label distance = -2.5cm]90:1.95\%}] (B) at (60:3cm) {};
        \node [venn circle = red, label = {[label distance = -3cm]340:0.00\%}] (C) at (0:3cm) {};
        \node[left] at (barycentric cs:A=1/2,B=1/1.5) {0.33\%}; 
        \node[below] at (barycentric cs:A=1/2,C=1/2) {0.05\%}; 
        \node[right] at (barycentric cs:B=1/1.5,C=1/2) {0.00\%}; 
        \node[below] at (barycentric cs:A=1/4,B=2/4,C=1/4){0.06\%};
        \node (Cocaine)[above left of=A, xshift=-5em, yshift=-7em]{\textbf{Cocaine}};
        \node (Marijuana)[above of=B, xshift=7em, yshift=3em]{\textbf{Marijuana}};
        \node (Basuco)[above of=C, xshift=7em, yshift=-8em]{\textbf{Basuco}};
    \end{tikzpicture}
    \end{adjustbox}
    \begin{tablenotes}
        \item \small{\textbf{Notes:} Total sample includes 49,439 individuals, representative of approximately 23.6 million, approximately the Colombian urban population aged between 12 and 65 years in 2019. Most of the individuals report that they do not consume any drug regularly. Individuals who only consume marijuana represent the largest share, followed by individuals who consume marijuana and cocaine, and then individuals who consume all three drugs.}
    \end{tablenotes}
\end{figure}

The previous analysis aggregated individuals according to their drug of choice to obtain market-level statistics. However, our analysis accounts for the fact that a user may consume multiple drugs. Figure \ref{Fig:Drug_Consumption_Bundles} shows the distribution of users across the consumption bundles implied by our considered drug choices (including irregular and non-consumers for reference). This figure reveals that 97.5\% of individuals report not regularly consuming any of the three drugs. Among the remaining 2.5\%, 2.34\% are regular marijuana users, 0.55\% consume cocaine, and 0.11\% consume basuco. The majority consume only marijuana (1.95\%), followed by those who consume both marijuana and cocaine regularly (0.33\%), those who exclusively consume cocaine (0.11\%), and those who consume all three drugs (0.06\%). Notably, the second-largest group consists of individuals who jointly consume marijuana and cocaine, suggesting potential complementarity between these substances on the extensive margin. In contrast, basuco users represent the smallest share, which aligns with expectations given its high addictiveness and severe adverse health effects.

Table \ref{Tab:Demographic_Statistics} presents descriptive statistics for demographic characteristics available in the survey. We see from this Table that the representative (modal) regular drug user is a male that has drug dealers in his neighborhood, lives in a low-socioeconomic stratum, his friends also use drugs, consumes alcohol and cigarettes jointly, is 29 years old, studied for 12 years, and is currently employed. This person spends on average USD 30.7 per month on drugs, has access to marijuana, cocaine, and basuco, has used marijuana per 11 years, reports having good physical and mental health, and has a high-risk perception about drugs.

\begin{table}[!htbp]
    \centering
    \caption{Descriptive statistics: individual characteristics}
    \label{Tab:Demographic_Statistics}
    \begin{adjustbox}{max totalsize = {\textwidth}{0.94\textheight}, center}
        \begin{threeparttable}
            \begin{tabular}{l c c c c }
                \toprule
                Variable & Mean & Std. Dev. & Min. & Max. \\
                \midrule
                \textbf{Access indicators} & & & & \\ 
                Marijuana & 0.99 & 0.05 & 0 & 1\\
                Cocaine & 0.79 &0.41  & 0 & 1 \\
                Basuco & 0.54 & 0.50 & 0 & 1\\
                Dealer presence & 0.51 & 0.50  & 0  & 1 \\
                Consumer in network & 0.94 &  0.24 & 0  & 1\\
                \textbf{Drug purchase sources indicators} & & & & \\
                Obtained marijuana online & 0.09 & 0.29 & 0 & 1 \\
                Obtained marijuana in person & 0.35 & 0.48 & 0 & 1 \\
                Obtained marijuana through friends & 0.60 & 0.49 & 0 & 1 \\
                Obtained cocaine online & 0.03 & 0.16 & 0 & 1 \\
                Obtained cocaine in person & 0.10 & 0.30 & 0 & 1 \\
                Obtained cocaine through friends & 0.10 & 0.30 & 0 & 1 \\
                Obtained basuco online & 0 & 0.03 & 0 & 1 \\
                Obtained basuco in person & 0.03 & 0.18 & 0 & 1 \\
                Obtained basuco through friends & 0.02 & 0.14 & 0 & 1 \\
                \textbf{Offering indicators} & & & & \\ 
                Marijuana & 0.89 &  0.31 & 0 & 1\\
                Cocaine & 0.58 & 0.49 & 0  & 1\\
                Basuco & 0.24 & 0.43 & 0  & 1\\
                \textbf{Risk perception indicators} & & & & \\
                Low risk & 0.02 & 0.14 & 0 & 1\\
                Medium risk & 0.05 & 0.22 & 0 & 1\\
                High risk & 0.93 & 0.26 & 0 & 1 \\
                \textbf{Time consuming drugs (years)} & & & & \\
                Marijuana & 10.8 & 9.80 & 0 & 50 \\
                Cocaine & 3.67 & 7.78 & 0 & 47\\
                Basuco & 1.44 & 6.12 & 0 & 45 \\
                \textbf{General and other demographics} \\
                Expenditure in drugs (monthly, 2019 USD) & 30.73 & 58.24 & 0.26 & 488.70\\
                Female & 0.25 & 0.43 & 0 & 1 \\
                Parents at home & 0.15 & 0.36 & 0 & 1 \\
                Years of education & 12.21 & 3.79 & 0 & 24\\
                Head of household & 0.39 & 0.49 & 0 & 1\\
                Working & 0.60 & 0.49 & 0 & 1 \\
                Age (years) & 28.8 & 10.43 & 13 & 65 \\
                Metropolitan area & 0.37 & 0.48 & 0 & 1\\
                Medicinal marijuana products & 0.39 & 0.49 & 0 & 1 \\
                Cigarette and alcohol consumption & 0.85 & 0.36 & 0 & 1 \\
                \textbf{Socioeconomic status (SES) indicators} & & & & \\ 
                Low SES & 0.57 & 0.49 & 0 & 1\\
                Medium SES & 0.37 & 0.48 & 0 & 1\\
                High SES & 0.06 & 0.23 & 0 & 1 \\
                \textbf{Health indicators} & & & & \\
                Good physical health & 0.78 & 0.41 & 0 & 1\\
                Good mental health & 0.69 & 0.46 & 0 & 1 \\
                \textbf{Geographically distance-weighted instruments} & & & \\
                Drug dealers captures in neighborhood (2018) & 481.67 & 659.76 & 0 & 3,587 \\
                Ellicited marijuana price (2019 USD) & 0.78 & 0.52 & 0.15 & 3.05 \\
                Ellicited cocaine price (2019 USD) & 2.94 & 1.45 & 0.76 & 9.16 \\
                \bottomrule
            \end{tabular}
        \begin{tablenotes}
            \item \footnotesize{\textbf{Notes:} Total sample size includes 1,236 consumers from ENCSPA survey in Colombia. The representative (modal) drug user has drug dealers in his neighborhood, lives in a low-socioeconomic stratum, his friends also use drugs, consumes alcohol and cigarettes, he is 29 years old, studied for 12 years, and works. This person spends on average USD 30.7 per month on drugs, and has access to marijuana, cocaine, and basuco. He seems to have a high-risk perception about drugs. This person has used marijuana per 11 years, and reports to have good physical and mental health.}
        \end{tablenotes}
    \end{threeparttable}
    \end{adjustbox}
\end{table}

The bottom set of variables in Table \ref{Tab:Demographic_Statistics} includes instruments derived from consumers’ geo-referenced locations and detailed geo-referenced data on drug dealer arrests from the previous year (2018). Using Gaussian kernels with a 1,000-meter bandwidth centered on each consumer’s coordinates, we compute a distance-weighted average of nearby arrests. We also construct average marijuana and cocaine prices using geo-referenced price data. On average, we find approximately 482 drug dealer captures, with a large standard deviation of 660 captures. This large number reflects the prevalence of drug dealing in Colombia and how concentrated it is in specific urban areas \citep{puerta2024spatial}. The marijuana and cocaine prices elicited from drug dens near consumers are very similar to those reported in Table \ref{Tab:Summary_Statistics}, but they are more precisely measured and are less prone to measurement error \citep{Zhen2014}. The role of these instruments is expanded upon in sections \ref{Sec:Methodology} and \ref{Sec:Results}.

To sum up, individuals in their twenties represent the largest market share, where marijuana is the most relevant expenditure (see Figure \ref{Fig:Drug_Expenditure}). Although there is some degree of heterogeneity in average consumption, where individuals in their thirties have the largest average consumption (see Figure \ref{Fig:Price_Radar}), it seems that the market composition is explained by the largest number of users being in their twenties (see Figure \ref{Fig:Drug_Users}).

\begin{figure}
    \centering
    \caption{Non-parametric Engel curves of demand for illicit drugs on full sample}
    \label{Fig:Engel_Descriptive}
    \begin{subfigure}{\textwidth}
        \centering
        \includegraphics[width = 0.58\textwidth, page = 1]{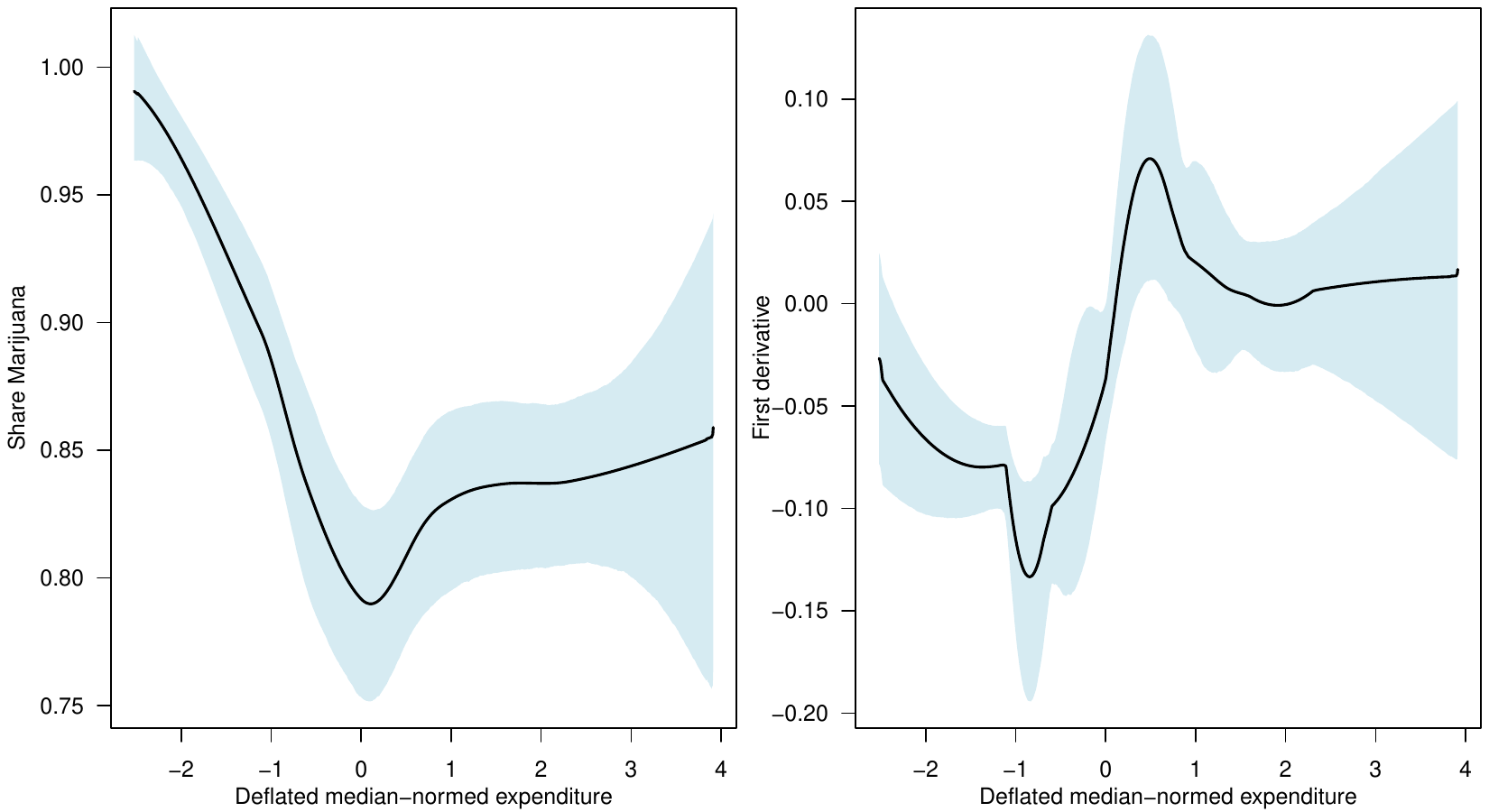}
        \caption{Marijuana}
        \label{Fig:Engel_Descriptive_Marijuana}
    \end{subfigure}
    
    \begin{subfigure}{\textwidth}
        \centering
        \includegraphics[width = 0.58\textwidth, page = 2]{Figures/Engel_Descriptive_FullSample.pdf}
        \caption{Cocaine}
        \label{Fig:Engel_Descriptive_Cocaine}
    \end{subfigure}
    
    \begin{subfigure}{\textwidth}
        \centering
        \includegraphics[width = 0.58\textwidth, page = 3]{Figures/Engel_Descriptive_FullSample.pdf}
        \caption{Basuco}
        \label{Fig:Engel_Descriptive_Basuco}
    \end{subfigure}
    \begin{tablenotes}
        \item \small{\textbf{Notes:} Engel curves estimated using a local regression procedure \citep{Cleveland2017}. First derivative computed using simple sorted differences along a grid of expenditure values. Confidence intervals are point-wise 95\% percentile-$t$ intervals from 5,000 bootstrap replications. Full sample Engel curves exhibit clear non-linearities that are not captured by limited-rank demand systems \citep{Lewbel2009}.}
    \end{tablenotes}
\end{figure}

Finally, we note that most parametric demand systems, such as the AID or QAID, place a restriction on the rank of the Engel curves (relationship between expenditure and consumption) such that only quadratic relationships can be modeled using these demand systems. This is not the case for the EASI demand system as it can deal with an arbitrarily large rank of the Engel curves. While this is a less restrictive assumption when studying the demand for more standard goods, the demand for illicit drugs presents non-linearities beyond the simple quadratic patterns in Engel curves recoverable by other demand systems. As shown in Figure \ref{Fig:Engel_Descriptive}, a non-parametric estimate of these Engel curves and their slopes using our full sample of consumers showcases highly non-linear patterns for drug consumption, which require flexible demand systems such as the EASI to be captured accurately.

\section{Econometric Framework}
\label{Sec:Methodology}
The EASI demand system \citep{Lewbel2009} is constructed to satisfy the standard microeconomic restrictions on demand functions. Specifically, this system satisfies the axioms of choice, such that additivity, homogeneity, and symmetry restrictions are easily imposed to perform estimation. Moreover, the rank in the function space spanned by the Engel curves can be more than three, such that the Engel curves may take flexible shapes as those found in granular demand data. This feature is absent in two of the most widely used demand systems in the literature: the almost ideal demand system \citep[AID;][]{Deaton1980} or quadratic AID \citep[QAID;][]{Banks1997}. These properties are particularly relevant in our case given the rank of Engel curves for illicit drugs as shown in Figure \ref{Fig:Engel_Descriptive} and the fact we work with micro-level data where variation in expenditure is not smoothed out by aggregation \citep{Blundell2007, Zhen2014}.

\subsection{Review: the EASI model}\label{sec:sumEASI}

Details of the EASI demand system can be found in \cite{Lewbel2009}. We outline the basic framework here for convenience of exposition. In our application, we simultaneously model $S = 3$ implicit Marshallian budget shares, given by $\bm{\omega}(\widetilde{\bm{p}}_i, y_i, \bm{h}_i, \widetilde{\bm{\varepsilon}}_i) \coloneqq \widetilde{\bm{w}}_i = [\text{marijuana}_i \, , \text{cocaine}_i \, , \text{basuco}_i ]\T$ for individuals $i=1,2,\dots, n$, where
\begin{align}
    \label{Eq:EASI_Full} \widetilde{\bm{w}}_i & = \sum_{r=0}^R \widetilde{\bm{b}}_r y_i^r + \widetilde{\bm{A}}_0 \widetilde{\bm{p}}_i + \sum_{m=1}^{M_p} \widetilde{\bm{A}}_m \widetilde{\bm{p}}_i h_{im}^p + \widetilde{\bm{B}} \widetilde{\bm{p}}_i y_{i} + \widetilde{\bm{C}} \bm{h}_i + \widetilde{\bm{D}} \bm{h}_i^y y_i + \widetilde{\bm{\varepsilon}}_i , \\
    \label{Eq:Nonlinear_Indirect_Utility}
    y_i & = \frac{e_i - \widetilde{\bm{p}}_i\T \widetilde{\bm{w}}_i - \sum_{m=0}^M \widetilde{\bm{p}}_i\T \widetilde{\bm{A}}_m \widetilde{\bm{p}}_i h_{im}^p/2}{1 - \widetilde{\bm{p}}_i\T \widetilde{\bm{B}} \widetilde{\bm{p}}_i/2}.
\end{align}
Implicit utility $y_i$ is an exact affine transformation of the (log) Stone index given by $\widetilde{\bm{p}}_i\T \widetilde{\bm{w}}_i$. Additionally, $e_i$ denotes log-nominal expenditure, and $\widetilde{\bm{p}}_i$ is the log-price vector faced by individual $i$.\footnote{We emphasize that our data allow us to obtain variability in individual price levels, which is not common as in most demand applications consumers face an aggregate measure of price instead.} The system of equations in \eqref{Eq:EASI_Full} can involve polynomials of arbitrary degree $R$ in $y$, providing flexibility for the Engel curves.

The EASI model naturally allows for sources of \emph{observable} heterogeneity through the inclusion of socioeconomic controls $\bm{h}_i$ and variables that interact with prices ($\bm{h}_i^p$) and implicit utility ($\bm{h}_i^y$). Each of these vectors of controls, with dimensions $M$, $M_p$, and $M_y$ respectively, can contain distinct variables from one another or could be empty (where $h_{i0} = h_{i0}^p = h_{i0}^y \coloneqq 1$ and these first elements are not included in the vector definitions). The stochastic error $\widetilde{\bm{\varepsilon}}_i$, on the other hand, can directly be interpreted as a source of \emph{unobserved} preference heterogeneity. However, this heterogeneity does not affect the elasticities or Engel curves, as these objects depend on parameters and observable characteristics. Table \ref{Tab:EASI_Summaries} presents a complete depiction of all price and expenditure effects that can be derived from the EASI demand system. 

To satisfy standard microeconomic regularity conditions in the EASI model, one can impose the following restrictions on the coefficients in system \eqref{Eq:EASI_Full}: $\widetilde{\bm{A}}_m\bm{1}_S = \widetilde{\bm{B}} \bm{1}_S = \bm{0}_S$ for cost function homogeneity; the unit-sum constraint of shares requires $\bm{1}_S\T \widetilde{\bm{b}}_0 = 1$, $\bm{1}_S\T \widetilde{\bm{b}}_r = 0$ for $r = 1, \ldots, R$, $\bm{1}_S\T \widetilde{\bm{A}}_m = \bm{0}_S\T$ for $m = 0, \ldots, M_p$, $\bm{1}_S\T \widetilde{\bm{B}} = \bm{0}_S\T$, $\bm{1}_S\T \widetilde{\bm{C}} = \bm{0}_{M}\T$, $\bm{1}_S\T \widetilde{\bm{D}} = \bm{0}_{M_y}\T$ and $\bm{1}_S\T \widetilde{\bm{\varepsilon}}_i = 0$ for $i = 1, \ldots, n$; Slutsky symmetry requires all $\widetilde{\bm{A}}_m$ matrices for $m = 0, \ldots, M_p$ and $\widetilde{\bm{B}}$ to be symmetric; strict cost monotonicity requires $\widetilde{\bm{p}}\T \left[\sum_{r=0}^R \widetilde{\bm{b}}_r r y^{r-1}+\widetilde{\bm{D}} \bm{h}^y + \widetilde{\bm{B}} \bm{p} / 2\right] + 1 > 0$; finally, a sufficient and necessary condition for concavity of the cost function is negative semi-definiteness of the normalized Slutsky matrix $\sum_{m=0}^{M_p} \widetilde{\bm{A}}_m h_m^p+\widetilde{\bm{B}} y + \widetilde{\bm{w}} \widetilde{\bm{w}}\T - \bm{W}$, where $\bm{W}$ is a diagonal matrix whose diagonal equals $\widetilde{\bm{w}}$, and $\bm{0}_S$ and $\bm{1}_S$ are $S$-dimensional vector of zeros and ones, respectively.


\subsection{Model specfication}
We begin by specifying the system of equations \eqref{Eq:EASI_Full} in latent form using the last share ($S$) as the base category or numeraire good (basuco in our case given its generally low expenditure share). These latent shares, denoted by $\widetilde{\bm{w}}_i^* = \left[w_{i1}^*, \ldots, w_{iS}^*\right]\T$, capture the un-normalized marginal utility of individual $i$ of consuming each of the goods considered (illicit drugs in our case). We can then recover observable shares $\bm{w}_i$ from latent ones \citep{Kasteridis2011, RamirezHassan2021}:
\begin{equation}
	w_{il} = \begin{cases}
		\frac{w_{il}^*}{\sum_{j \in L_i} w_{ij}^*} & \text{ if } w_{il}^* > 0 \, , \\
		0 & \text{ if } w_{il}^* \leq 0  \, ,
	\end{cases} \quad \text{for } l = 1, \ldots, S \, ,
\end{equation}
where $L_i = \{l: w_{il}^* > 0\} = \{l: w_{il} > 0\}$ is the set of drugs with positive consumption for individual $i$. Imposing the previously discussed microeconomic restrictions onto \eqref{Eq:EASI_Full} yields an EASI demand system specified directly on latent shares:
\begin{equation}
    \label{Eq:EASI_reduced}
    \bm{w}_i^* = \sum_{r=0}^R \bm{b}_r y_i^r + \sum_{m=0}^{M_p} \bm{A}_m \bm{p}_i h_{im}^p + \bm{B} \bm{p}_i y_{i} + \bm{C} h_i + \bm{D} \bm{h}_i^y y_i + \bm{\varepsilon}_i,
\end{equation}
where these variables and coefficients are defined from the original quantities as
\begin{align*}
    \begin{gathered}
        \underset{(S \times 1)}{\widetilde{\bm{w}}_i} \coloneqq \begin{bmatrix} \bm{w}_i \\ w_{iS} \end{bmatrix}, \quad  \underset{(S \times 1)}{\widetilde{\bm{w}}_i^*} \coloneqq \begin{bmatrix} \bm{w}_i^* \\ w_{iS}^* \end{bmatrix} , \quad  \underset{(S \times 1)}{\widetilde{\bm{\varepsilon}}_i} \coloneqq \begin{bmatrix} \bm{\varepsilon}_i \\ \varepsilon_{iS} \end{bmatrix}, \quad \underset{(s \times 1)}{\bm{p}_i} \coloneqq \begin{bmatrix} p_{i1} - p_{iS} \\ \vdots \\ p_{is} - p_{iS} \end{bmatrix} \, , \\
        \underset{(S \times 1)}{\widetilde{\bm{b}}_r} \coloneqq \begin{bmatrix} \bm{b}_r \\ b_{rS} \end{bmatrix}, \text{ for } r = 0, \ldots, R, \quad \underset{(S \times S)}{\widetilde{\bm{A}}_m} \coloneqq \begin{bmatrix} \bm{A}_m & \bm{A}_{m,S1} \\ \bm{A}_{m,1S} & A_{m,SS} \end{bmatrix}, \text{ for } m = 0, \ldots, M_p \, , \\
        \underset{(S \times S)}{\widetilde{\bm{B}}} \coloneqq \begin{bmatrix} \bm{B} & \bm{B}_{S1} \\ \bm{B}_{1S} & B_{SS} \end{bmatrix}, \quad \underset{(S \times M)}{\widetilde{\bm{C}}} \coloneqq \begin{bmatrix} \bm{C} \\ \bm{C}_S\T \end{bmatrix}, \quad \text{and} \quad \underset{(S \times M_y)}{\widetilde{\bm{D}}} \coloneqq \begin{bmatrix} \bm{D} \\ \bm{D}_S\T \end{bmatrix} \, .
    \end{gathered}
\end{align*}
Here, $\bm{p}_i$ represents the vector of relative log prices with respect to the base price $p_{iS}$, and Slutsky symmetry also implies $\bm{A}_m = \bm{A}_m\T$ for $m = 0, \ldots, M_p$ and $\bm{B} = \bm{B}\T$. Note that the unit-sum, cost function homogeneity, and Slutsky symmetry restrictions allow us to recover the share and coefficients from the base category in terms of the information from remaining goods. That is, once we impose these restrictions, we only need to model $s \coloneqq S - 1$ of the shares (marijuana and cocaine).

Our estimation framework also takes endogeneity issues into account. A first source of endogeneity arises mechanically from \eqref{Eq:Nonlinear_Indirect_Utility}, with budget shares $\bm{w}_i$ used to construct indirect utility $y_i$, creating simultaneous causality. However, simple valid instruments can be constructed and it has been documented in the literature that this endogeneity is numerically negligible \citep{Lewbel2009, Zhen2014}. Reverse causality between prices and quantities is also not of general concern when working with micro-level data, as one can argue individual purchase decisions should not affect aggregate market prices.

A final source of endogeneity comes through the relative prices $\bm{p}_i$ due to omitted variables and measurement errors; likely to be more relevant in a micro setting as these are not averaged out when aggregated. For instance, omitted variables can arise through the strategic search of consumers when looking for drug providers, particularly in markets where individuals declare to have easy access to drugs, which translates to relatively good price information and potentially many providers \citep[see][for a structural market model for illicit drugs]{Galenianos2017}. Additionally, the prices provided in our data are self-reported by individuals, meaning they can be subject to non-classical measurement errors created by recall, socially desirable responses, or other self-reporting biases \citep{Embree1993, Johnson2005, Fadnes2009, Steenkamp2010, Rosenman2011}.

Let $\bm{x}_i \coloneqq [1, y_i, \ldots, y_i^R, \bm{h}_i\T, \bm{h}_i^{y\top} y_i]\T$ collect all exogenous variables in a vector of dimension $1+R+M+M_y$ and $\bm{p}_i^* \coloneqq [\bm{p}_i\T h_{i0}^p, \ldots, \bm{p}_i\T h_{iM_p}^p, \bm{p}_i\T y_i]\T$ collect all endogenous variables in a vector of dimension $d^* \coloneqq s (M_p + 2)$. To counteract these sources of endogeneity, we assume we have access to an $\ell$-dimensional vector $\bm{z}_i$ of excluded instruments that are uncorrelated with $\bm{\varepsilon}_i$, are relevant to predict $\bm{p}_i^*$, and we have enough instruments to satisfy $\ell \geq d^*$ as a necessary condition for identification.\footnote{The endogenous $\bm{p}_i^*$ is composed of prices $\bm{p}_i$ and cross-products of $\bm{p}_i$ with exogenous variables $\bm{h}_i^{p}$ and $y_i$. If we have $l$ credible instruments for prices (denoted by $\widetilde{\bm{z}}_i$) with $l \geq s$, then $\widetilde{\bm{z}}_i h_{01}^p, \ldots, \widetilde{\bm{z}}_i h_{iM_p}^p$, and $\widetilde{\bm{z}}_iy_i$ are valid instruments as well as long as they contain sufficient variation across individuals.} We group all structural information in equation \eqref{Eq:EASI_reduced} into an $s \times d_{\beta}$ matrix $\bm{F}_i \coloneqq [\bm{I}_s \otimes x_i\T , \, (\bm{I}_s \otimes \bm{p}_i \T)\bm{D}_s h_{i0}^p, \, \cdots , \, (\bm{I}_s \otimes \bm{p}_i\T) \bm{D}_s h_{iM_p}^p, \, (\bm{I}_s \otimes \bm{p}_i\T) \bm{D}_s y_i]$ and all first-stage information into an $d^* \times d_{\gamma}$ matrix $\bm{G}_i \coloneqq [\bm{I}_{d^*} \otimes x_i\T, \, \bm{I}_{d^*} \otimes z_i\T]$.\footnote{$\bm{D}_s$ is defined as the $s^2 \times s(s+1)/2$ duplication matrix such that $\bm{D}_s\vecv(\bm{A}) = \vech(\bm{A})$ for any symmetric $s \times s$ matrix $\bm{A}$. Post-multiplication by this matrix comes from the Slutsky symmetry restriction. Additional details for deriving these stacked expressions are provided in Appendix \ref{apx_sec:easi_stacked} and \ref{apx_sec:easi_computational}.} This results in the following EASI system with endogeneity:
\begin{align}
    \label{Eq:Joint_Linear_System}
	\begin{aligned}
		\bm{w}_i^* & = \bm{F}_i \bm{\beta} + \bm{\varepsilon}_i \, , \\
		\bm{p}_i^* & = \bm{G}_i \bm{\gamma} + \bm{u}_i \, ,
	\end{aligned}
\end{align}
where $\bm{\beta}$ is the structural coefficient of interest with dimension $d_{\beta} \coloneqq s(1+R+M+M_y) + d^*(s+1)/2$, and $\bm{\gamma}$ are the first-stage coefficients with dimension $d_{\gamma} \coloneqq d^*(1+R+M+M_y+\ell)$. To reproduce the endogeneity in the system, we place a multivariate normal distribution with non-diagonal covariance matrix $\bm{\Sigma}$ between $\bm{\varepsilon}_i$ and $\bm{u}_i$.

To acknowledge further \emph{unobserved} heterogeneity, we allow for individuals to differ according to unobserved types, where drug demand responses vary across types and are similar for all individuals of the same type. That is, we are assuming that our sample is representative of a population composed by sub-populations, with homogeneous drug preferences within each group and heterogeneity across them. Introduce an individual cluster indicator $\psi_i \in \{1, \ldots, J\}$ such that $\psi_i = j$ means the observation belongs to cluster $C_j$, for $j = 1, \ldots, J$. We will assume $J$, the number of clusters, to be known and experiment with its value in our applications.

As the structural coefficients capture the drug preferences of individuals, we allow for all elements in the structural equation ($\bm{\beta}$ and the relevant components of $\bm{\Sigma}$) to vary at the cluster-level. However, we do not allow for the first-stage parameters ($\bm{\gamma}$ and $\bm{\Sigma}_{u u}$) to vary with the clusters as it is unlikely that the same population segments driving heterogeneity in drug preferences also drive heterogeneity in the reduced form. Additionally, using cluster-specific first-stage regressions do not allow the model to exploit the full variability in instruments to recover these coefficients, leading to artificial issues of weak instruments and larger uncertainty. We incorporate this into our model by assuming
\begin{equation}
    \label{Eq:Error_Distribution}
	\left. \begin{bmatrix} \bm{\varepsilon}_i \\ \bm{u}_i \end{bmatrix} \right| \psi_i = j \overset{iid}{\sim} \N_{s(M_p+3)}(\bm{0}_{s(M_p+3)}, \bm{\Sigma}_j) \quad \text{with} \quad \bm{\Sigma}_j = \begin{bmatrix} \bm{\Sigma}_{j, \varepsilon \varepsilon} & \bm{\Sigma}_{j, \varepsilon u} \\ \bm{\Sigma}_{j, u \varepsilon} & \bm{\Sigma}_{u u} \end{bmatrix}
\end{equation}
independently across individuals, where $\N(\cdot\mid\mu, \bm{\Sigma})$ represents the density function of a normally distributed variable with mean $\mu$ and variance $\bm{\Sigma}$. Finally, we note that by also including priors for both the cluster indicators ($\psi_1, \ldots, \psi_n$) and assignment probabilities ($\psi_1, \ldots, \psi_J$), their updated posterior produce a data-driven probabilistic assignment of individuals to clusters. This is an important policy tool, as we will see that our model meaningfully classifies individuals according to their drug-use behavior and risk. Additionally, as we correlate the characteristics of these individuals to the model's assignment, it allows us to preemptively identify individuals who are highly at-risk due to their drug use.

\subsection{Bayesian estimation}

We implement a Bayesian inferential framework that allows us to simultaneously handle all key elements of this framework: heterogeneity in structural preferences, censoring at corner solutions, endogeneity, and imposition of microeconomic restrictions. Collect all model parameters into $\bm{\theta} \coloneqq (\bm{\beta}_1, \ldots, \bm{\beta}_J, \bm{\gamma}, \bm{\Sigma}_1, \ldots, \bm{\Sigma}_J, \psi_1, \ldots, \psi_n, \phi_1, \ldots, \phi_J, \bm{w}^*_1, \ldots, \bm{w}^*_n)$, augmented with latent shares. The point of departure for Bayesian analysis is a prior probabilistic belief about unknown parameters, which is updated using sample information $\mathcal{D} = \{\bm{w}_i, \bm{p}^*_i, \bm{F}_i, \bm{G}_i\}_{i=1}^{n}$. Letting $p(\mathcal{D} \mid \bm{\theta})$ represent the likelihood function and $\pi(\bm{\theta})$ be the prior distribution, we can use Bayes' rule to obtain the posterior distribution as
\begin{equation*}
    \pi(\bm{\theta}|\mathcal{D}) \propto p(\mathcal{D} \mid \bm{\theta}) \times \pi(\bm{\theta}) \, .
\end{equation*}
Under the distribution for the error terms provided in \eqref{Eq:Error_Distribution}, the contribution to the likelihood by individual $i$ assigned to an arbitrary cluster $\psi_i = j$ is a function of cluster-specific parameters $\bm{\beta}_j, \bm{\Sigma}_j$ and $\bm{\gamma}$ given by the joint distribution over latent shares $\bm{w}_i^*$ and endogenous variables $\bm{p}_i^*$:
\begin{equation}
	\label{Eq:LatentLik}
	p(\bm{w}_i^*, \bm{p}_i^* \mid \psi_i = j, \bm{\beta}_j, \bm{\gamma}, \bm{\Sigma}_j) \propto |\bm{\Sigma}_j|^{-\frac{1}{2}} \exp\left\{-\frac{1}{2} \begin{bmatrix} \bm{w}_i^* - \bm{F}_i \bm{\beta}_j \\ \bm{p}_i^* - \bm{G}_i \bm{\gamma} \end{bmatrix}\T 
    \begin{bmatrix} \bm{\Sigma}_{j, \varepsilon \varepsilon} & \bm{\Sigma}_{j, \varepsilon u} \\ \bm{\Sigma}_{j, u \varepsilon} & \bm{\Sigma}_{u u} \end{bmatrix}^{-1} \begin{bmatrix} \bm{w}_i^* - \bm{F}_i \bm{\beta}_j \\ \bm{p}_i^* - \bm{G}_i \bm{\gamma} \end{bmatrix} \right\}.
\end{equation}
The likelihood function $p(\mathcal{D} \mid \bm{\theta})$ is then the product over all contributions \eqref{Eq:LatentLik} across individuals $i = 1, \ldots, n$. For a full Bayesian specification, we are simply left with providing priors for estimation. Motivated by the structure of our problem, we assume conditionally conjugate priors with an added twist to the specification: we allow for homogeneous first-stage coefficients while maintaining heterogeneous structural demand preferences (see Eq. \ref{Eq:Priors} and surrounding discussion for details).\footnote{We also implement heterogeneous first-stage regressions at the same level as the structural equations and provide results below as a comparison.} Combining the likelihood and priors results in the following conditional posterior distributions (where the notation $\bm{\theta}_{-\delta}$ represents the vector of parameters $\bm{\theta}$ with component $\bm{\delta}$ removed):
\begin{align}
    \bm{\beta}_j \mid \bm{\theta}_{-\beta_j}, \mathcal{D} & \overset{iid}{\sim} \N_{d_{\beta}}(\Bar{\bm{\beta}}_j, \Bar{\bm{B}}_j) \, , \quad j = 1, \ldots, J,  \label{Eq:Posteriors1} \\
    \bm{\gamma} \mid \bm{\theta}_{-\gamma}, \mathcal{D} & \sim \N_{d_{\gamma}}(\Bar{\bm{\gamma}}, \Bar{\bm{\Gamma}}) \, , \label{Eq:Posteriors2} \\
    \psi_i \mid \bm{\theta}_{-\psi_i}, \mathcal{D} & \overset{iid}{\sim} \Cat(\Bar{\bm{\phi}}_i) \, , \quad i = 1, \ldots, n, \label{Eq:Posteriors3} \\
    \bm{\phi} \mid \bm{\theta}_{-\phi}, \mathcal{D} & \sim \Dir(\Bar{\bm{\alpha}}) \, , \label{Eq:Posteriors4} \\
    \pi(\bm{\Sigma} \mid \bm{\theta}_{-\Sigma}, \mathcal{D}) & = \IW(\bm{\Sigma}_{u u} \mid \Bar{\nu}_{u u}, \Bar{\bm{R}}_{u u}) \times \prod_{j=1}^{J} \IW(\bm{\Sigma}_{j, \varepsilon \varepsilon \cdot u} \mid \Bar{\nu}_j, \Bar{\bm{R}}_{j}) \mathcal{MN}(\bm{\Sigma}_{j, u \cdot u \varepsilon} \mid \Bar{\bm{M}}_j, \Bar{\bm{U}}_j, \bm{\Sigma}_{j, \varepsilon \varepsilon \cdot u}) \, . \label{Eq:Posteriors5}
\end{align}
The posterior hyperparameters in each of these expressions are given by:
\begin{align}
    & \begin{aligned}
        \label{Eq:Posterior_Pars_beta}
        \Bar{\bm{B}}_j & \coloneqq \left( \uBar{\bm{B}}_j^{-1} + \sum_{i \in C_j} \bm{F}_i\T \bm{\Sigma}_{j, \varepsilon \varepsilon \cdot u}^{-1} \bm{F}_i \right)^{-1} \, , \quad j = 1, \ldots, J \\
        \Bar{\bm{\beta}}_j & \coloneqq \Bar{\bm{B}}_j \left\{ \uBar{\bm{B}}_j^{-1} \uBar{\bm{\beta}}_j + \sum_{i \in C_j} \bm{F}_i\T \bm{\Sigma}_{j, \varepsilon \varepsilon \cdot u}^{-1} \left[ \bm{w}_i^* - \bm{\Sigma}_{j, u \cdot u \varepsilon}\T (\bm{p}_i^* - \bm{G}_i \bm{\gamma}) \right] \right\}, \quad j = 1, \ldots, J \, , \\
        \bm{\Sigma}_{j, \varepsilon \varepsilon \cdot u} & \coloneqq \bm{\Sigma}_{j, \varepsilon \varepsilon} - \bm{\Sigma}_{j, \varepsilon u} \bm{\Sigma}_{u u}^{-1} \bm{\Sigma}_{j, \varepsilon u}\T, \quad j = 1, \ldots, J \, ,
    \end{aligned} \\
    & \begin{aligned}
        \label{Eq:Posterior_Pars_gamma}
        \Bar{\bm{\Gamma}} & \coloneqq \left[\uBar{\bm{\Gamma}}^{-1} + \sum_{j = 1}^{J} \sum_{i \in C_j} \bm{G}_i\T \bm{\Sigma}_{j, u u \cdot \varepsilon}^{-1} \bm{G}_i \right]^{-1}, \\
        \Bar{\bm{\gamma}} & \coloneqq \Bar{\bm{\Gamma}} \left\{ \uBar{\bm{\Gamma}}^{-1} \uBar{\bm{\gamma}} + \sum_{j = 1}^{J} \sum_{i \in C_j} \bm{G}_i\T \bm{\Sigma}_{j, u u \cdot \varepsilon}^{-1} \left[\bm{p}_i^* - \bm{\Sigma}_{j, \varepsilon \cdot \varepsilon u}\T (\bm{w}_i^* - \bm{F}_i \bm{\beta}_j) \right] \right\}, \\
        \bm{\Sigma}_{j, \varepsilon \varepsilon \cdot u} & \coloneqq \bm{\Sigma}_{u u} - \bm{\Sigma}_{j, \varepsilon u}\T \bm{\Sigma}_{j, \varepsilon \varepsilon}^{-1} \bm{\Sigma}_{j, \varepsilon u}, \quad j = 1, \ldots, J \, ,
    \end{aligned} \\
    & \begin{aligned}
        \label{Eq:Posterior_Pars_mixture}
        \Bar{\phi}_{ij} & \coloneqq \frac{\phi_j \mathcal{N}_{ij}}{\sum_{j = 1}^J \phi_j \mathcal{N}_{ij}}, \quad i = 1, \ldots, n, \quad j = 1, \ldots, J \, , \\
        \mathcal{N}_{ij} & \coloneqq |\bm{\Sigma}_j|^{-1/2} \exp \left\{-\frac{1}{2} \begin{bmatrix} \bm{w}_i^* - \bm{F}_i \bm{\beta}_j \\ \bm{p}_i^* - \bm{G}_i \bm{\gamma} \end{bmatrix}\T \bm{\Sigma}_j^{-1} \begin{bmatrix} \bm{w}_i^* - \bm{F}_i \bm{\beta}_j \\ \bm{p}_i^* - \bm{G}_i \bm{\gamma} \end{bmatrix} \right\} , \quad i = 1, \ldots, n, \quad j = 1, \ldots, J \, , \\
        \Bar{\alpha}_j & \coloneqq \uBar{\alpha}_j + n_j, \quad j = 1, \ldots, J \, ,
    \end{aligned} \\
    & \begin{aligned}
        \label{Eq:Posterior_Pars_Sigma}
        \Bar{\nu}_{u u} & \coloneqq \uBar{\nu}_{u u} - s + n \, , \\
        \Bar{\bm{R}}_{u u} & \coloneqq \uBar{\bm{R}}_{u u} + \sum_{i=1}^{n} (\bm{p}_i^* - \bm{G}_i \bm{\gamma}) (\bm{p}_i^* - \bm{G}_i \bm{\gamma})\T \, , \\
        \Bar{\nu}_j & \coloneqq \uBar{\nu}_j + n_j, \quad j = 1, \ldots, J \, , \\
        \Bar{\bm{R}}_j & \coloneqq \uBar{\bm{R}}_{j, \varepsilon \varepsilon \cdot u} - \Bar{\bm{M}}_j\T \Bar{\bm{U}}_j^{-1} \Bar{\bm{M}}_j + \sum_{i \in C_j} (\bm{w}_i^* - \bm{F}_i \bm{\beta}_j) (\bm{w}_i^* - \bm{F}_i \bm{\beta}_j)\T, \quad j = 1, \ldots, J \, , \\
        \Bar{\bm{U}}_j & \coloneqq \left[ \uBar{\bm{R}}_{u u} + \sum_{i \in C_j} (\bm{p}_i^* - \bm{G}_i \bm{\gamma}) (\bm{p}_i^* - \bm{G}_i \bm{\gamma})\T \right]^{-1}, \quad j = 1, \ldots, J \, , \\
        \Bar{\bm{M}}_j & \coloneqq \Bar{\bm{U}}_j \left[\uBar{\bm{R}}_{j, u \cdot u \varepsilon} + \sum_{i \in C_j} (\bm{p}_i^* - \bm{G}_i \bm{\gamma}) (\bm{w}_i^* - \bm{F}_i \bm{\beta}_j)\T \right], \quad j = 1, \ldots, J \, .
    \end{aligned}
\end{align}
Based on these expressions, we provide a Gibbs sampler that can obtain draws from the joint posterior of all parameters. First, conditional on a given value of the latent shares, we obtain a new draw of the model parameters using (\ref{Eq:Posteriors1})-(\ref{Eq:Posteriors5}). We then draw the latent shares with zero consumption from conditional on those with positive consumption using \eqref{Eq:PosteriorShares} and re-compute all latent shares. Repeating this process $S$ times leaves us with a chain of posterior draws $(\bm{\theta}^{(1)}, \ldots, \bm{\theta}^{(S)})$ that we can use to summarize model estimates. Additional details on the computational implementation of our algorithm can be found in Appendix \ref{Sec:Appendix_Computation}.

\section{Results}
\label{Sec:Results}
We now explore the results of applying our previously described methodology to the analysis of demand for illicit drugs in Colombia. We use the full sample of 1,236 consumers contained in the nationally representative 2019 \textit{ENCSPA} survey to provide Bayesian inference of the EASI demand system specified by equations \eqref{Eq:Joint_Linear_System}, \eqref{Eq:Error_Distribution} and \eqref{Eq:Posteriors1} to \eqref{Eq:Posteriors5}. In the specification, we include as exogenous information all the variables whose descriptive statistics are presented in Table \ref{Tab:Demographic_Statistics} except for the geographically distance-weighted variables. To save on degrees of freedom, and as recommended by \cite{Lewbel2009}, we include these demographics directly in the equation and only explore interactions between price and indirect utility, rather than including additional observed heterogeneity that can greatly increase the size of the estimated demand system.

We tackle potential issues of endogeneity in our application by using instrumental variables based on two sources of information. First, the number of drug-related captures in the neighborhood of the individual allows us to consider supply-side effects exploiting random variation in availability of drug dealers. Second, we use a secondary source of prices to deal with potential search effects (interactions between consumers and dealers) and misreporting by individuals, where these instruments are prices imputed as geographically-weighted averages of the prices in drug dens. As the prices in these locations is both standardized and less prone to measurement error, we can control for additional demand-side variation in consumer prices \citep[a similar strategy is used in][]{Zhen2014}. 

In our implementation, we set the prior hyperparameters to standard non-informative values: for $j = 1, \ldots, J$, $\uBar{\bm{\beta}}_j = \bm{0}_{d_{\beta}}$, $\uBar{\bm{B}}_j = 1000 \bm{I}_{d_{\beta}}$, $\uBar{\bm{\gamma}} = \bm{0}_{d_{\gamma}}$, $\uBar{\bm{\Gamma}} = 1000 \bm{I}_{d_{\gamma}}$, $\uBar{\alpha} = (1/J) \bm{1}_J$, $\uBar{\nu}_j = s(M_p+3)$, $\uBar{\nu}_{uu} = s(M_p+3)$, and $\uBar{\bm{R}} = \bm{I}_{s(M_p+3)}$. We initially set three clusters in our application (\( J=3 \)). However, we found that one cluster disappeared after discarding the burn-in iterations. Thus, we fixed the number of components at \( J = 2 \), as supported by the variability in the data and instruments. We do not implement any random permutation of the cluster identifiers \citep{Fruhwirth2006}, as this was shown to hinder convergence of our Gibbs sampler in simulation exercises. We also verify label-switching is not an issue in our application by considering the consistency of individual segmentation of posterior chains (results available upon request).

Using our coefficient draws $(\bm{\theta}^{(1)}, \ldots, \bm{\theta}^{(S)})$ from the EASI specification (allowing or not for unobserved heterogeneity clusters), we can use the posterior draws to compute all relevant microeconomic summaries previously derived in Table \ref{Tab:EASI_Summaries}. We are then left with posterior draws of these summaries, such that inference on these highly non-linear quantities is a simple by-product of the estimation algorithm; a key feature of Bayesian inference. The coefficients of the EASI model itself are usually not of direct interest, so we provide the full estimates in the Appendix tables \ref{Tab:First_Stage}, \ref{Tab:Structural} and \ref{Tab:Correlation}. Nonetheless, we highlight that these estimates provide evidence for: (i) the importance of including demographic variables to deal with observed heterogeneity in consumer preferences; (ii) the chosen instruments being relevant and jointly significant in explaining additional variation in drug prices aside from the demographics; and (iii) prices being endogenous due to the significant correlation to consumed shares in their latent disturbances.

We then turn to studying the unobserved heterogeneity clusters recovered by our method, and provide evidence that these identify two clear consumer segments: ``soft'' and ``hard'' users. This classification is completely data-driven and automatically identified by the estimation strategy, leading to a valuable policy targeting mechanism in addition to demand system estimates. We provide balance tests across survey variables to further understand the differences between these population segments. Our results confirm that access to and use of harder drugs (among other key demographic variables) are drivers of the classification into one cluster or another, providing further rationale behind our cluster labels. Additionally, we show how the classification into these population segments is correlated to addiction indicators that can be obtained from the survey and that one of the identified segments has similar drug consumption patterns to the population of homeless individuals whom showcase large levels of drug consumption. Heterogeneous Engel curves and income effects are explored in Appendix \ref{apx_sec:engel}.

An important aspect of our Bayesian inferential framework is that it allows us to test the microeconomic restrictions imposed for estimation of the EASI demand system. As discussed in Section \ref{sec:sumEASI}, while the unit-sum restrictions are imposed due to a mechanical property of expenditure shares, the constraints from Slutsky symmetry, strict cost monotonicity and cost concavity are not innocuous as a way to regularize demand behavior towards microeconomic theory predictions. As Slutsky symmetry implies an equality (or point) restriction, we can use the Savage--Dickey density ratio to calculate the Bayes factor in favor of the restriction.\footnote{Let $M_1$ represent the EASI model imposing Slutsky symmetry and $M_2$ the unrestricted model. Recall that Slutsky symmetry imposes $\widetilde{\bm{A}}_m = \widetilde{\bm{A}}_m\T$ for $m = 0, \ldots, M_p$ and $\widetilde{\bm{B}} = \widetilde{\bm{B}}\T$, meaning $M_1$ imposes an equality restriction on the model parameters of the form $\bm{\theta} = \bm{\theta}_0$. The Bayes factor comparing models $M_1$ and $M_2$ can then be computed using only the unrestricted model according to the Savage--Dickey density ratio \citep[see][]{kass1995bayes}: $BF_{1, 2} = p(\bm{\theta} = \bm{\theta}_0 \mid \mathcal{D}, M_2) / p(\bm{\theta} = \bm{\theta}_0 \mid M_2).$} For cost monotonicity and concavity, which imply inequality (or set) restrictions, we can directly calculate the Bayes factor according to the posterior probability that the constraints are satisfied in our Markov Chain Monte Carlo (MCMC) algorithm draws. Computing these measures for our specification of interest results in very strong evidence in favor of Slutsky symmetry ($2 \log BF_{1,2} = 17.58$), strong evidence in favor of strict cost monotonicity ($2 \log BF_{1,2} = 8.56$), and only slight evidence in favor of cost concavity ($2 \log BF_{1,2} = 1.84$). Evidence generally points towards microeconomic restrictions being satisfied in-sample using the Bayes factor scale provided in \citet{kass1995bayes}.



\subsection{Demand-price elasticities}

Our first set of results considers the price effects on demand of illicit drugs. As we model the demand of drugs in a fully structural system, we can obtain not just the own-price elasticities of each drug (as usually done in the literature), but also the cross-price elasticities of one drug onto another. This is of particular importance in the face of potential marijuana legalization policies that are likely to have large impacts through price, as the price of one drug will face clear changes due to policy intervention whereas the price paths of other drugs will likely remain fixed in the short-term. In this way, we will be able to study how the effect of a price change due to legalization translates into re-balancing of a consumer's drug bundle.

The main story behind the results is presented in Table \ref{Tab:Results_Mixture}, which provides estimates from all our Bayesian specifications. The panels of the Table are arranged in the way we were led by our exploration of the different specifications and understanding of the data. We begin by considering estimation of the EASI model accounting for corner solutions (censoring) due to zero consumption of drugs by many consumers, before augmenting the model with price endogeneity and using external instruments to recover identification (two top panels in Table \ref{Tab:Results_Mixture}). Both models do not account for unobserved heterogeneity. In the endogenous specification, the relevance of endogeneity is evident when comparing the two sets of results, as the inclusion of instruments highlights the high levels of price sensitivity in cocaine and basuco. However, this comes at the cost of a remarkably reduced precision. These results align with previous literature that has found marijuana to be an inelastic product in countries such as Australia \citep{jacobi2016marijuana}, Colombia \citep{ramirez2023marijuana}, South Africa \citep{Riley2020}, Thailand \citep{Sukharomana2017}, and the United States \citep{Davis2016}. On the other hand, the demand for more harmful drugs, like cocaine and basuco, has shown greater elasticity \citep{jofre2008trading,Gallet2014}. Regarding cross-price elasticities, limited evidence finds some complementarity between marijuana and cocaine \citep{jofre2008trading}.\footnote{Tables \ref{Tab:Results_FullSample} (full sample) and \ref{Tab:Results_SubSample} (``soft'' cluster) in the Appendix show similar patterns using least squares (LS) and two-stage LS. However, these estimators do not account for censoring nor unobserved heterogeneity.}

Our next two specifications introduce unobserved heterogeneity by considering two clusters. The results presented are for the subsample classified as ``soft'' users, whose parameter estimates are the most stable (those for the ``hard'' cluster are highly unstable or not updated through the data due to its small sample size). Again, we observe the relevance of endogeneity when comparing these two sets of posterior estimates. We improve the precision of the posterior estimates by fixing the parameters of the first stage to be homogeneous across clusters and, consequently, fully exploiting the variability in the instruments, leaving only unobserved heterogeneity in the structural demand equation.\footnote{The two bottom panels in Table \ref{Tab:Results_SubSample} in the Appendix show the results assuming heterogeneity in the price equations. We observe greater variability in the results.}

\begin{figure}[htbp]
    \centering
    \caption{Density plots of price elasticities of demand implied by EASI system estimates}
    \label{Fig:Density_DemandPrice}
    \includegraphics[width = \textwidth, page = 2]{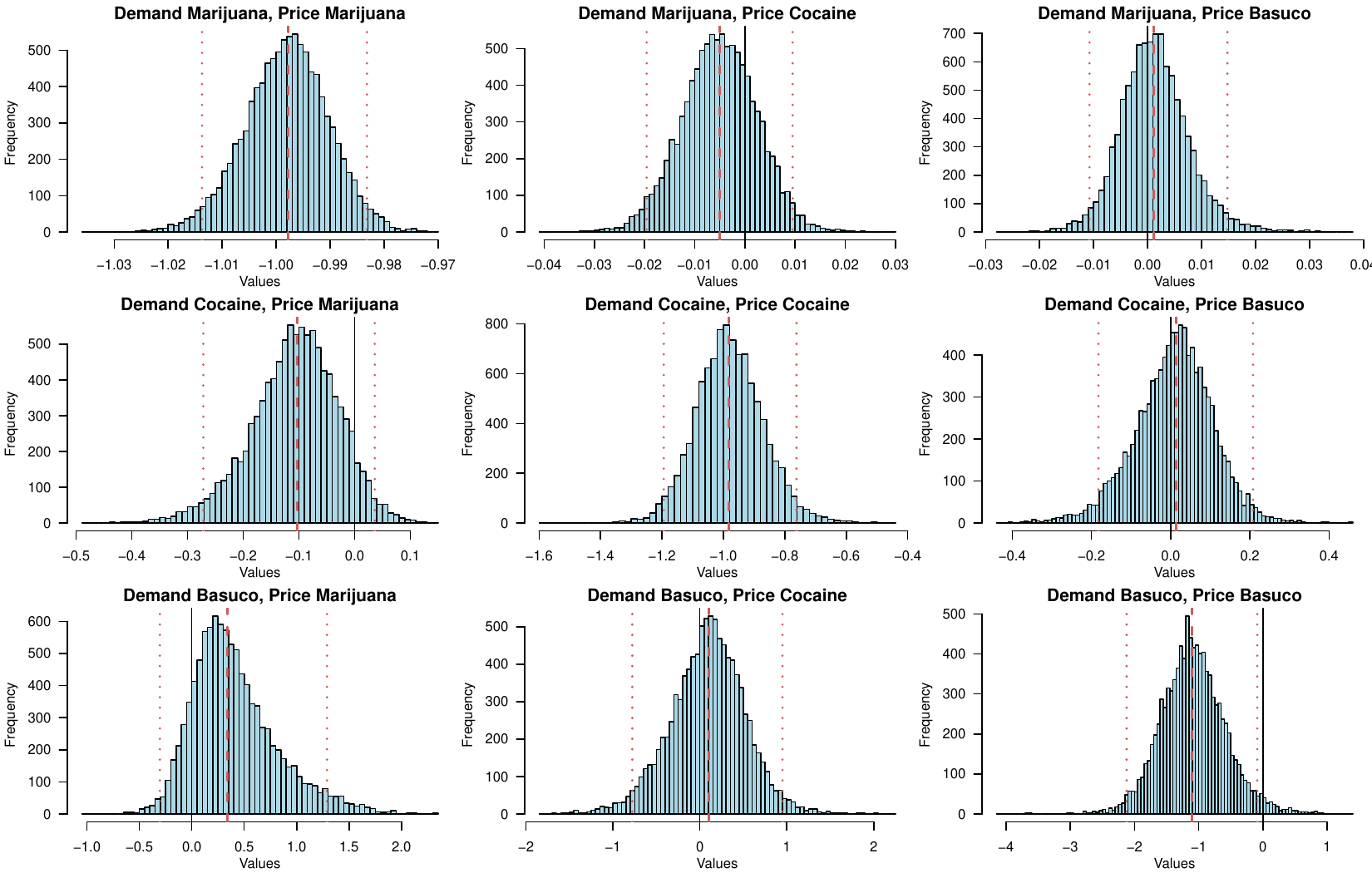}
    \begin{tablenotes}
        \item \small{\textbf{Notes:} Full posterior density over price elasticity of demand parameters (own and crossed between illicit drugs). Dashed line represents median of chains, dotted line provides the 95\% HPD for each parameter, and a solid line at 0 is drawn for reference. Single chain run of 10,000 iterations after a burn-in window of 5,000 and thinning every 10 draws.}
    \end{tablenotes}
\end{figure}

We focus the remaining analysis of the results based on the last specification with a homogeneous first-stage. This specification identifies a total of 1,069 ``soft'' consumers in our sample, representative of 551,507 drug consumers nationwide (the remaining 167 in-sample ``hard'' consumers represent 81,983 national consumers using the expansion factors from the survey). The full marginal posterior distributions and 95\% highest posterior density intervals of the demand price elasticities from this specification can be visualized in Figure \ref{Fig:Density_DemandPrice}. Additional visualizations are provided in Figures \ref{Fig:Trace_DemandPrice} (trace plots) and \ref{Fig:ACF_DemandPrice} (autocorrelation plots), with draws shown to satisfy standard posterior convergence diagnostics such that posterior expectations should be accurately computed (results available upon request).

The demand-price elasticity estimates from this specification suggest that ``soft'' users are unit-elastic to the price of illicit drugs (a 1\% increase in price implies a 1\% decrease in consumption), such that they are highly responsive and their consumption follows the rational law of demand. These elasticities are statistically significant, as the 95\% credible intervals do not include zero. The similarity of these own-price elasticities is novel in comparison to previous literature, which clearly identifies greater price sensitivity for harmful drugs than for marijuana. See the panel \textit{Bayesian Censored with Endogeneity but without Mixtures} in Table \ref{Tab:Results_Mixture}. The difference arises because prior research overlooks unobserved heterogeneity, combining individuals with different unobserved preferences. In particular, we show below that the ``soft'' cluster spends the most on marijuana, with a small percentage spent on cocaine and basuco. As a result, price variations in these two drugs do not have a high impact on their spending, and consequently the price elasticity exhibits similar values to those found for marijuana. Unobserved preference heterogeneity is therefore crucial for identifying meaningful price effects.

Additionally, as we have the cross-price elasticities of demand, we can determine that both harder drugs, cocaine and basuco, are complementary to marijuana in the sense that it follows the same direction as the own-price effect of marijuana (a 1\% increase in the price of marijuana results in a decrease of 0.03\% in the quantity of cocaine and 0.11\% for basuco). Similarly, marijuana responds in the same direction to changes in the price of harder drugs, suggesting further complementarity, though the magnitude of this effect is one or two orders of magnitude smaller, making the complementarity asymmetric. Finally, we note that cocaine and basuco are instead substitutes, as can be expected from basuco, being a lower-quality product and considered to be inferior to cocaine. Our estimates suggest a 1\% increase in the price of cocaine results in an increase in the consumption of basuco of 0.14\%, whereas a 1\% increase in the price of basuco results in an increased cocaine consumption of only 0.03\%. This is in line with the much larger average price per gram of cocaine, such that a 1\% increase in the price of cocaine has larger level effects on expenditure than the corresponding increase for basuco. However, the uncertainty associated with the cross-price elasticities varies and the 95\% credible intervals include zero, with a high probability of a complementary effect between marijuana and cocaine remaining as the key effect, which is consistent with previous literature. Further results are provided in Appendix \ref{apx_sec:additional_easi_estimates}.

\begin{table}[!htbp]
    \centering
    \caption{Price elasticities of demand obtained from Bayesian estimates}
    \label{Tab:Results_Mixture}
    \begin{threeparttable}
        \begin{adjustbox}{max totalsize={\textwidth}{.83\textheight}, center}
            \begin{tabular}{cccc}
                \toprule
                Good demand & Price Marijuana & Price Cocaine & Price Basuco \\
                \midrule
                \multicolumn{4}{c}{Bayesian Censored without Mixtures} \\
                \midrule
                \multirow{2}{*}{Marijuana} & $-0.9747$ & $-0.0118$ & $-0.0096$ \\
                      & $(-0.9986, -0.9522)$ & $(-0.0291, 0.0065)$ & $(-0.0203, 0.0007)$ \\
                \multirow{2}{*}{Cocaine} & $-0.1401$ & $-0.9715$ & $0.0637$ \\
                      & $(-0.3017, 0.0196)$ & $(-1.1213, -0.8265)$ & $(-0.0115, 0.1415)$ \\
                \multirow{2}{*}{Basuco} & $-0.2857$ & $0.2980$ & $-0.9379$ \\
                      & $(-0.7106, 0.1122)$ & $(-0.0447, 0.6471)$ & $(-1.3188, -0.5238)$ \\
                \midrule
                \multicolumn{4}{c}{Bayesian Censored with Endogeneity but  without Mixtures} \\
                \midrule
                \multirow{2}{*}{Marijuana} & $-0.9734$ & $-0.0182$ & $-0.0042$ \\
                      & $(-1.0109, -0.9365)$ & $(-0.0641, 0.3063)$ & $(-0.2155, 0.0235)$ \\
                \multirow{2}{*}{Cocaine} & $-0.2315$ & $-1.2466$ & $0.4256$ \\
                      & $(-0.5874, 0.7247)$ & $(-7.1771, -0.6637)$ & $(-0.0241, 3.8440)$ \\
                \multirow{2}{*}{Basuco} & $0.0747$ & $1.9327$ & $-2.8755$ \\
                      & $(-3.2705, 1.0507)$ & $(-0.0268, 18.0337)$ & $(-11.1489, -0.6727)$ \\
                \midrule
                \multicolumn{4}{c}{Bayesian Censored including Mixtures (``soft'' cluster)} \\
                \midrule
                \multirow{2}{*}{Marijuana} & $-1.0114$ & $-0.0524$ & $-0.0323$ \\
                      & $(-1.0566, -0.9718)$ & $(-0.0864, -0.0075)$ & $(-0.0645, -0.0032)$ \\
                \multirow{2}{*}{Cocaine} & $-0.2613$ & $-0.7202$ & $0.0766$ \\
                      & $(-0.6248, 0.1276)$ & $(-1.1246, -0.3807)$ & $(-0.1409, 0.2683)$ \\
                \multirow{2}{*}{Basuco} & $2.0414$ & $0.6582$ & $-0.1528$ \\
                      & $(-0.5318, 3.5122)$ & $(-0.2391, 1.3353)$ & $(-1.4840, 1.4202)$ \\
                \midrule
                \multicolumn{4}{c}{Bayesian Censored with Endogeneity} \\
                \multicolumn{4}{c}{including Mixtures (``soft'' cluster)} \\
                \midrule
                \multirow{2}{*}{Marijuana} & $-0.9863$ & $-0.0115$ & $0.0019$ \\
                      & $(-1.0151, -0.9580)$ & $(-0.0402, 0.0153)$ & $(-0.0137, 0.0182)$ \\
                \multirow{2}{*}{Cocaine} & $-0.1362$ & $-0.8395$ & $-0.0668$ \\
                      & $(-0.3599, 0.0849)$ & $(-1.1893, -0.4436)$ & $(-0.3952, 0.2458)$ \\
                \multirow{2}{*}{Basuco} & $0.1126$ & $-0.2893$ & $-0.7685$ \\
                      & $(-0.4680, 0.6756)$ & $(-1.7641, 1.1189)$ & $(-2.2193, 0.8167)$ \\
                \midrule
                \multicolumn{4}{c}{Bayesian Censored with Endogeneity and Mixtures,} \\
                \multicolumn{4}{c}{Homogeneous Price Equations (``soft'' cluster)} \\
                \midrule
                \multirow{2}[1]{*}{Marijuana} & $-0.9978$ & $-0.0050$ & $0.0011$ \\
                      & $(-1.0137, -0.9832)$ & $(-0.0196, 0.0095)$ & $(-0.0108, 0.0148)$ \\
                \multirow{2}[0]{*}{Cocaine} & $-0.1028$ & $-0.9830$ & $0.0136$ \\
                      & $(-0.2718, 0.0363)$ & $(-1.1942, -0.7621)$ & $(-0.1825, 0.2082)$ \\
                \multirow{2}[1]{*}{Basuco} & $0.3403$ & $0.1056$ & $-1.1034$ \\
                      & $(-0.3045, 1.2899)$ & $(-0.7776, 0.9500)$ & $(-2.1234, -0.0881)$ \\
                \bottomrule
            \end{tabular}
        \end{adjustbox}
        \begin{minipage}{\textwidth}
            \begin{tablenotes}
            \item \small{\textbf{Notes:} Identified ``soft'' cluster includes 1,069 consumers. Final panel uses full sample of 1,236 consumers for the first-stage regression. Point-estimate presented is the median of chains across 10,000 iterations, after a burn-in window of 5,000 and thinning every 10 draws. 95\% highest posterior density (HPD) intervals of chains provided in parenthesis.}
        \end{tablenotes}
        \end{minipage}
    \end{threeparttable}
\end{table}

\subsection{Consumer segments by drug preferences}

The finite mixture framework used in this paper allows us to provide posterior estimates of both cluster indicators and probability of belonging to each cluster for all individuals in our sample. We first provide evidence that the clusters identified by our algorithm are stable, in the sense that the classification of individuals is sharp and consistent across posterior draws. Figure \ref{Fig:Inclusion_Probability} showcases the posterior inclusion probability to $C_1$, normalized to be the ``soft'' cluster of consumers, denoted as $\Bar{\phi}_{i, 1}$ in \eqref{Eq:Posterior_Pars_mixture}.

Observe that there is a sharp edge that distinguishes the probability of inclusion of individuals to the ``soft'' and ``hard'' clusters, with a few individuals falling on the frontier between these two states. This provides a way for the model to identify individuals that can be assigned to either cluster, which can be interpreted in our application as those at risk of moving from being a soft consumer towards harder use. That is, these users are identified as the ones with the highest potential of fulfilling the gateway hypothesis, where their marijuana consumption is large but skewing towards cocaine and basuco.

To add further evidence to the rationale behind our cluster labels, we evaluate the average differences between all relevant covariates collected in the survey. Figure \ref{Fig:Balance_Tstats} showcases the sorted T-statistics obtained from a mean-difference test that takes into account the different cluster sizes (see numerical p-values and intervals in Table \ref{Tab:Balance_Table}). The largest differences between users in the ``hard'' and ``soft'' clusters arise due to the former having higher access to and consumption of both cocaine and basuco, including addiction-related questions contained in the original survey. Other key differences in terms of demographic characteristics include larger probabilities that individuals are consumers of substances in their personal networks, have access to a drug dealer, and use tobacco and alcohol jointly for the ``hard'' cluster. Additionally, individuals in this cluster are older, report feeling mentally healthy less often, have less education, and are more likely to be male.

\begin{figure}
    \centering
    \caption{Shares of consumption of illicit drugs across clusters}
    \label{Fig:Cluster_Shares_Simplex}
    \begin{adjustbox}{max totalsize={0.9\textwidth}{\textheight}, center}
        \includegraphics[width = \linewidth, trim={60pt 35pt 60pt 35pt}, clip]{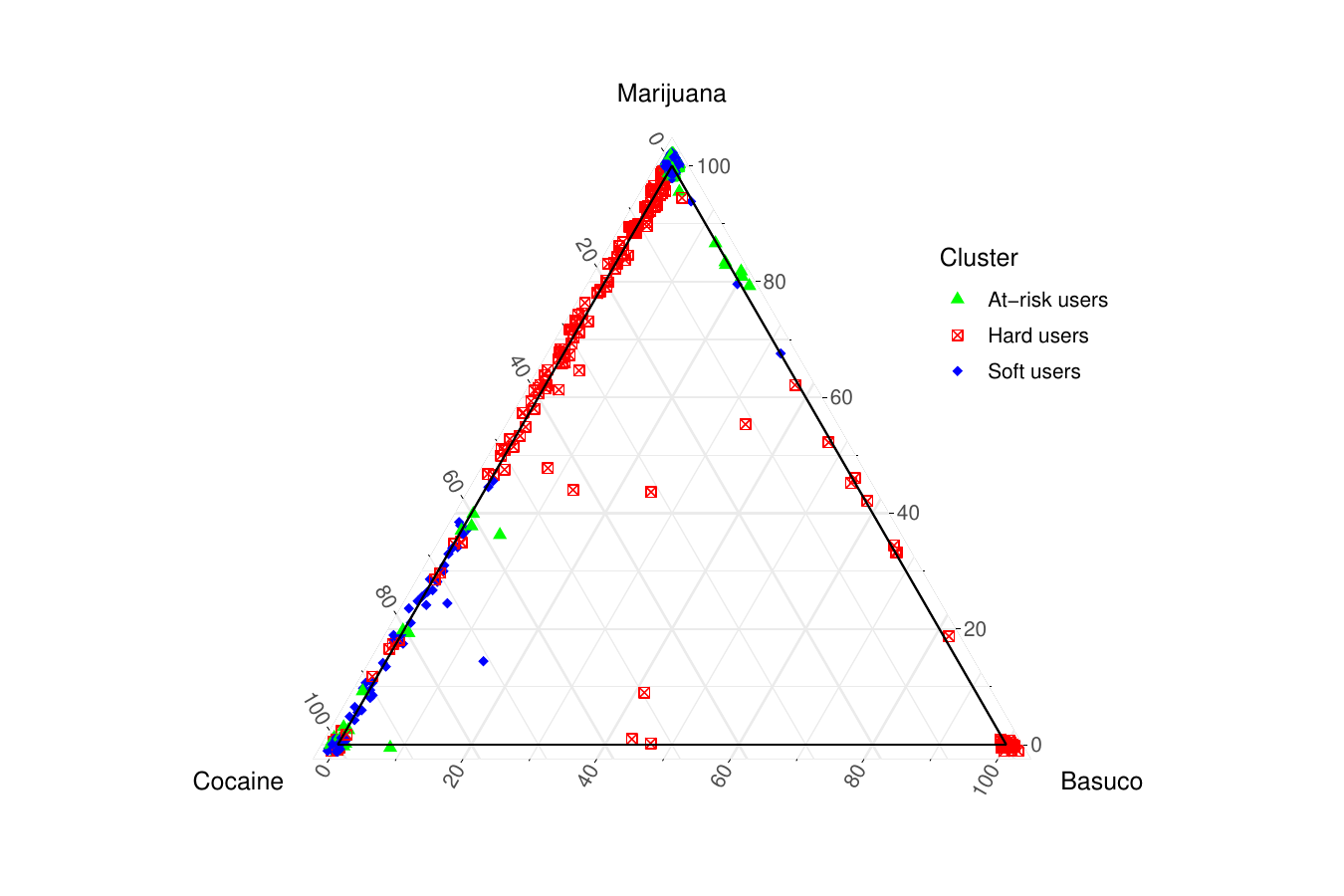}
    \end{adjustbox}
    \begin{minipage}{\textwidth}
        \begin{tablenotes}
            \item \small{\textbf{Notes:} Consumption shares for each identified individual of the ``soft'' and ``hard'' clusters of consumers. We additionally highlight a subgroup of ``at-risk'' individuals who are not currently in the ``hard'' cluster but exhibit similar characteristics that might lead to switching from a ``soft'' drug consumer. Random jitter is added to the consumption shares to highlight concentration of consumers at the edges and borders of the simplex. These individuals present zero consumption of at least one illicit drug.}
        \end{tablenotes}
    \end{minipage}
\end{figure}

Based on this classification, we can probe the differences between the characteristics of individuals that are sorted into either cluster. Figure \ref{Fig:Cluster_Shares_Simplex} presents the consumption shares of individuals classified into the cluster of either ``soft'' or ``hard'' users. One of the main reason behind this labeling is the fact that consumers classified into the second cluster have larger shares of both cocaine and basuco compared with the share of marijuana. Given the more addictive nature of these substances, when individuals include them in their consumption bundle, it is more likely that this will be correlated to negative outcomes in terms of their covariates.

Particular attention must be given to individuals belonging to the ``hard" cluster, as their drug consumption patterns are similar to those of homeless individuals, who report basuco as the most-consumed drug according to the Census of Homeless Individuals in Colombia (2019). We also examined the answers provided by the ``hard" cluster to questions related to drug-related risks, such as issues with friends, family, or work due to drug consumption. We find that the hard-drug user group, on average, responded positively to 5 out of 8 questions, whereas this rate is measured to less than 1 in 8 for those in the ``soft'' cluster. Using this procedure, we identify a subset of ``soft'' consumers classified as ``at-risk" in our model (representing 5,716 individuals) who show indication of moving towards harder drug use (green triangles in Figure \ref{Fig:Cluster_Shares_Simplex}).


\section{Marijuana legalization counterfactuals}
\label{Sec:Legalization}
In this section, we use our estimated drug demand behavior to conduct counterfactual exercises of the potential effects of a marijuana legalization policy. Following the large political backing that such a policy has already amassed in Colombia ---as well as the legalization experience of Uruguay, a country both geographically and economically similar--- it becomes crucial to understand its potential effects. This is emphasized by the idiosyncrasies of the Colombian drug market with its large production amounts, low costs, and salient demographics, as well as the large sources of heterogeneity in drug demand by consumers found in our estimation exercise.

Legalization policies will directly affect consumers' access to the drug and prices faced for its legal purchase. The final effect on marijuana price in Colombia is uncertain as legalization could entail conflicting forces in the market. Accounting for direct production costs for marijuana of US\textcent 1 per potency-adjusted gram \citep{velez2021medicinal}, as well as average logistical costs and the rate of return, the lower bound on the tax-free price of a gram of marijuana should be US\textcent 2. This matches the average tax-free price of a cigarette in Colombia \citep{ramirez2023marijuana}. The survey we use shows the average market price of marijuana in Colombia is US\textcent 83 (see Table \ref{Tab:Summary_Statistics}), with the difference between this and the tax-free rate likely attributable to the illegality margin. By reducing such large illegality margins, legalization in Colombia is potentially likely to reduce the price of marijuana. On the other hand, marijuana legalization would introduce additional taxes, operating and regulatory costs, which could potentially dominate the final price \citep{caulkins2015options}.\footnote{Optimal taxation for ``sin goods'' such as marijuana requires additional information or models on dynamic formation of addictive behaviour \citep{Gruber2001}, substitutability within and between sin and non-sin goods \citep{Arnabal2021, Fleissig2021}, demand for potency \citep{Hansen2020}, or preferences over inequality \cite{Lockwood2017}, among others. As such information is not available in the survey, this exercise is outside the scope of the current paper and left for future research.}

We therefore consider several scenarios based on the potential price changes of marijuana that such a legalization policy might entail (leaving all other drug prices fixed), considering a representative agent that did not have access to a dealer pre-implementation but now does have access due to legalization.\footnote{We fix the values of the covariates for the representative agent to their means for continuously-distributed covariates (i.e., ``average'' agent) and to the mode for discretely-distributed covariates (i.e., ``modal'' agent). These values are fixed throughout the counterfactual exercises to make all results comparable.} Figure \ref{Fig:Legalization_Price_Distribution} showcases the average post-legalization price of marijuana implied by each scenario. All prices occur naturally within the distribution faced by individuals in the sample, showing potential for external validity of the provided legalization scenarios.
\begin{enumerate}
    \item[\textbf{Scenario 1.}] Our most likely scenario is one of a 50\% decrease in prices of marijuana, given the low production and distribution costs of marijuana in Colombia. This scenario mirrors Uruguay’s recent experience following marijuana legalization, where the price of marijuana was reduced to incentivize demand in the newly legal market \citep{pluasbeyond}.
    \item[\textbf{Scenario 2.}] The potential for legalization to decrease prices even further, setting the final price at 10\% of the initial value. This would be comparable to the current tax-included price of a cigarette in Colombia.
    \item[\textbf{Scenario 3.}] The overhead costs of legalization might outweigh its cost benefits leading to a 25\% net \emph{increase} in the price of marijuana, similar to the counterfactual exercise proposed by \cite{jacobi2016marijuana}.
\end{enumerate}

We present results showcasing the counterfactual effects of this path of legalization policy on drug consumption, consumer welfare (as measured by the absolute value of Equivalent Variation), government revenue, and illegal drug market size. Estimating the effects of such price changes and implied taxes on aggregate outcomes requires assumptions on both demand and supply forces in the market \citep{Angrist1996, Imbens2014}. Given the structure of the drug market in Colombia and low marginal cost of production for marijuana, it is reasonable to approximate the supply curve for marijuana as perfectly elastic at traded prices. Under such assumption, price effects can be directly identified from our demand curve estimates that rely on instrumental variables. A limitation of our approach is the inability to include dynamic or post-legalization effects, given that the survey has not been conducted since its initial wave and legalization itself has not been implemented. Therefore, our results can provide short-run effects without an increase on the user base, which can be interpreted as lower bounds for government revenue and upper bounds for dealer losses. Providing results \emph{ex-ante} to the legalization is crucial as it allows policy makers to consider desirable scenarios prior to implementation.

As most demand system models are based on expenditure shares, they cannot identify changes to total expenditure nor changes in total quantities. Therefore, we approximate percentage changes to the consumed quantity of drug $l$ ($q_l$) given a change to the price of drug $j$ ($p_j$) as $\%\Delta q_l \approx \exp \{\widetilde{\varepsilon}^{M}_{lj} \cdot \Delta \log(p_j) \} - 1$, where $\widetilde{\varepsilon}^{M}_{lj}$ are the estimated demand-price elasticities (see Table \ref{Tab:EASI_Summaries}). Given our Bayesian framework, we are able to provide full posterior distributions over all policy quantities involved in the analysis, meaning inference also comes as a by-product of estimation. The uncertainty associated to our Bayesian results for the ``soft'' cluster of users implies highly accurate estimates of all quantities considered here.

\begin{figure}
    \centering
    \caption{Posterior distribution of policy summaries following a price change in marijuana as implied by EASI system estimates from ``soft'' cluster of users}
    \label{Fig:Legalization_Scenario50}
    
    \begin{subfigure}{0.49\textwidth}
        \includegraphics[scale = 0.29, page = 2]{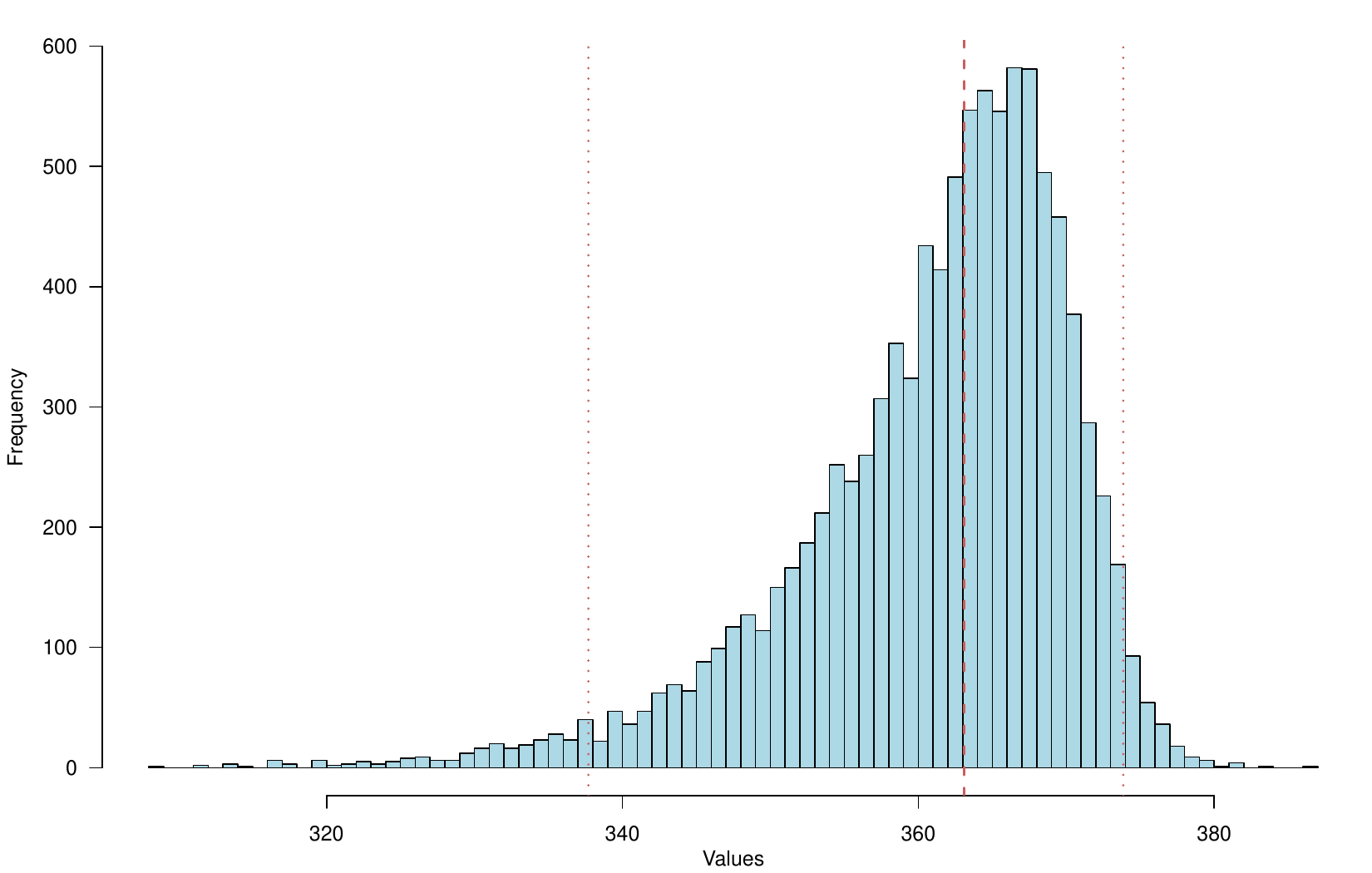}
        \caption{Equivalent Variation}
        \label{Fig:Density_EV_Scenario50}
    \end{subfigure}
    \begin{subfigure}{0.49\textwidth}
        \includegraphics[scale = 0.29, page = 2]{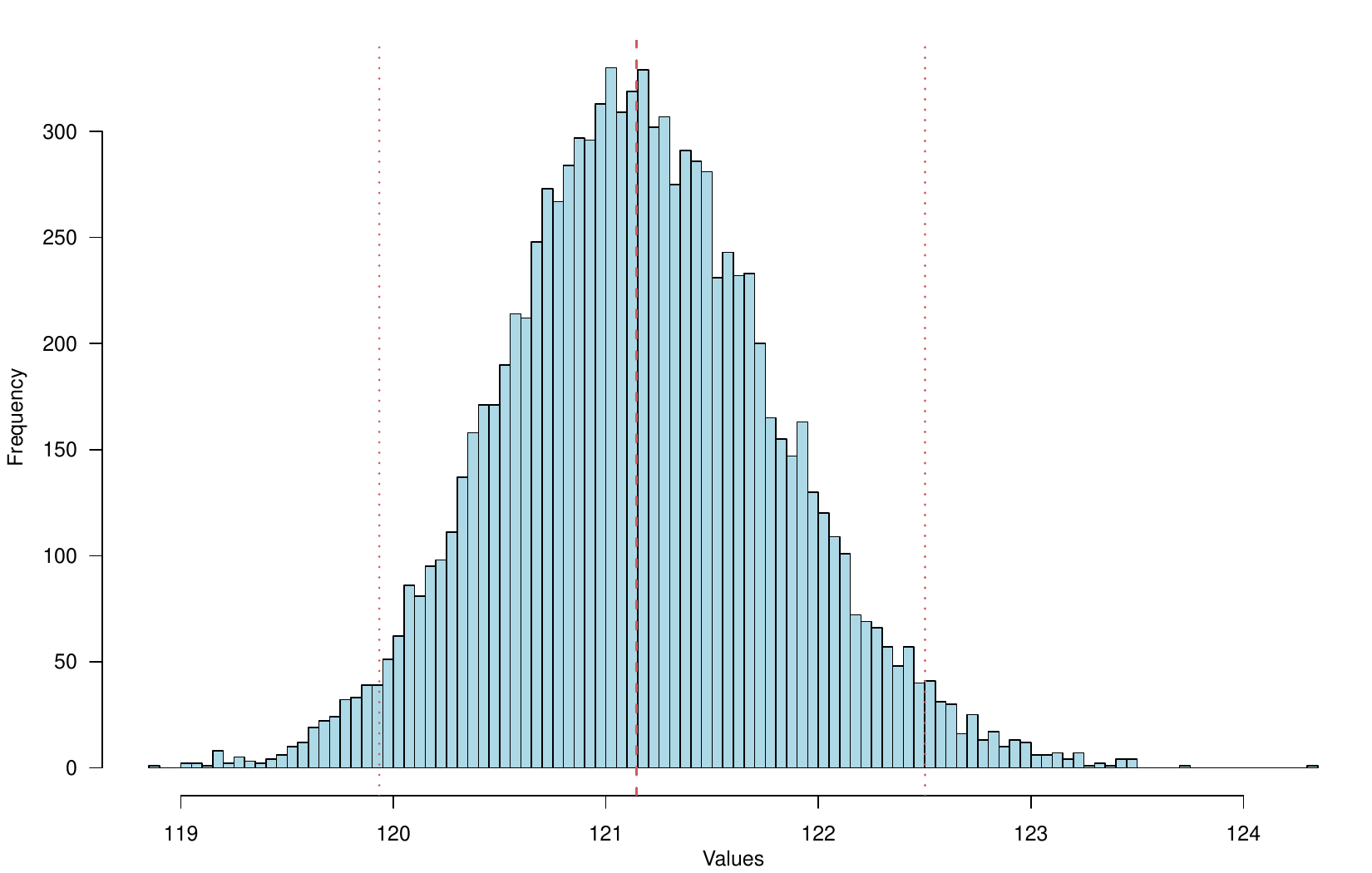}
        \caption{Change in Government Revenue (Millions)}
        \label{Fig:Density_GovernmentRevenue_Scenario50}
    \end{subfigure}

    \begin{subfigure}{0.49\textwidth}
        \includegraphics[scale = 0.29, page = 2]{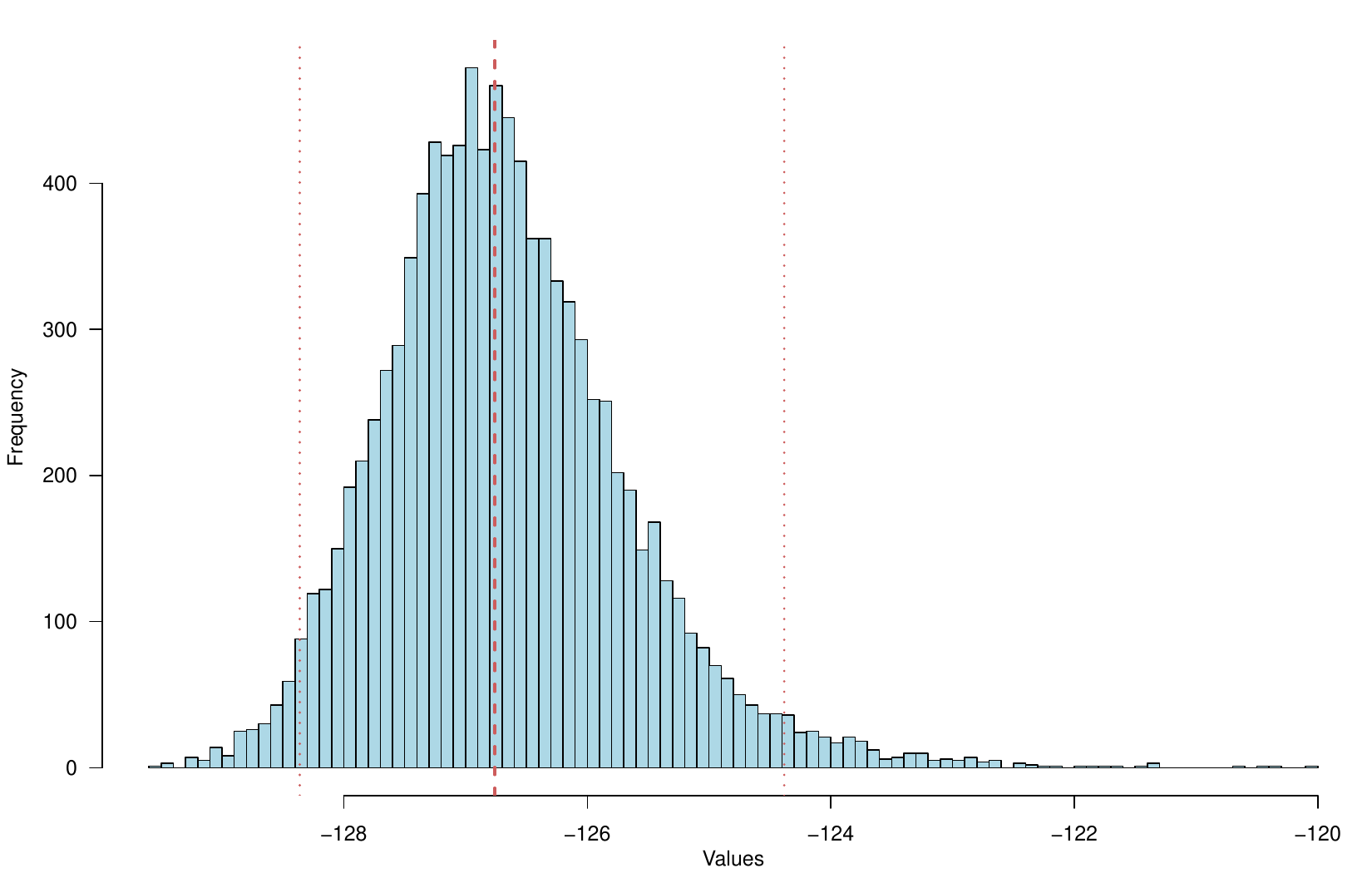}
        \caption{Change in Dealer Revenue (Millions)}
        \label{Fig:Density_RevenueChange_Scenario50}
    \end{subfigure}
    \begin{subfigure}{0.49\textwidth}
        \includegraphics[scale = 0.29, page = 2]{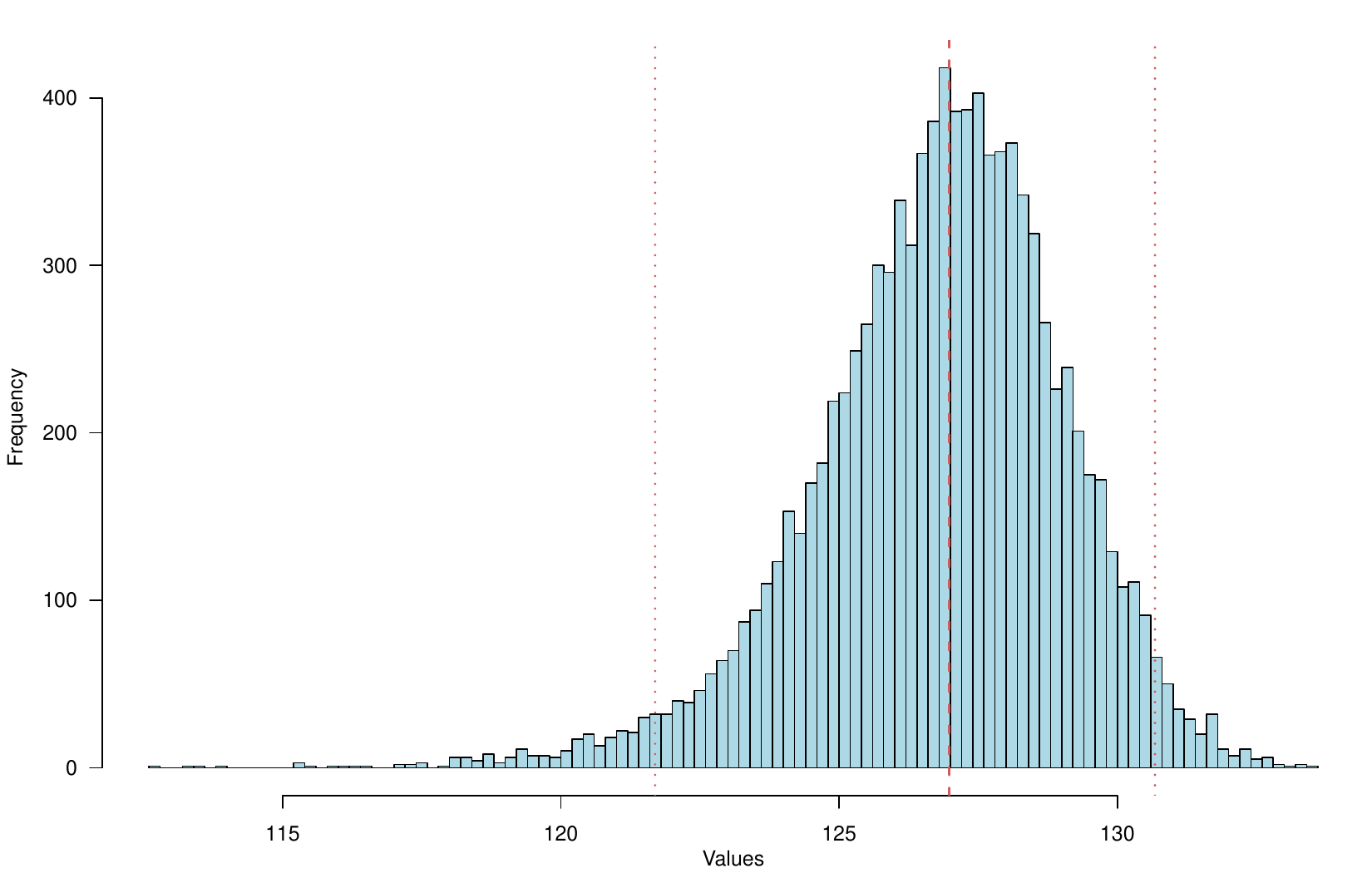}
        \caption{Users for Revenue Offset}
        \label{Fig:Density_UserChange_Scenario50}
    \end{subfigure}
    \begin{minipage}{\textwidth}
        \begin{tablenotes}
            \item \small{\textbf{Notes:} Full posterior density over policy quantities calculated in 2019 USD\$ after a 50\% decrease in the price of marijuana and complete access to dealers following legalization of marijuana. Dashed line represents median of chains, dotted line provides the 95\% HPD for each parameter, and a solid line at 0 is drawn for reference. Single chain run keeping 10,000 iterations after a burn-in window of 5,000 and thinning every 10 draws.}
        \end{tablenotes}
    \end{minipage}
\end{figure}

We provide a full description of the results for the 50\% decrease in the body of the paper, exploring alternative scenarios in Appendix \ref{apx_sec:additional_legalization_results}. Figure \ref{Fig:Density_EV_Scenario50} shows the posterior distribution of the equivalent variation (EV) in annual terms following the price change, as deduced from the EASI model \citep{RamirezHassan2024}:
\begin{align*}\label{refM:eq11}
	EV & = e \left[\exp \left\{-\sum_{l=1}^S \left(\widetilde{w}_l^1\widetilde{p}^{1}_l - \widetilde{w}_l^0\widetilde{p}^{0}_l \right) + \frac{1}{2} \sum_{l=1}^S \sum_{j=1}^S \widetilde{a}_{lj}(\widetilde{p}_l^1 \widetilde{p}_j^1-\widetilde{p}_l^0\widetilde{p}_j^0) \right\} - 1 \right] \times 12,
\end{align*}
where $e$ is the total monetary expenditure in the three drugs, $\widetilde{w}_s^0$ and $\widetilde{w}_s^1$ are the pre- and post-legalization expenditure shares, $\widetilde{p}_s^0$ and $\widetilde{p}_s^1$ are pre and post ($\log$) prices, and $\widetilde{a}_{sj}$ is the $sj$-th element of matrix $\widetilde{\bm A}_0$, respectively. The result suggests that the representative ``soft'' consumer perceives a change in their utility that is equivalent to a median change in total expenditure of approximately \$363 USD, with a 95\% credible interval of $(338, 374)$. This change is approximately equivalent to the annual average drug expenditure, meaning consumer welfare effects are considerable and largely due to an increased marijuana consumption that offsets the price decrease, with the remaining drugs experiencing negligible changes.

After the legalization is implemented, previous studies suggest only a fraction of users will switch to the legal marijuana market, leaving approximately 34\% of the illegal market active \citep{ramirez2023marijuana, pluasbeyond},\footnote{For instance, the size of the black market for cigarettes in Colombia is 34\% (see the study \href{https://www.semana.com/salud/articulo/consumo-de-cigarrillos-34-de-cada-100-fueron-de-contrabando-esto-revelo-la-federacion-nacional-de-departamentos/202315/}{``Consumo de cigarrillos ilegales Colombia'' 2022}). Similarly, 1 out 3 marijuana consumers in Uruguay reports obtaining marijuana from an illicit provider after legalization.} whereas dealers maintain sole control of the cocaine and basuco markets due to their continuing illegal status. Figure \ref{Fig:Density_GovernmentRevenue_Scenario50} shows the posterior distribution of the estimated government tax revenue from the ``soft'' consumers, with a median value of approximately \$121.5 million USD and a 95\% credible interval of approximately (120, 123) million USD. The posterior distribution is derived from a tax of approximately US\textcent 39 per marijuana joint (calculated as the average price of US\textcent 41 minus the tax-free price of US\textcent 2) and the predicted monthly number of marijuana joints consumed by individuals in this cluster, again considering expansion factors. 




Additionally, Figure \ref{Fig:Density_RevenueChange_Scenario50} shows that dealers would face a median revenue loss of \$127 million USD from the ``soft'' consumers, the 95\% credible interval is $(-129, -125)$ million USD (similar figures for marijuana profits that subtract marginal cost are provided in Figure \ref{Fig:Density_ProfitChange_OtherScenarios}). In order to make up for such losses after legalization, drug dealers will need the total number of drug users to increase. The annual weighted average expenditure in drugs post-legalization is estimated to be approximately \$157 USD (using the fraction of expansion factor over the sum of total users as weights). Based on this value, we can calculate the required number of average users required to offset a revenue change equal to the one found in Figure \ref{Fig:Density_RevenueChange_Scenario50}. This provides a full posterior distribution over the number of required users given in each price scenario. Figure \ref{Fig:Density_UserChange_Scenario50} showcases that drug dealers would need an approximately 130\% increase in the number of users spending at the annual average rate of \$157 USD to offset the \$127 million USD total revenue decrease. This corresponds to approximately 825,000 new users, which would imply a jump in the regular use rate from 2.5\% to 5.8\% of Colombians aged between 12 and 65 years. No state or country in the world has exhibited such an increase following legalization \citep{anderson2023public}. The estimated increase in Colombia would be approximately closer to 30\% after legalization \citep{ramirez2023marijuana}, which is similar to the figures found in the United States after legalization in different states \citep{anderson2023public}.

\begin{table}[!htbp]
    \centering
    \caption{Bayesian point estimates and credible intervals of key policy summaries post-legalization}
    \label{Tab:Policy_Quantity_Bayes}
    \begin{adjustbox}{max totalsize = {\textwidth}{\textheight}}
        \begin{threeparttable}
            \begin{tabular}{ccccc}
            \toprule
            Scenario & \multicolumn{4}{c}{Policy Quantity} \\
            \cmidrule{2-5}
            \multirow{2}{*}{Marijuana} & Equivalent & Government & Dealer Revenue & Users for \\ 
            \multirow{2}{*}{Price Change} & Variation & Revenue & Change & Revenue Offset \\ 
             & (USD\$) & (Millions USD\$) & (Millions USD\$) & (\% change) \\ 
            \midrule
            \multirow{2}{*}{50\% decrease} & $363.11$ & $121.14$ & $-126.76$ & $126.98$ \\ 
             & $(337.70, 373.85)$ & $(119.93, 122.50)$ & $(-128.36, -124.38)$ & $(121.69, 130.68)$ \\ 
            \multirow{2}{*}{90\% decrease} & $3204.33$ & $95.01$ & $-122.80$ & $118.32$ \\ 
             & $(2801.66, 3394.66)$ & $(91.90, 98.60)$ & $(-129.22, -106.50)$ & $(88.69, 132.71)$ \\ 
            \multirow{2}{*}{25\% increase} & $-73.17$ & $125.26$ & $-127.96$ & $129.73$ \\ 
             & $(-75.91, -68.76)$ & $(124.82, 125.67)$ & $(-128.41, -127.45)$ & $(128.55, 130.79)$ \\ 
            \bottomrule
        \end{tabular}
            \begin{tablenotes}
                \item \small{\textbf{Notes:} Key policy quantities calculated in 2019 USD\$ after a 50\% decrease in the price of marijuana and complete access to dealers following legalization of marijuana and using results from the ``soft'' user segment. While consumer welfare depends on the sign of the price change, the estimated elasticites and legalization effort imply there is always a gain for the government and a loss for suppliers.}
            \end{tablenotes}
        \end{threeparttable}
    \end{adjustbox}
\end{table}

Table \ref{Tab:Policy_Quantity_Bayes} summarizes the point estimates and credibility intervals of the policy-relevant quantities across all legalization price scenarios considered. We see that the magnitude and sign of the effect on consumer welfare as measured by the EV depends on the price change, with price decreases having a larger compensating effect on consumers compared to the unlikely price increase. On the other hand, the estimated revenue collected by the government and losses experienced by the dealers remains consistent across the scenarios, reflecting that the largest portion of the policy effect can be explained by the large loss to marijuana market shares experienced by dealers. In summary, a legalization policy applied to the Colombian context is estimated to increase marijuana consumption, increase welfare for consumers through standardized pricing and access, and reallocate considerable revenue sources to local governments from illegal drug dealers. Legalization thus has the potential to decrease drug profitability and disincentivize the use of violence to control the domestic black market, which is the current a large source of violent crime in Colombia.

\section{Policy recommendations and conclusion}
\label{Sec:Conclusions}
This paper models the joint demand for illicit drugs in Colombia, introducing a Gaussian mixture of endogenous EASI demand systems estimated via Bayesian methods. We provide evidence that unobserved heterogeneity is a key driver of drug demand, and using our data-driven cluster assignment, we simultaneously classify individuals into either a ``soft'' or ``hard'' drug consumer segment and estimate the drug demand behavior of each segment. We control for both demand- and supply-based endogenous sources of consumer price variation using as instruments geo-referenced, distance-weighted averages of drug-related captures and elicited prices. Tailoring the priors to suit the setting of structural demand modeling and to take full advantage of the variability in instruments, our Bayesian results are able to provide accurate estimates of EASI coefficients and their demand implications. In particular, the results from the ``soft'' cluster of consumers provide accurate descriptions of drug demand patterns that represent over 550,000 consumers across Colombia.

The framework emphasizes the importance of modeling drug demand jointly, which is not usually available given data limitations in other sources. This importance is reflected in our estimates, where complementarity and substitution effects between drugs arise. Specifically, results for our preferred specification suggest that marijuana and cocaine are complementary in an asymmetric way, such that increases to marijuana prices decrease both marijuana and cocaine consumption, but increases in cocaine price only have sizable effects on the consumption of cocaine, leaving marijuana consumption statistically unchanged. Basuco is instead found to be an inferior substitute for cocaine, though largely unrelated to marijuana. Additionally, our estimates provide a good fit to the implied Engel curves in the data, showcasing the importance of accounting unobserved heterogeneity and price endogeneity to model drug demand in the Colombian setting.

Finally, we used our estimates to evaluate potential effects of a marijuana legalization policies based on different price scenarios the policy could entail. Given Colombia's drug production structure and previous experiences in legalization in developing countries, the most likely scenario of a 50\% decrease in the price of marijuana creates sizable gains for consumers and the government as taxing legal providers of marijuana, with heavy losses incurred by actual illegal drug suppliers. These losses account for the fact that an illegal market of marijuana will continue existing alongside the legal purchasing system, but at a greatly reduced size compared to pre-legalization. Specifically, estimates suggest that revenue gains to the government will be close to \$120 million USD, with suppliers losing upwards of \$127 million USD in revenue. These results suggest that the profitability of the illicit drug market would decrease, and consequently, reduced incentives to control the domestic black market would lead to a decrease in violent crimes associated with local drug trafficking. In addition, the estimated unit-elasticity of marijuana and cocaine in Colombia implies that a legalization policy resulting in a 50\% decrease in marijuana prices would keep the total size of the drug market largely unchanged, shifting from \$226.3 million USD to an estimated \$228.0 million USD post-legalization. This would come at the cost of higher consumption levels among existing regular users and a higher likelihood of attracting new consumers, as suggested by the literature on the effects of marijuana legalization. Nevertheless, we find that these new consumers would not considerably compensate for the lost income of drug dealers.

Given the largely mixed effects of marijuana consumption on both consumer health and job market outcomes found in the literature, our results imply that while policy makers would clearly benefit from a legalization policy and dealers would be clearly negatively impacted, the effects on consumers are not clear cut. On one hand, while consumption of marijuana is expected to almost double for the actual representative agent (given the 50\% price decrease at unit-elasticity), this is valued by consumers at approximately \$363 annual USD of utility-equivalent expenditure, which represents 100\% of current weighted average total drug expenditure, and around 12\% of the yearly minimum wage in 2019. On the other hand, the increased consumption would have additional heterogeneous indirect effects on public health, productivity, educational attainment, etc. Therefore, it is key that a legalization policy is accompanied by additional targeted policy efforts for each population segment, particularly for individuals in the ``hard" and ``at-risk" user groups.

\bibliographystyle{apalike}
\bibliography{Main}

\clearpage

\section*{Appendices}
\begin{appendices}
    \numberwithin{equation}{section}
    \numberwithin{figure}{section}
    \numberwithin{table}{section}
    
    \section{EASI model details}
    \label{Sec:Appendix_EASI}
    \subsection{Stacked EASI equations}
\label{apx_sec:easi_stacked}

\begin{table}[!htbp]
\caption{Summary of price and income effects obtained from the EASI demand system}
\label{Tab:EASI_Summaries}
\centering
\adjustbox{width = \textwidth}{
    \begin{tabular}{ll}
    \hline
    Type & \multicolumn{1}{c}{Hicksian} \\
    \midrule
    Share-price semi-elasticities & $\nabla_{\widetilde{\bm p}}\bm\omega^*(\widetilde{\bm p}, y, \widetilde{\bm h}, \widetilde{\boldsymbol{\varepsilon}}) \coloneqq \bm\Gamma =\sum_{m=0}^{M_p} \widetilde{\bm A}_m h_m^p+\widetilde{\bm B} y$ \\
    & \\
    Income-share semi-elasticities & $\nabla_y\bm\omega^*(\widetilde{\bm p}, y, \widetilde{\bm h}, \widetilde{\boldsymbol{\varepsilon}})=\sum_{r=1}^R \widetilde{\bm b}_r r y^{r-1}+\widetilde{\bm D}\bm h^y+\widetilde{\bm B} \widetilde{\bm p}$ \\
    & \\
    Demand-price elasticities & $\epsilon_{lj}^H \coloneqq -\mathbbm{1}(l=j)+\frac{1}{w_l}\nabla_{\widetilde{\bm p}}\bm\omega^*(\widetilde{\bm p}, y, \bm h, \widetilde{\boldsymbol{\varepsilon}})_{lj}+w_j$ \\
    & \\
    Normalized Slutsky matrix & $\bm{S} \coloneqq \bm\Gamma+\widetilde{\bm w} \widetilde{\bm w}^{\top}-\bm W$\\
    \hline
    Type & \multicolumn{1}{c}{Marshallian} \\
    \hline
    \addlinespace[1mm]
    Share-price semi-elasticities &  $\nabla_{\widetilde{\bm p}}\widetilde{\bm{w}}^*(\widetilde{\bm p}, e, \bm{h}, \widetilde{\boldsymbol{\varepsilon}})=
    \nabla_{\widetilde{\bm p}}\bm \omega^*(\widetilde{\bm{p}},y,\bm {h},\widetilde{\boldsymbol
    {\varepsilon}}) -
    \nabla_{e}\widetilde{\bm{w}}^*(\widetilde{\bm{p}}, e, \bm{h}, \widetilde{\boldsymbol{\varepsilon}}) \bm{\omega}^*(\widetilde{\bm{p}},\bm{h},y,\widetilde{\boldsymbol{\varepsilon}})^{\top}$\\
    & \\
    Income-share semi-elasticities & $\nabla_{e}\widetilde{\bm{w}}^*(\widetilde{\bm p}, e, \bm{h}, \widetilde{\boldsymbol{\varepsilon}})=\left(\bm{I}_{J} + \frac{\nabla_{y}\bm \omega^*(\widetilde{\bm{p}}, \bm h,y,\widetilde{\boldsymbol{\varepsilon}})\widetilde{\bm{p}}^{\top}}{1-\widetilde{\bm{p}}^{\top} \widetilde{\bm{B}} \widetilde{\bm p}/2}\right)^{-1}\left(\frac{\nabla_{y}\bm\omega^*( \widetilde{\bm{p}}, \bm{h},y,\widetilde{\boldsymbol{\varepsilon}})}{1-\widetilde{\bm{p}}^{\top} \widetilde{\bm{B}} \widetilde{\bm p}/2}\right)$\\
    & \\
    Demand-price elasticities & $\widetilde{\boldsymbol{\varepsilon}}_{lj}^M \coloneqq -\mathbbm{1}(l=j) + \frac{1}{w_l}\nabla_{\widetilde{\bm p}}\bm\omega^*(\widetilde{\bm p}, y, \bm h, \widetilde{\boldsymbol{\varepsilon}})_{lj}-\frac{w_j}{w_l}\nabla_{e}\widetilde{\bm{w}}^*(\widetilde{\bm p}, e, \bm{h}, \widetilde{\boldsymbol{\varepsilon}})_l$\\
    & \\
    Demand-income elasticities & $\eta_j^M \coloneqq \frac{1}{w_j}\nabla_{e}\widetilde{\bm{w}}^*(\widetilde{\bm p}, e, \bm{h}, \widetilde{\boldsymbol{\varepsilon}})_j+1$\\
    & \\
    Marshallian Engel curve & $\widetilde{\bm{w}}_i^* = \sum_{r=0}^R\widetilde{\bm{b}}_r e^r_i+\widetilde{\bm{C}}\bm h_i+\widetilde{\bm{D}} \bm{h}_i^y e_i+\widetilde{\boldsymbol{\varepsilon}}_i$\\
    \hline
    \end{tabular}
}
    \begin{tablenotes}
    \item \small{\textbf{Notes:} Price and income effects from exact affine Stone index (EASI) model \citep{Lewbel2009}.}
    \end{tablenotes}
\end{table}

Each equation in the reduced system \eqref{Eq:EASI_reduced} can be written as
\begin{equation}
    \label{Eq:EASI_Equation-by-Equation}
    w_{il}^* = \bm{x}_i\T \bm{\beta}_l^{(1)} + \bm{p}_i^{*\top} \bm{\delta}_l + \varepsilon_{il} \, ,
\end{equation}
where $\bm{x}_i \coloneqq [1, y_i, \ldots, y_i^R, \bm{h}_i\T, \bm{h}_i^{y\top} y_i]\T$ collects all exogenous variables in a vector of dimension $1+R+M+M_y$, $\bm{p}_i^* \coloneqq [\bm{p}_i\T h_{i0}^p, \ldots, \bm{p}_i\T h_{iM_p}^p, \bm{p}_i\T y_i]\T$ collects all endogenous variables in a vector of dimension $s (M_p + 2)$, $\bm{\beta}_l^{(1)} \coloneqq [b_{0l}, \ldots, b_{Rl}, \bm{C}_l\T, \bm{D}_l\T]\T$, and $\bm{\delta}_l \coloneqq [\bm{A}_{0l}\T, \ldots, \bm{A}_{M_p l}\T, \bm{B}_{l}\T]\T$. Stacking across the $l = 1, \ldots, s$ equations yields some redundancy in $\bm{\delta} \coloneqq [\bm{\delta}_1\T, \ldots, \bm{\delta}_s\T]\T$ due to the Slutsky symmetry constraint. Let $P_{s,M_p}$ be a $s^2(M_p+2) \times s^2(M_p+2)$ permutation matrix such that $\bm{\delta}\T = [\vecv(\bm{A}_0)\T, \ldots, \vecv(\bm{A}_{M_p})\T, \vecv(\bm{B})\T] \bm{P}_{s,M_p}\T$ and $\bm{D}_s$ be the $s^2 \times s(s+1)/2$ duplication matrix such that $\bm{D}_s\vecv(\bm{A}) = \vech(\bm{A})$ for any symmetric $s \times s$ matrix $\bm{A}$. Define $\bm{X}_i \coloneqq \bm{I}_s \otimes \bm{x}_i\T$ and
\begin{align*}
    \bm{P}_i^* & \coloneqq (\bm{I}_s \otimes \bm{p}_i^{*\top})\bm{P}_{s,M_p}(\bm{I}_{M_p+2} \otimes \bm{D}_s) \\
    & = \begin{bmatrix} (\bm{I}_s \otimes {\bm{p}_i}\T)\bm{D}_s & (\bm{I}_s \otimes {\bm{p}_i}\T)h_{i1}^p\bm{D}_s & \cdots & (\bm{I}_s \otimes {\bm{p}_i}\T)h_{iM_p}^p\bm{D}_s & (\bm{I}_s \otimes {\bm{p}_i}\T)y_i\bm{D}_s\end{bmatrix}.
\end{align*}
Denote $\bm{\beta}^{(1)} \coloneqq [\bm{\beta}_1^{(1)\top}, \ldots, \bm{\beta}_s^{(1)\top}]\T$ and $\bm{\beta}^{(2)} \coloneqq [\vech(\bm{A}_0)\T, \ldots, \vech(\bm{A}_{M_p})\T, \vech(\bm{B})\T]\T$. With these definitions at hand, we can stack across equations $l = 1, \ldots, s$ in \eqref{Eq:EASI_Equation-by-Equation} to obtain
\begin{align}
    \label{Eq:EASI_LinearReg}
    \bm{w}_i^* = \bm{X}_i \bm{\beta}^{(1)} + \bm{P}_i^* \bm{\beta}^{(2)} + \bm{\varepsilon}_i \, .
\end{align}
The first-stage equation is similarly defined, though it requires less definitions and manipulations as there are no cross-equation restrictions to impose, leading to
\begin{align}
    \label{Eq:FirstStage_LinearReg}
    \bm{p}_i^* = [\bm{I}_{s(M_p+2)} \otimes \bm{x}_i\T] \bm{\gamma}^{(1)} + [\bm{I}_{s(M_p+2)} \otimes \bm{z}_i\T] \bm{\gamma}^{(2)} + u_i \, .
\end{align}
We can finally define $\bm{F}_i \coloneqq [\bm{X}_i, \, \, \bm{P}_i^*]$, $\bm{G}_i \coloneqq [\bm{I}_{s(M_p+2)} \otimes \bm{x}_i\T, \, \, \bm{I}_{s(M_p+2)} \otimes \bm{z}_i\T]$, $\bm{\beta} \coloneqq [\bm{\beta}^{(1) \top}, \bm{\beta}^{(2)\top}]\T$, and $\bm{\gamma} \coloneqq [\bm{\gamma}^{(1) \top}, \bm{\gamma}^{(2)\top}]\T$ to arrive at the compact specification provided in \eqref{Eq:Joint_Linear_System}. 

\subsection{Computational representation}
\label{apx_sec:easi_computational}

An alternative representation that proves useful for the derivations and computational implementation of the system of EASI equation with endogeneity is as follows. Let
\begin{equation*}
    \check{\bm{\beta}}^{(1)} \coloneqq
    \begin{bmatrix}
        \bm{\beta}_1^{(1)} & \cdots & \bm{\beta}_s^{(1)}
    \end{bmatrix} = \begin{bmatrix}
        \bm{b}_0\T \\ \vdots \\ \bm{b}_R\T \\ \bm{C}\T \\ \bm{D}\T 
    \end{bmatrix} \quad \text{and} \quad  \check{\bm{\beta}}^{(2)} \coloneqq
    \begin{bmatrix}
        \bm{A}_0 \\ \vdots \\ \bm{A}_M \\ \bm{B},
    \end{bmatrix}
\end{equation*}
such that we can express \eqref{Eq:EASI_Equation-by-Equation} in row vectors instead of column vectors to obtain
\begin{equation}  
    \label{Eq:EASI_LinearTransposed}
    \bm{w}_i^{*\top} = \bm{x}_i\T \check{\bm{\beta}}^{(1)} + \bm p_i^{*\top} \check{\bm{\beta}}^{(2)} + \varepsilon_i\T.
\end{equation}
Stacking \eqref{Eq:EASI_LinearTransposed} across individuals gives a matrix representation of the EASI  equations with endogeneity:
\begin{align}    
    \mathbf{W}^* & = \mathbf{X} \check{\bm{\beta}}^{(1)} + \mathbf{P}^*\check{\bm{\beta}}^{(2)} + \bm{\varepsilon} = \mathbf{F} \check{\bm{\beta}} + \bm{\varepsilon}
    \label{Eq:EASI_matrix} \, ,
\end{align}
where $\mathbf{F} \coloneqq [\mathbf{X} ,  \mathbf{P}^*]$, $\check{\bm{\beta}} \coloneqq [\check{\bm{\beta}}^{(1)\top}, \check{\bm{\beta}}^{(2)\top}]\T$, and we define
\begin{equation*}
    \underset{(n \times s)}{\mathbf{W}^*} \coloneqq \begin{bmatrix}
        \bm{w}_1^{*\top} \\ \vdots \\ \bm{w}_n^{*\top}
    \end{bmatrix} \, , \quad \underset{(n \times (1+R+M+M_y))}{\mathbf{X}} \coloneqq \begin{bmatrix}
        \bm{x}_1\T \\ \vdots \\ \bm{x}_n\T
    \end{bmatrix} \, , \quad \underset{(n \times s(M_p+2))}{\mathbf{P}^*} \coloneqq \begin{bmatrix}
        \bm p_1^{*\top} \\ \vdots \\ \bm p_n^{*\top}
    \end{bmatrix} \quad \text{and} \quad \underset{(n \times s)}{\bm{\varepsilon}} \coloneqq \begin{bmatrix}
        \varepsilon_1\T \\ \vdots \\ \varepsilon_n\T
    \end{bmatrix}.
\end{equation*}
This is a more efficient matrix representation memory-wise compared to that provided in \eqref{Eq:EASI_LinearReg}, in the sense that $\mathbf{X}$ and $\mathbf{P}^*$ are data matrices with no padding structure in contrast to the rows of $\bm{F}_1, \ldots, \bm{F}_n$. We use these representations in our implementations of the Bayesian sampling algorithms outlined above. Additionally, we can relate the coefficients obtained by both representations since $\vecv(\check{\bm{\beta}}^{(1)}) = \bm{\beta}^{(1)}$ and $\vecv(\check{\bm{\beta}}^{(2)\top}) = (\bm{I}_{M_p+2} \otimes \bm{D}_s)\bm{\beta}^{(2)}$. The corresponding representation for the first stage equation is
\begin{equation}
    \label{Eq:Endog_matrix}
    \mathbf{P}^* = \mathbf{G} \check{\bm{\gamma}} + \mathbf{u},
\end{equation}
where $\check{\bm{\gamma}} \coloneqq [\check{\bm{\gamma}}_1, \ldots, \check{\bm{\gamma}}_s]$, and we define similar matrices for the first-stage:
\begin{equation*}
    \underset{(n \times (1+R+M+M_y+\ell))}{\mathbf{G}} \coloneqq \begin{bmatrix}
        \bm{x}_1\T & \bm{z}_1\T \\ \vdots & \vdots \\ \bm{x}_n\T & \bm{z}_n\T
    \end{bmatrix} \quad \text{and} \quad
    \underset{(n \times s(M_p + 2))}{\mathbf{u}} \coloneqq \begin{bmatrix}
        \bm{u}_1\T \\ \vdots \\ \bm{u}_n\T
    \end{bmatrix}.
\end{equation*}
An additional benefit of this representation --- and the motivation behind our derivation --- is that we can more efficiently compute sums across units $i$ where each element is of the form $\bm{F}_i\T \bm{S} \bm{F}_i$ or $\bm{F}_i \bm{S} \bm{w}_i^*$, for any $s \times s$ symmetric matrix $\bm{S}$ (where $\bm{F}_i$ is as defined in Eq. \ref{Eq:Joint_Linear_System} and has $s$ rows). Specifically, one can easily check the following equalities hold:
\begin{align}
    \label{Eq:Kronecker_Sum_Representation}
    \begin{aligned}
        \sum_{i=1}^{n} \bm{F}_i\T \bm{S} \bm{F}_i & = \bm{S} \otimes \left( \mathbf{F}\T \mathbf{F} \right) \, , \\
        \sum_{i=1}^{n} \bm{F}_i\T \bm{S} \bm{w}_i^{*} & = \left[ \bm{I}_s \otimes \left( \mathbf{F}\T \mathbf{W}^* \right) \right] \vecv (\bm{S}) \, .
    \end{aligned}
\end{align}
In standard packages, these expressions create bottlenecks for computation as they are computed within a loop over $n$ instead of using matrix operations. As the sample size $n$ is large in microeconomic datasets, the memory taken by the stacked padded $\bm{F}_i$ can also be much larger than that of $\mathbf{F}$ if not stored as sparse, and computations need to be adjusted to take this into account. By using simple matrix operations, we achieve both speed and potential memory gains with respect to the commonly implemented non-sparse approaches.
    
    \section{Bayesian implementation}
    \label{Sec:Appendix_Computation}
    Bayesian estimation requires the specification of a generative likelihood for the model of interest and a prior structure reflecting researcher information and problem constraints. The full likelihood in our setup is given by the product of contributions \eqref{Eq:LatentLik} across individuals $i = 1, \ldots, n$. Note that the likelihood is proportional to the density of a normal distribution on $(\bm{w}_i^*, \bm{p}_i^*)$ with mean $\mu_{ij} \coloneqq [(\bm{F}_i \bm{\beta}_j)\T, (\bm{G}_i \bm{\gamma})\T]\T$ and covariance matrix $\bm{\Sigma}_j$. This normality does not formally hold since $\bm{F}_i$ contains the endogenous $\bm{p}_i^*$ and so the mean of this distribution would depend on the value of the random variable. However, the determinant of the Jacobian of the inverse transformation from $(\bm{\varepsilon}_i, \bm{u}_i)$ to $(\bm{w}_i^*, \bm{p}_i^*)$ is 1, such that we can use the normal representation for all our estimation purposes.

We note how this unobserved clustering structure leads directly to a finite mixture approach. Let $\phi_j \coloneqq \Pr(\psi_i = j)$ for all clusters $j = 1, \ldots, J$ and individuals $i = 1, \ldots, n$. Then, a large number of individuals having similar high probabilities of belonging to same clusters suggests that these clusters represent similar sets of preferences. Collecting $\bm{\phi} = (\phi_1, \ldots, \phi_J)$, our current setup implies a mixture distribution of EASI regressions after marginalizing the cluster indicators:
\begin{equation*}
	p(\bm{w}_i^*, \bm{p}_i^* \mid \bm{\beta}_1, \ldots, \bm{\beta}_J, \bm{\gamma}, \bm{\Sigma}_1, \ldots, \bm{\Sigma}_J, \bm{\phi}) \propto \sum_{j=1}^{J} \phi_j \N_{s(M_p+3)}(\bm{w}_i^*, \bm{p}_i^* \mid \bm\mu_{ij}, \bm{\Sigma}_j) \, .
\end{equation*}
Our prior uses standard conjugate distributions for the location and scale parameters due to their tractability and robustness properties.
\begin{align}
    \label{Eq:Priors}
    \begin{aligned}
		\bm{\beta}_j & \overset{iid}{\sim} \N_{d_{\beta}}(\uBar{\bm{\beta}}_j, \uBar{\bm{B}}_j), \quad j = 1, \ldots, J \\
        \bm{\gamma} & \sim \N_{d_{\gamma}}(\uBar{\bm{\gamma}}, \uBar{\bm{\Gamma}}) \\
		\psi_i \mid \bm{\phi} & \overset{iid}{\sim} \Cat(\bm{\phi}), \quad i = 1, \ldots, n \\
		\bm{\phi} & \sim \Dir(\uBar{\bm{\alpha}}) \\
        \pi(\bm{\Sigma}) & = \IW(\bm{\Sigma}_{u u} \mid \uBar{\nu}_{u u} - s, \uBar{\bm{R}}_{u u}) \times \prod_{j = 1}^{J} \IW(\bm{\Sigma}_{j, \varepsilon \varepsilon \cdot u} \mid \uBar{\nu}_j, \uBar{\bm{R}}_{j, \varepsilon \varepsilon \cdot u}) \mathcal{MN}(\bm{\Sigma}_{j, u \cdot u \varepsilon} \mid \uBar{\bm{R}}_{j, u \cdot u \varepsilon}, \uBar{\bm{R}}_{u u}^{-1}, \bm{\Sigma}_{j, \varepsilon \varepsilon \cdot u}) \, ,
	\end{aligned}
\end{align}
where $\IW$ is the inverse-Wishart distribution and $\mathcal{MN}$ is the matrix-normal distribution. The factored prior on $\bm{\Sigma}$ takes into account the model structure by connecting the homogeneous first-stage equation with the cluster-specific structural equations. This is important from an economic perspective as the groups driving similarity in preferences are not generally the same that drive similarities between individuals in the first-stage equation. One could additionally incorporate separate clustering schemes at no additional computational cost, only requiring more complicated notation to define. Combining the likelihood with the joint prior using Bayes' rule, the joint posterior for the augmented parameters can be expressed as
\begin{align}
    \label{Eq:FullPosterior}
    \begin{aligned}
        \pi(\bm{\theta} \mid \mathcal{D}) & \propto \prod_{i=1}^{n} \prod_{l=1}^{s} \left[\mathbbm{1}(w_{il} = 0) \mathbbm{1}(w_{il}^* \leq 0) + \mathbbm{1}\left(w_{il} = \frac{w_{il}^*}{\sum_{j \in L_i} w_{ij}^*} \right)\mathbbm{1}(w_{il}^* > 0) \right] \\
        & \phantom{{} = {}} \times \prod_{j=1}^{J} \mid\bm{\Sigma}_j\mid^{-\frac{n_j}{2}} \exp \left\{ -\frac{1}{2} \sum_{i \in C_j} \begin{bmatrix} \bm{w}_i^* - \bm{F}_i \bm{\beta}_j \\ \bm{p}_i^* - \bm{G}_i \bm{\gamma} \end{bmatrix}\T \begin{bmatrix} \bm{\Sigma}_{j, \varepsilon \varepsilon} & \bm{\Sigma}_{j, \varepsilon u} \\ \bm{\Sigma}_{j, u \varepsilon} & \bm{\Sigma}_{u u} \end{bmatrix}^{-1} \begin{bmatrix} \bm{w}_i^* - \bm{F}_i \bm{\beta}_j \\ \bm{p}_i^* - \bm{G}_i \bm{\gamma} \end{bmatrix} \right\} \\
        & \phantom{{} = {}} \times \prod_{j=1}^{J} \pi(\bm{\beta}_j) \times \prod_{i=1}^{n} \pi(\psi_i \mid \bm{\phi}) \pi(\bm{\phi}) \times \pi(\bm{\gamma}) \pi(\bm{\Sigma}) \, ,
    \end{aligned}
\end{align}
where, for $j = 1,\ldots, J$, $C_j = \{i: \psi_i = j\}$ is the $j$-th cluster of observations and $n_j = |C_j| = \sum_{i=1}^{n} \mathbbm{1}(\psi_i = j)$ is the number of individuals in the cluster. For simplicity in the notation that follows, we assume that for each individual the share variables are ordered such that the first set is all the goods with no consumption and the remaining are for the goods with positive consumption. That is, for each individual $i$ there are $N_i = s - |L_i|$ shares with $w_{il} = 0$ for $l = 1, \ldots, N_i$ and the remaining are positive; $w_{il} > 0$ for $l = N_i+1, \ldots, s$. We can then partition
\begin{equation}
    \label{Eq:PartitionShares}
	\bm{F}_i \bm{\beta} = \begin{bmatrix} \bm{F}_i^{-} \bm{\beta}_j^{-} \\ \bm{F}_i^{+} \bm{\beta}_j^{+} \end{bmatrix} \quad \text{and} \quad \bm{\Sigma}_j = \begin{bmatrix} \bm{\Sigma}_{j, \bm{\varepsilon}_i^{-} \bm{\varepsilon}_i^{-}} & \bm{\Sigma}_{j, \bm{\varepsilon}_i^{-} \tau_i} \\ \bm{\Sigma}_{j, \tau_i \bm{\varepsilon}_i^{-}} & \bm{\Sigma}_{j, \tau_i \tau_i} \end{bmatrix},
\end{equation}
where $\bm{\varepsilon}_i^{-} = [\varepsilon_{i1}, \ldots, \varepsilon_{iN_i}]$, $\bm{\varepsilon}_i^{+} = [\varepsilon_{i(N_i+1)}, \ldots, \varepsilon_{is}]$, $\bm{\tau}_i = [\bm{\varepsilon}_i^{+\top}, \bm{u}_i\T]\T$, and $\bm{F}_i^{-}$, $\bm{F}_i^{+}$, $\bm{\beta}_j^{-}$, and $\bm{\beta}_j^{+}$ are appropriate partitions of the variables and coefficients, respectively. For each individual $i$, we now provide the posterior of the latent shares $\bm{w}_i^*$ conditional on model parameters and data. Using the partitions given in \eqref{Eq:PartitionShares}, and assuming the individual belongs to an arbitrary cluster such that $i \in C_j$, the relevant conditional posterior is
\begin{multline}
    \label{Eq:PosteriorShares}
    \bm{w}_i^{*-} \mid \theta_{-w_i^{*-}}, \mathcal{D} \sim  \\
    \text{TM}\N_{(-\infty, 0]^{N_i}}\left(\bm{F}_i^{-} \bm{\beta}_j^{-} + \bm{\Sigma}_{j, \bm{\varepsilon}_i^{-} \bm{\tau}_i} \bm{\Sigma}_{j, \tau_i \tau_i}^{-1} \begin{bmatrix}
        \bm{w}_i^{+} - \bm{F}_i^{+} \bm{\beta}_j^{+} \\
        \bm{p}_i^* - \bm{G}_i \bm{\gamma}
    \end{bmatrix}, \bm{\Sigma}_{j, \bm{\varepsilon}_i^{-} \bm{\varepsilon}_i^{-}} - \bm{\Sigma}_{j, \bm{\varepsilon}_i^{-} \tau_i} \bm{\Sigma}_{j, \tau_i \tau_i}^{-1} \bm{\Sigma}_{j, \tau_i \bm{\varepsilon}_i^{-}} \right) \, ,
\end{multline}
where $\text{TM}\N_\mathcal{S}$ represents a multivariate normal distribution truncated to region $\mathcal{S}$. In our case, $\mathcal{S} = \times_{j = 1}^{N_i} (-\infty, 0]$ so that the truncation region is simply the Cartesian product of $N_i$ negative half-open intervals to 0. For computational stability, we center the posterior at the observed positive shares ($\bm{w}_i^{+}$) rather than the latent ones ($\bm{w}_i^{*+}$). Finally, conditional on the drawn negative latent shares, the positive share components have a degenerate density that places all mass at $w_{il}^{*+} = (1 - \sum_{j=1}^{N_i}  w_{ij}^{*-} ) w_{i(N_i + l)}$ for each $l = 1, \ldots, |L_i|$.

    \section{Survey reliability}
    \label{Sec:Appendix_Survey_Reliability}
    In the main body of the text, we use data from the National Survey of the Consumption of Psychoactive Substances performed in 2019 (\textit{ENCSPA}). This
is a nationally representative survey aiming to measure the consumption of legal and illegal psychoactive
substances. The survey design was mindful of the potential for individuals to misrepresent sensitive information related to illicit drug use, implementing a scheme where enumerators privately performed the survey and returned multiple times to homes where selected participants were absent. Descriptive statistics and model estimates like price-elasticities of demand obtained from the full sample are within expected range according to literature and anecdotal evidence. This reduces potential concerns of data reliability or substantial bias stemming from under-reporting \citep{Needle1983, Harrison1993, Johnson2005, Fadnes2009}.

In this Appendix, we examine reliability of survey responses by conducting an analysis of similar summaries as in the main text on a sub-sample of single-dweller households. As individuals in these households live alone, they have a reduced incentive to distort their responses or provide inaccurate information. By comparing responses of consumers in single-dweller vs multiple-dweller households, we obtain further reassurance on the reliability of survey data and estimates obtained therefrom. While this analysis does not rule additional sources of under-reporting, our extensive use of covariates in the demand estimation system and additional unobserved heterogeneity should capture similar sources of variation controlling for additional reliability effects.

Figure \ref{Tab:Summary_Statistics_SingleHousehold} presents aggregate consumption and price statistics for the single-dweller household subsample. Comparing to Table \ref{Tab:Summary_Statistics} in the main body, consumption shares and total quantities present similar distributions. Reassuringly, even the rate of non-consumption across both samples is similar, with the proportions of zeros across remaining stable. The prices paid for marijuana and basuco are slightly higher in comparison, though combined with the fact quantities remain stable, this points to a potential effect of higher demanded quality or potency that is also correlated with age.

\begin{table}[!htbp]
    \begin{threeparttable}
    \centering
    \caption{Summary statistics for shares of drug consumption and prices of single individual households}
    \label{Tab:Summary_Statistics_SingleHousehold}
    \begin{tabular}{lcccc}
    \toprule
        Shares (percentage points, pp.) & Mean & Std. Dev. & \multicolumn{2}{c}{Proportion with no expenditure} \\
        \midrule
        Marijuana & 84.91 & 31.20 & \multicolumn{2}{c}{7.50} \\
        Cocaine & 13.06 & 29.59 & \multicolumn{2}{c}{77.08} \\
        Basuco & 2.03 & 11.69 & \multicolumn{2}{c}{95.42} \\
        \midrule
        Quantity (grams per month) & Mean & Std. Dev. & Min. & Max. \\
        \midrule
        Marijuana & 34.01 & 62.02 & 0.00 & 400.00 \\
        Cocaine & 1.28 & 4.67 & 0.00 & 40.00 \\
        Basuco & 0.68 & 5.10 & 0.00 & 60.00 \\
        \midrule
        Prices (cents per gram, USD \textcent/gr.) & Mean & Std. Dev. & Min. & Max. \\
        \midrule
        Marijuana & 91.91 & 59.89 & 15.23 & 304.66 \\
        Cocaine & 335.12 & 180.44 & 60.93 & 913.97 \\
        Basuco & 73.14 & 38.81 & 30.47 & 304.66 \\
      \bottomrule
    \end{tabular}
    \begin{tablenotes}
    \item \small{\textbf{Notes:} Total sample size includes 240 consumers from ENCSPA survey in Colombia. Monthly quantity is conditional on consumption. Prices are converted to dollars using the average exchange rate, which is equivalent to COP/USD 3282.39 in 2019. Trends for single-individual households remain similar to those of the full sample.}
    \end{tablenotes}
    \end{threeparttable}
\end{table}

To provide a more thorough overview of the differences between the two samples, we conduct mean-difference tests between all relevant consumption and demographic characteristics in the survey. Unsurprisingly, an individual living alone is classified as head of household at a larger rate and does not have their parents at home. These individuals are also older, more likely to be employed, more likely to be of higher socioeconomic status, have higher education, and have been using drugs for longer; all effects likely correlated to age. A majority of sensitive variables that individuals could misconstrue in their survey interviews, such as consumption amounts or interactions with peers, are found to be statistically indistinguishable across samples.

\begin{figure}
    \centering
    \caption{Balance test between single and multiple-dweller households}
    \label{Fig:Balance_Tstats_SingleHouseholds}
    \begin{adjustbox}{max totalsize = {\textwidth}{\textheight}, center}    
        \includegraphics[page = 1]{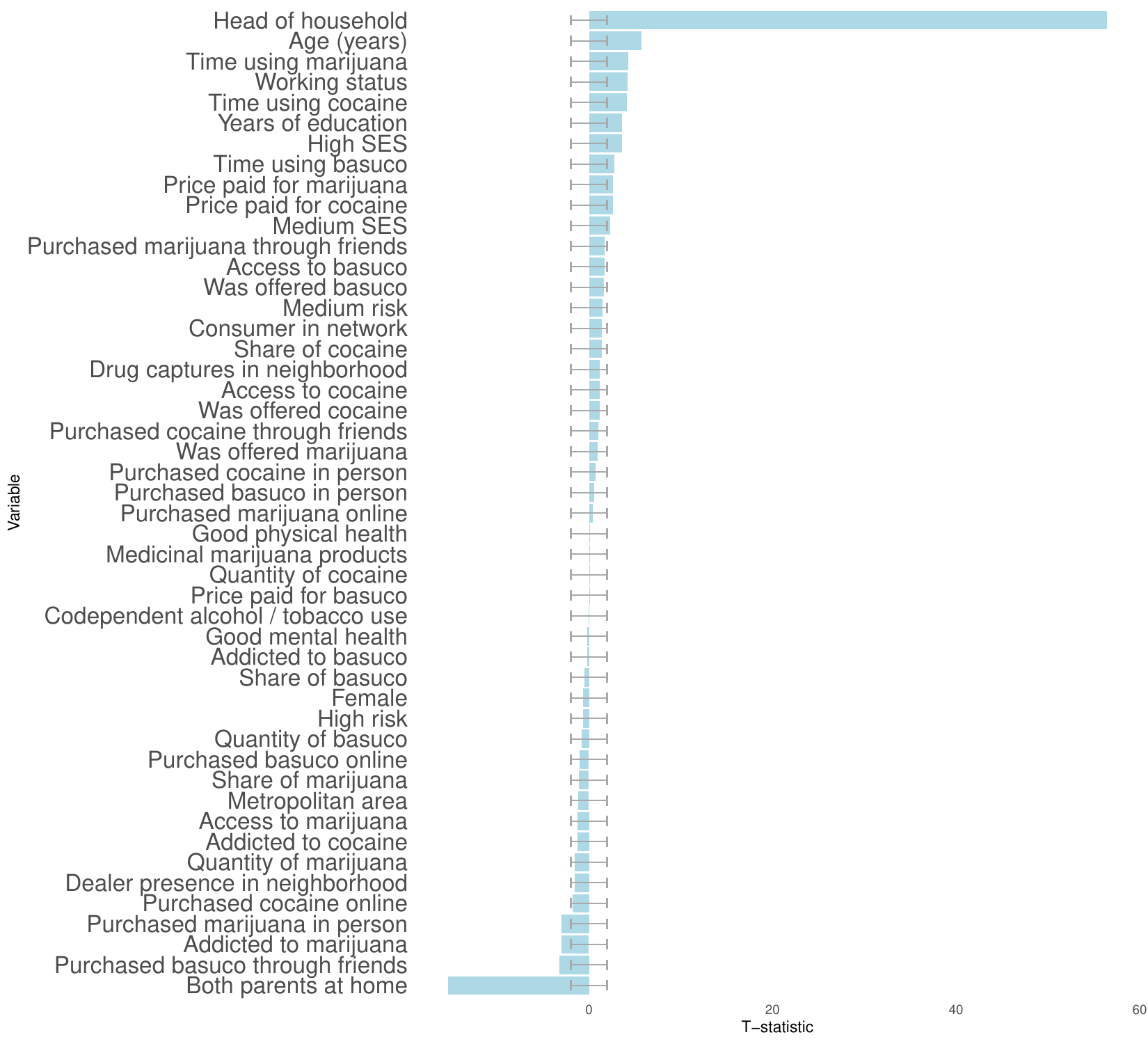}
    \end{adjustbox}
    \begin{minipage}{\textwidth}
        \begin{tablenotes}
            \item \small{\textbf{Notes:} T-statistic of a two-sided mean-difference test between households with one or multiple dwellers. Plotted interval provides critical values for significance of the mean-difference at 5\% level. No large noticeable differences aside from variables that naturally relate to individuals living on their own, such as the being head of household by construction, being older, more likely to work, and using drugs for longer. Some remaining differences in prices paid for drugs remain but not for quantities consumed, likely explained by quality or search effects on the consumer side.}
        \end{tablenotes}
    \end{minipage}
\end{figure}

\begin{figure}[htbp]
    \centering
    \caption{Shares of total expenditure on drugs and users from single-dweller households in Colombia by age groups and drugs}
    \label{Fig:SingleHousehold_Expenditure_Users}
    \begin{subfigure}{0.49\linewidth}
        \includegraphics[width = \linewidth, trim={18pt 20pt 20pt 38pt}, clip]{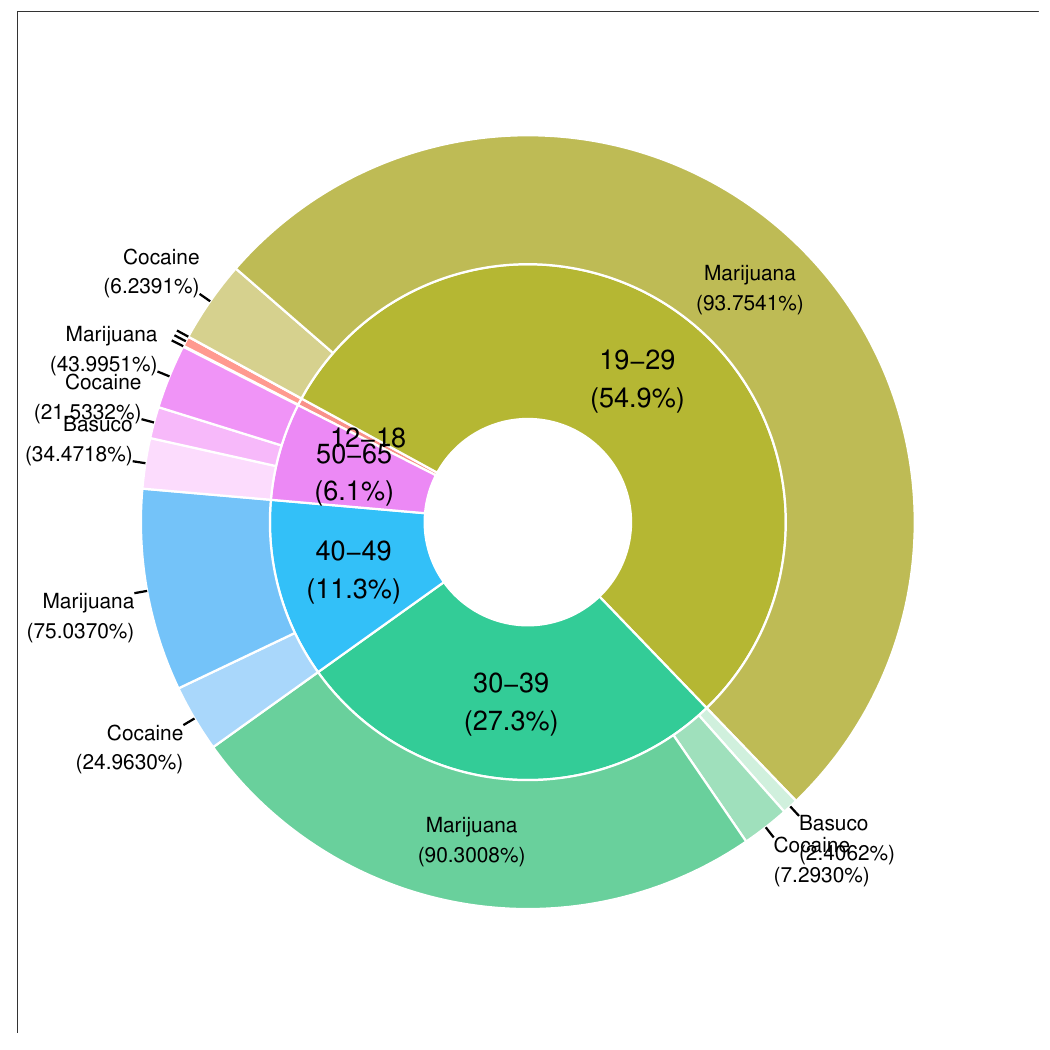}
        \caption{Expenditures}
        \label{Fig:SingleHousehold_Expenditure}
    \end{subfigure}
    \begin{subfigure}{0.49\linewidth}
        \includegraphics[width = \linewidth, trim={18pt 20pt 20pt 38pt}, clip]{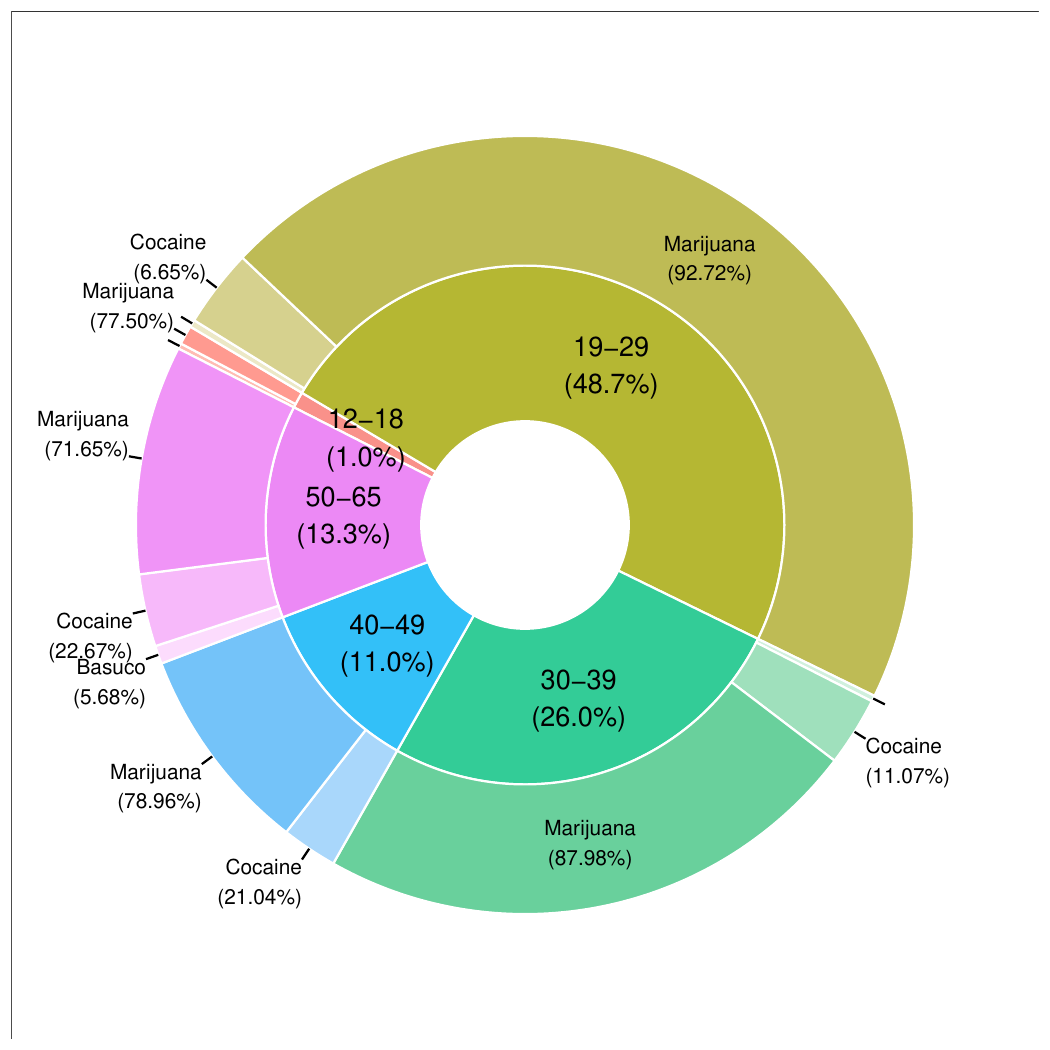}
        \caption{Users}
        \label{Fig:SingleHousehold_Users}
    \end{subfigure}
    \begin{tablenotes}
        \item \small{\textbf{Notes:} In 2019, total drug market expenditure in Colombia was approximately 150.2 million USD out of the total 225.7, all from users in single-dweller households amounting to 49,884 out of the total 633,490 users. We classify individuals according to their most consumed drug.}
    \end{tablenotes}
\end{figure}

Finally, Figure \ref{Fig:SingleHousehold_Expenditure_Users} showcase the shares of users and expenditure from single-dweller households, akin to Figure \ref{Fig:Drug_Expenditure_Users}. Aside from the reduction of the share from individuals under 18 years old, the distribution of both variables remains consistent with the full sample distribution.

    \section{Additional results}
    \label{Sec:Appendix_Results}
\subsection{EASI Demand System Estimates}
\label{apx_sec:additional_easi_estimates}

\begin{longtable}[c]{lcc}
    \caption{Bayesian estimates of first-stage coefficients \label{Tab:First_Stage}} \\
    \toprule
    \multirow{2}{*}{Covariate} & \multicolumn{2}{c}{Dependent Variable} \\
    \cmidrule{2-3}      & Relative price marijuana & Relative price cocaine \\
    \midrule
    \multicolumn{3}{c}{Included instruments} \\
    \midrule
    \multirow{2}{*}{Constant} & $-0.997$ & $-1.419$ \\
          & $(-1.311, -0.684)$ & $(-1.846, -1.020)$ \\
    \multirow{2}{*}{Dealer presence} & $0.051$ & $0.094$ \\
          & $(-0.012, 0.111)$ & $(0.014, 0.179)$ \\
    \multirow{2}{*}{Medium risk} & $-0.040$ & $-0.148$ \\
          & $(-0.281, 0.201)$ & $(-0.469, 0.177)$ \\
    \multirow{2}{*}{High risk} & $-0.050$ & $-0.183$ \\
          & $(-0.258, 0.156)$ & $(-0.452, 0.089)$ \\
    \multirow{2}{*}{Consumer in network} & $-0.150$ & $-0.131$ \\
          & $(-0.275, -0.024)$ & $(-0.292, 0.039)$ \\
    \multirow{2}{*}{Female} & $-0.017$ & $0.017$ \\
          & $(-0.087, 0.053)$ & $(-0.080, 0.111)$ \\
    \multirow{2}{*}{Both parents at home} & $-0.030$ & $-0.065$ \\
          & $(-0.119, 0.059)$ & $(-0.183, 0.052)$ \\
    \multirow{2}{*}{Medicinal marijuana products} & $-0.015$ & $-0.055$ \\
          & $(-0.077, 0.045)$ & $(-0.137, 0.027)$ \\
    \multirow{2}{*}{Good physical health} & $-0.029$ & $-0.012$ \\
          & $(-0.107, 0.047)$ & $(-0.113, 0.090)$ \\
    \multirow{2}{*}{Good mental health} & $0.008$ & $0.002$ \\
          & $(-0.058, 0.075)$ & $(-0.088, 0.092)$ \\
    \multirow{2}{*}{Years of education} & $0.015$ & $0.037$ \\
          & $(0.006, 0.024)$ & $(0.024, 0.050)$ \\
    \multirow{2}{*}{Head of household} & $-0.029$ & $-0.040$ \\
          & $(-0.098, 0.041)$ & $(-0.132, 0.052)$ \\
    \multirow{2}{*}{Medium SES} & $0.266$ & $0.290$ \\
          & $(0.197, 0.337)$ & $(0.197, 0.383)$ \\
    \multirow{2}{*}{High SES} & $-0.302$ & $0.422$ \\
          & $(-0.441, -0.162)$ & $(0.236, 0.606)$ \\
    \multirow{2}{*}{Working status} & $-0.043$ & $-0.000003$ \\
          & $(-0.109, 0.023)$ & $(-0.086878, 0.086823)$ \\
    \multirow{2}{*}{Age minus 30} & $0.004$ & $0.0003$ \\
          & $(-0.001, 0.010)$ & $(-0.0065, 0.0073)$ \\
    \multirow{2}{*}{Codependent substance use} & $0.032$ & $0.059$ \\
          & $(-0.052, 0.116)$ & $(-0.052, 0.167)$ \\
    \multirow{2}{*}{Metropolitan area} & $-0.426$ & $-0.433$ \\
          & $(-0.494, -0.358)$ & $(-0.522, -0.344)$ \\
    \multirow{2}{*}{Time using marijuana} & $-0.005$ & $-0.002$ \\
          & $(-0.010, 0.000)$ & $(-0.010, 0.005)$ \\
    \multirow{2}{*}{Time using cocaine} & $0.003$ & $0.003$ \\
          & $(-0.002, 0.008)$ & $(-0.004, 0.010)$ \\
    \multirow{2}{*}{Time using basuco} & $-0.002$ & $-0.004$ \\
          & $(-0.008, 0.005)$ & $(-0.012, 0.005)$ \\
    \multirow{2}{*}{Was offered marijuana} & $0.011$ & $0.022$ \\
          & $(-0.089, 0.113)$ & $(-0.110, 0.163)$ \\
    \multirow{2}{*}{Was offered cocaine} & $0.052$ & $0.066$ \\
          & $(-0.021, 0.126)$ & $(-0.035, 0.165)$ \\
    \multirow{2}{*}{Was offered basuco} & $-0.013$ & $-0.028$ \\
          & $(-0.100, 0.069)$ & $(-0.141, 0.083)$ \\
    \multirow{2}{*}{Obtained marijuana online} & $0.103$ & $0.193$ \\
          & $(-0.002, 0.209)$ & $(0.055, 0.333)$ \\
    \multirow{2}{*}{Obtained marijuana in person} & $-0.006$ & $0.033$ \\
          & $(-0.083, 0.074)$ & $(-0.074, 0.140)$ \\
    \multirow{2}{*}{Obtained marijuana through friends} & $-0.010$ & $0.021$ \\
          & $(-0.088, 0.068)$ & $(-0.082, 0.124)$ \\
    \multirow{2}{*}{Obtained marijuana other / missing} & $-0.030$ & $-0.062$ \\
          & $(-0.207, 0.141)$ & $(-0.296, 0.168)$ \\
    \multirow{2}{*}{Obtained cocaine online} & $0.062$ & $0.378$ \\
          & $(-0.134, 0.258)$ & $(0.115, 0.637)$ \\
    \multirow{2}{*}{Obtained cocaine in person} & $-0.017$ & $-0.180$ \\
          & $(-0.130, 0.096)$ & $(-0.330, -0.031)$ \\
    \multirow{2}{*}{Obtained cocaine through friends} & $0.039$ & $-0.090$ \\
          & $(-0.069, 0.146)$ & $(-0.233, 0.054)$ \\
    \multirow{2}{*}{Obtained cocaine other / missing} & $0.305$ & $1.501$ \\
          & $(-0.283, 0.898)$ & $(0.726, 2.284)$ \\
    \multirow{2}{*}{Obtained basuco online} & $0.250$ & $-0.301$ \\
          & $(-0.784, 1.286)$ & $(-1.680, 1.090)$ \\
    \multirow{2}{*}{Obtained basuco in person} & $-0.042$ & $0.139$ \\
          & $(-0.240, 0.156)$ & $(-0.116, 0.398)$ \\
    \multirow{2}{*}{Obtained basuco through friends} & $-0.160$ & $0.058$ \\
          & $(-0.394, 0.072)$ & $(-0.241, 0.367)$ \\
    \multirow{2}{*}{Indirect utility ($y$)} & $-0.187$ & $-0.026$ \\
          & $(-0.257, -0.117)$ & $(-0.117, 0.066)$ \\
    \multirow{2}{*}{Indirect utility$^2$} & $0.023$ & $-0.002$ \\
          & $(0.014, 0.032)$ & $(-0.015, 0.010)$ \\
    \multirow{2}{*}{Indirect utility$^3$} & $-0.005$ & $-0.003$ \\
          & $(-0.008, -0.002)$ & $(-0.007, 0.002)$ \\
    \midrule
    \multicolumn{3}{c}{Excluded instruments} \\
    \midrule
    \multirow{2}{*}{$\log(1 + \text{Drug captures})$} & $0.063$ & $0.125$ \\
          & $(0.041, 0.085)$ & $(0.095, 0.155)$ \\
    \multirow{2}{*}{Distance-weighted price of marijuana} & $0.0003$ & $-0.00001$ \\
          & $(0.0003, 0.0003)$ & $(-0.00003, 0.00002)$ \\
    \multirow{2}{*}{Distance-weighted price of cocaine} & $0.000004$ & $0.000009$ \\
          & $(-0.000002, 0.000011)$ & $(0.000000, 0.000017)$ \\
    \multirow{2}{*}{$\log(1 + \text{Drug captures}) \times y$} & $-0.006$ & $0.001$ \\
          & $(-0.016, 0.004)$ & $(-0.012, 0.014)$ \\
    \multirow{2}{*}{$\text{Distance-weighted price of marijuana} \times y$} & $0.00007$ & $0.000007$ \\
          & $(0.00006, 0.00008)$ & $(-0.000008, 0.000022)$ \\
    \multirow{2}{*}{$\text{Distance-weighted price of cocaine} \times y$} & $0.0000007$ & $-0.0000007$ \\
          & $(-0.0000023, 0.0000037)$ & $(-0.0000047, 0.0000032)$ \\
    \midrule
    \multirow{2}{*}{Covariate} & \multicolumn{2}{c}{Dependent Variable} \\
    \cmidrule{2-3}      & $\text{Relative price marijuana} \times y$ & $\text{Relative price cocaine} \times y$ \\
    \midrule
    \multicolumn{3}{c}{Included instruments} \\
    \midrule
    \multirow{2}{*}{Constant} & $-0.003$ & $-0.078$ \\
          & $(-0.436, 0.429)$ & $(-0.555, 0.382)$ \\
    \multirow{2}{*}{Dealer presence} & $-0.019$ & $0.051$ \\
          & $(-0.126, 0.088)$ & $(-0.075, 0.177)$ \\
    \multirow{2}{*}{Medium risk} & $-0.034$ & $-0.034$ \\
          & $(-0.361, 0.299)$ & $(-0.387, 0.323)$ \\
    \multirow{2}{*}{High risk} & $-0.076$ & $-0.058$ \\
          & $(-0.366, 0.216)$ & $(-0.369, 0.256)$ \\
    \multirow{2}{*}{Consumer in network} & $-0.130$ & $0.059$ \\
          & $(-0.327, 0.069)$ & $(-0.173, 0.285)$ \\
    \multirow{2}{*}{Female} & $-0.032$ & $-0.067$ \\
          & $(-0.149, 0.085)$ & $(-0.207, 0.070)$ \\
    \multirow{2}{*}{Both parents at home} & $-0.058$ & $-0.084$ \\
          & $(-0.194, 0.076)$ & $(-0.245, 0.073)$ \\
    \multirow{2}{*}{Medicinal marijuana products} & $-0.029$ & $0.042$ \\
          & $(-0.137, 0.075)$ & $(-0.085, 0.166)$ \\
    \multirow{2}{*}{Good physical health} & $0.048$ & $-0.079$ \\
          & $(-0.075, 0.173)$ & $(-0.227, 0.072)$ \\
    \multirow{2}{*}{Good mental health} & $-0.045$ & $0.121$ \\
          & $(-0.153, 0.066)$ & $(-0.008, 0.254)$ \\
    \multirow{2}{*}{Years of education} & $0.013$ & $0.025$ \\
          & $(-0.006, 0.030)$ & $(0.003, 0.047)$ \\
    \multirow{2}{*}{Head of household} & $-0.072$ & $-0.035$ \\
          & $(-0.187, 0.042)$ & $(-0.177, 0.102)$ \\
    \multirow{2}{*}{Medium SES} & $-0.069$ & $0.025$ \\
          & $(-0.187, 0.048)$ & $(-0.121, 0.160)$ \\
    \multirow{2}{*}{High SES} & $0.226$ & $-0.081$ \\
          & $(0.029, 0.430)$ & $(-0.304, 0.143)$ \\
    \multirow{2}{*}{Working status} & $-0.004$ & $0.051$ \\
          & $(-0.112, 0.103)$ & $(-0.078, 0.181)$ \\
    \multirow{2}{*}{Age minus 30} & $0.014$ & $0.00001$ \\
          & $(0.002, 0.024)$ & $(-0.01493, 0.01387)$ \\
    \multirow{2}{*}{Codependent substance use} & $0.023$ & $-0.027$ \\
          & $(-0.109, 0.159)$ & $(-0.187, 0.128)$ \\
    \multirow{2}{*}{Metropolitan area} & $-0.016$ & $-0.148$ \\
          & $(-0.129, 0.101)$ & $(-0.285, -0.007)$ \\
    \multirow{2}{*}{Time using marijuana} & $-0.014$ & $-0.008$ \\
          & $(-0.025, -0.003)$ & $(-0.021, 0.006)$ \\
    \multirow{2}{*}{Time using cocaine} & $-0.002$ & $-0.015$ \\
          & $(-0.013, 0.009)$ & $(-0.028, -0.001)$ \\
    \multirow{2}{*}{Time using basuco} & $0.006$ & $0.010$ \\
          & $(-0.009, 0.020)$ & $(-0.008, 0.028)$ \\
    \multirow{2}{*}{Was offered marijuana} & $0.002$ & $0.015$ \\
          & $(-0.154, 0.157)$ & $(-0.165, 0.193)$ \\
    \multirow{2}{*}{Was offered cocaine} & $0.047$ & $0.016$ \\
          & $(-0.075, 0.174)$ & $(-0.130, 0.161)$ \\
    \multirow{2}{*}{Was offered basuco} & $0.033$ & $0.003$ \\
          & $(-0.096, 0.163)$ & $(-0.145, 0.155)$ \\
    \multirow{2}{*}{Obtained marijuana online} & $-0.052$ & $0.201$ \\
          & $(-0.209, 0.114)$ & $(0.016, 0.394)$ \\
    \multirow{2}{*}{Obtained marijuana in person} & $-0.106$ & $-0.008$ \\
          & $(-0.235, 0.021)$ & $(-0.161, 0.139)$ \\
    \multirow{2}{*}{Obtained marijuana through friends} & $-0.081$ & $0.065$ \\
          & $(-0.204, 0.040)$ & $(-0.083, 0.207)$ \\
    \multirow{2}{*}{Obtained marijuana other / missing} & $-0.178$ & $0.253$ \\
          & $(-0.443, 0.097)$ & $(-0.054, 0.560)$ \\
    \multirow{2}{*}{Obtained cocaine online} & $0.026$ & $0.058$ \\
          & $(-0.220, 0.279)$ & $(-0.206, 0.335)$ \\
    \multirow{2}{*}{Obtained cocaine in person} & $0.077$ & $-0.240$ \\
          & $(-0.088, 0.241)$ & $(-0.431, -0.059)$ \\
    \multirow{2}{*}{Obtained cocaine through friends} & $0.060$ & $-0.119$ \\
          & $(-0.096, 0.218)$ & $(-0.299, 0.057)$ \\
    \multirow{2}{*}{Obtained cocaine other / missing} & $-0.935$ & $1.182$ \\
          & $(-1.648, -0.214)$ & $(0.468, 1.956)$ \\
    \multirow{2}{*}{Obtained basuco online} & $0.222$ & $-0.299$ \\
          & $(-0.997, 1.425)$ & $(-1.481, 0.886)$ \\
    \multirow{2}{*}{Obtained basuco in person} & $-0.111$ & $0.141$ \\
          & $(-0.360, 0.146)$ & $(-0.125, 0.415)$ \\
    \multirow{2}{*}{Obtained basuco through friends} & $0.262$ & $-0.222$ \\
          & $(-0.022, 0.551)$ & $(-0.523, 0.077)$ \\
    \multirow{2}{*}{Indirect utility ($y$)} & $-0.173$ & $-0.179$ \\
          & $(-0.274, -0.070)$ & $(-0.299, -0.065)$ \\
    \multirow{2}{*}{Indirect utility$^2$} & $-0.138$ & $-0.090$ \\
          & $(-0.159, -0.118)$ & $(-0.116, -0.066)$ \\
    \multirow{2}{*}{Indirect utility$^3$} & $-0.009$ & $-0.017$ \\
          & $(-0.016, -0.001)$ & $(-0.026, -0.008)$ \\
    \midrule
    \multicolumn{3}{c}{Excluded instruments} \\
    \midrule
    \multirow{2}{*}{$\log(1 + \text{Drug captures})$} & $-0.002$ & $0.016$ \\
          & $(-0.032, 0.028)$ & $(-0.016, 0.049)$ \\
    \multirow{2}{*}{Distance-weighted price of marijuana} & $0.0001$ & $-0.00009$ \\
          & $(0.0001, 0.0002)$ & $(-0.00013, -0.00005)$ \\
    \multirow{2}{*}{Distance-weighted price of cocaine} & $0.000004$ & $0.000002$ \\
          & $(-0.000004, 0.000013)$ & $(-0.000007, 0.000010)$ \\
    \multirow{2}{*}{$\log(1 + \text{Drug captures}) \times y$} & $-0.026$ & $0.033$ \\
          & $(-0.039, -0.013)$ & $(0.019, 0.048)$ \\
    \multirow{2}{*}{$\text{Distance-weighted price of marijuana} \times y$} & $0.0002$ & $-0.0001$ \\
          & $(0.0002, 0.0002)$ & $(-0.0001, -0.0001)$ \\
    \multirow{2}{*}{$\text{Distance-weighted price of cocaine} \times y$} & $0.0000007$ & $-0.000002$ \\
          & $(-0.0000033, 0.0000046)$ & $(-0.000006, 0.000002)$ \\
    \bottomrule
    \caption*{\small \textbf{Notes:} Posterior median coefficients from the Bayesian first-stage regressions on each of the endogenous variables of the EASI system. 95\% posterior quantile credibility intervals are presented underneath the coefficients. There is only one set of results as our main specification maintains homogeneous coefficients in the first stage.}
\end{longtable}

\begin{longtable}[c]{lcc}
    \caption{Bayesian estimates of EASI coefficients across clusters \label{Tab:Structural}} \\
    \toprule
    \multirow{2}{*}{Covariate} & \multicolumn{2}{c}{``Soft'' Cluster} \\
    \cmidrule{2-3}      & Share of marijuana & Share of cocaine \\
    \midrule
    \multicolumn{3}{c}{Exogenous} \\
    \midrule
    \multirow{2}{*}{Constant} & $0.953$ & $0.043$ \\
          & $(0.907, 1.001)$ & $(-0.007, 0.093)$ \\
    \multirow{2}{*}{Dealer presence} & $-0.005$ & $0.004$ \\
          & $(-0.014, 0.005)$ & $(-0.005, 0.014)$ \\
    \multirow{2}{*}{Medium risk} & $-0.005$ & $0.004$ \\
          & $(-0.041, 0.032)$ & $(-0.032, 0.041)$ \\
    \multirow{2}{*}{High risk} & $0.003$ & $-0.002$ \\
          & $(-0.027, 0.034)$ & $(-0.033, 0.028)$ \\
    \multirow{2}{*}{Consumer in network} & $-0.007$ & $0.006$ \\
          & $(-0.025, 0.012)$ & $(-0.012, 0.025)$ \\
    \multirow{2}{*}{Female} & $0.000$ & $0.002$ \\
          & $(-0.011, 0.010)$ & $(-0.009, 0.013)$ \\
    \multirow{2}{*}{Both parents at home} & $0.002$ & $-0.001$ \\
          & $(-0.011, 0.015)$ & $(-0.014, 0.013)$ \\
    \multirow{2}{*}{Medicinal marijuana products} & $0.004$ & $-0.004$ \\
          & $(-0.006, 0.013)$ & $(-0.014, 0.006)$ \\
    \multirow{2}{*}{Good physical health} & $-0.003$ & $0.004$ \\
          & $(-0.015, 0.009)$ & $(-0.008, 0.016)$ \\
    \multirow{2}{*}{Good mental health} & $0.001$ & $-0.001$ \\
          & $(-0.010, 0.011)$ & $(-0.012, 0.010)$ \\
    \multirow{2}{*}{Years of education} & $0.000$ & $0.000$ \\
          & $(-0.002, 0.001)$ & $(-0.001, 0.002)$ \\
    \multirow{2}{*}{Head of household} & $0.000$ & $0.002$ \\
          & $(-0.011, 0.010)$ & $(-0.009, 0.013)$ \\
    \multirow{2}{*}{Medium SES} & $0.002$ & $-0.001$ \\
          & $(-0.008, 0.013)$ & $(-0.013, 0.012)$ \\
    \multirow{2}{*}{High SES} & $0.010$ & $-0.008$ \\
          & $(-0.013, 0.033)$ & $(-0.032, 0.016)$ \\
    \multirow{2}{*}{Working status} & $-0.003$ & $0.002$ \\
          & $(-0.013, 0.007)$ & $(-0.008, 0.013)$ \\
    \multirow{2}{*}{Age minus 30} & $-0.002$ & $0.002$ \\
          & $(-0.003, -0.001)$ & $(0.001, 0.003)$ \\
    \multirow{2}{*}{Codependent substance use} & $0.006$ & $-0.006$ \\
          & $(-0.006, 0.018)$ & $(-0.019, 0.006)$ \\
    \multirow{2}{*}{Metropolitan area} & $0.005$ & $-0.005$ \\
          & $(-0.005, 0.016)$ & $(-0.017, 0.006)$ \\
    \multirow{2}{*}{Time using marijuana} & $0.003$ & $-0.002$ \\
          & $(0.002, 0.004)$ & $(-0.003, -0.001)$ \\
    \multirow{2}{*}{Time using cocaine} & $0.000$ & $0.000$ \\
          & $(-0.001, 0.001)$ & $(-0.001, 0.001)$ \\
    \multirow{2}{*}{Time using basuco} & $-0.001$ & $0.001$ \\
          & $(-0.002, 0.000)$ & $(0.000, 0.002)$ \\
    \multirow{2}{*}{Was offered marijuana} & $0.006$ & $-0.007$ \\
          & $(-0.010, 0.021)$ & $(-0.021, 0.009)$ \\
    \multirow{2}{*}{Was offered cocaine} & $-0.010$ & $0.009$ \\
          & $(-0.020, 0.001)$ & $(-0.002, 0.021)$ \\
    \multirow{2}{*}{Was offered basuco} & $0.009$ & $-0.010$ \\
          & $(-0.004, 0.023)$ & $(-0.024, 0.004)$ \\
    \multirow{2}{*}{Obtained marijuana online} & $0.017$ & $-0.017$ \\
          & $(0.000, 0.034)$ & $(-0.035, 0.000)$ \\
    \multirow{2}{*}{Obtained marijuana in person} & $0.041$ & $-0.041$ \\
          & $(0.027, 0.055)$ & $(-0.056, -0.027)$ \\
    \multirow{2}{*}{Obtained marijuana through friends} & $0.046$ & $-0.047$ \\
          & $(0.032, 0.060)$ & $(-0.061, -0.033)$ \\
    \multirow{2}{*}{Obtained marijuana other / missing} & $0.042$ & $-0.044$ \\
          & $(0.016, 0.069)$ & $(-0.071, -0.017)$ \\
    \multirow{2}{*}{Obtained cocaine online} & $0.871$ & $-0.870$ \\
          & $(0.813, 0.927)$ & $(-0.928, -0.811)$ \\
    \multirow{2}{*}{Obtained cocaine in person} & $-0.876$ & $0.878$ \\
          & $(-0.909, -0.842)$ & $(0.845, 0.911)$ \\
    \multirow{2}{*}{Obtained cocaine through friends} & $-0.891$ & $0.891$ \\
          & $(-0.918, -0.864)$ & $(0.864, 0.918)$ \\
    \multirow{2}{*}{Obtained cocaine other / missing} & $-0.225$ & $0.225$ \\
          & $(-0.360, -0.089)$ & $(0.089, 0.362)$ \\
    \multirow{2}{*}{Obtained basuco online} & $-0.612$ & $0.120$ \\
          & $(-61.668, 61.339)$ & $(-62.616, 62.290)$ \\
    \multirow{2}{*}{Obtained basuco in person} & $-0.114$ & $-0.032$ \\
          & $(-0.170, -0.059)$ & $(-0.090, 0.024)$ \\
    \multirow{2}{*}{Obtained basuco through friends} & $1.277$ & $-1.245$ \\
          & $(1.119, 1.396)$ & $(-1.360, -1.127)$ \\
    \multirow{2}{*}{Indirect utility ($y$)} & $0.000$ & $0.002$ \\
          & $(-0.006, 0.007)$ & $(-0.005, 0.010)$ \\
    \multirow{2}{*}{Indirect utility$^2$} & $0.001$ & $0.001$ \\
          & $(-0.002, 0.004)$ & $(-0.003, 0.004)$ \\
    \multirow{2}{*}{Indirect utility$^3$} & $0.000$ & $0.000$ \\
          & $(0.000, 0.001)$ & $(-0.001, 0.001)$ \\
    \midrule
    \multicolumn{3}{c}{Endogenous} \\
    \midrule
    \multirow{2}{*}{Relative price marijuana} & $0.004$ & $-0.004$ \\
          & $(-0.008, 0.016)$ & $(-0.015, 0.008)$ \\
    \multirow{2}{*}{Relative price cocaine} & $-0.004$ & $-0.001$ \\
          & $(-0.015, 0.008)$ & $(-0.021, 0.019)$ \\
    \multirow{2}{*}{$\text{Relative price marijuana} \times y$} & $0.002$ & $0.006$ \\
          & $(-0.009, 0.013)$ & $(-0.008, 0.021)$ \\
    \multirow{2}{*}{$\text{Relative price cocaine} \times y$} & $0.006$ & $0.005$ \\
          & $(-0.008, 0.021)$ & $(-0.014, 0.025)$ \\
    \midrule
    Observations & \multicolumn{2}{c}{1,069} \\
    \midrule
    \multirow{2}{*}{Covariate} & \multicolumn{2}{c}{``Hard'' Cluster} \\
    \cmidrule{2-3}      & Share of marijuana & Share of cocaine \\
    \midrule
    \multicolumn{3}{c}{Exogenous} \\
    \midrule
    \multirow{2}{*}{Constant} & $0.286$ & $6.340$ \\
          & $(-52.328, 53.268)$ & $(-48.412, 60.330)$ \\
    \multirow{2}{*}{Dealer presence} & $-1.007$ & $-2.305$ \\
          & $(-34.590, 34.177)$ & $(-38.548, 36.414)$ \\
    \multirow{2}{*}{Medium risk} & $11.974$ & $10.874$ \\
          & $(-36.213, 61.629)$ & $(-39.186, 60.963)$ \\
    \multirow{2}{*}{High risk} & $-6.132$ & $-3.588$ \\
          & $(-51.800, 40.220)$ & $(-52.805, 44.320)$ \\
    \multirow{2}{*}{Consumer in network} & $-1.869$ & $2.539$ \\
          & $(-49.216, 44.859)$ & $(-47.214, 51.802)$ \\
    \multirow{2}{*}{Female} & $-12.844$ & $-14.108$ \\
          & $(-48.805, 24.229)$ & $(-53.966, 25.876)$ \\
    \multirow{2}{*}{Both parents at home} & $-23.123$ & $-24.736$ \\
          & $(-63.325, 17.750)$ & $(-67.517, 18.437)$ \\
    \multirow{2}{*}{Medicinal marijuana products} & $5.353$ & $7.925$ \\
          & $(-30.008, 39.526)$ & $(-30.302, 44.349)$ \\
    \multirow{2}{*}{Good physical health} & $-9.401$ & $1.902$ \\
          & $(-48.692, 30.096)$ & $(-40.552, 44.496)$ \\
    \multirow{2}{*}{Good mental health} & $12.226$ & $11.535$ \\
          & $(-21.394, 46.615)$ & $(-24.906, 48.869)$ \\
    \multirow{2}{*}{Years of education} & $6.243$ & $5.645$ \\
          & $(-0.674, 12.560)$ & $(-1.844, 12.787)$ \\
    \multirow{2}{*}{Head of household} & $-13.831$ & $-18.654$ \\
          & $(-50.278, 22.952)$ & $(-57.086, 20.366)$ \\
    \multirow{2}{*}{Medium SES} & $-23.720$ & $-23.838$ \\
          & $(-63.235, 13.978)$ & $(-64.642, 15.916)$ \\
    \multirow{2}{*}{High SES} & $4.255$ & $29.499$ \\
          & $(-43.674, 49.875)$ & $(-20.071, 78.260)$ \\
    \multirow{2}{*}{Working status} & $9.931$ & $14.661$ \\
          & $(-24.325, 43.701)$ & $(-22.970, 50.572)$ \\
    \multirow{2}{*}{Age minus 30} & $1.606$ & $2.316$ \\
          & $(-3.139, 5.679)$ & $(-2.524, 7.020)$ \\
    \multirow{2}{*}{Codependent substance use} & $-5.617$ & $-10.641$ \\
          & $(-46.911, 34.951)$ & $(-55.738, 32.869)$ \\
    \multirow{2}{*}{Metropolitan area} & $-2.989$ & $2.068$ \\
          & $(-37.143, 33.256)$ & $(-34.840, 39.213)$ \\
    \multirow{2}{*}{Time using marijuana} & $-2.106$ & $-4.533$ \\
          & $(-6.028, 2.265)$ & $(-9.041, 0.249)$ \\
    \multirow{2}{*}{Time using cocaine} & $-4.773$ & $-2.381$ \\
          & $(-8.769, -0.587)$ & $(-6.548, 2.260)$ \\
    \multirow{2}{*}{Time using basuco} & $2.899$ & $2.100$ \\
          & $(-2.708, 8.394)$ & $(-4.274, 7.403)$ \\
    \multirow{2}{*}{Was offered marijuana} & $15.949$ & $-2.721$ \\
          & $(-26.390, 57.873)$ & $(-46.761, 41.754)$ \\
    \multirow{2}{*}{Was offered cocaine} & $-0.386$ & $9.001$ \\
          & $(-38.625, 39.271)$ & $(-31.552, 49.825)$ \\
    \multirow{2}{*}{Was offered basuco} & $7.018$ & $13.131$ \\
          & $(-30.462, 44.731)$ & $(-28.141, 52.844)$ \\
    \multirow{2}{*}{Obtained marijuana online} & $21.331$ & $-1.895$ \\
          & $(-21.058, 63.386)$ & $(-45.750, 42.966)$ \\
    \multirow{2}{*}{Obtained marijuana in person} & $-5.253$ & $-38.941$ \\
          & $(-43.469, 32.479)$ & $(-81.007, 1.816)$ \\
    \multirow{2}{*}{Obtained marijuana through friends} & $17.866$ & $-15.044$ \\
          & $(-19.675, 55.641)$ & $(-54.970, 25.497)$ \\
    \multirow{2}{*}{Obtained marijuana other / missing} & $11.016$ & $-6.409$ \\
          & $(-42.669, 65.789)$ & $(-62.444, 50.340)$ \\
    \multirow{2}{*}{Obtained cocaine online} & $-10.386$ & $31.744$ \\
          & $(-55.445, 33.324)$ & $(-14.425, 77.491)$ \\
    \multirow{2}{*}{Obtained cocaine in person} & $-36.220$ & $2.257$ \\
          & $(-76.435, 4.051)$ & $(-40.470, 45.501)$ \\
    \multirow{2}{*}{Obtained cocaine through friends} & $-23.087$ & $9.198$ \\
          & $(-62.550, 18.830)$ & $(-33.817, 53.913)$ \\
    \multirow{2}{*}{Obtained cocaine other / missing} & $6.944$ & $0.659$ \\
          & $(-46.995, 60.127)$ & $(-54.177, 56.094)$ \\
    \multirow{2}{*}{Obtained basuco online} & $5.904$ & $-6.570$ \\
          & $(-52.398, 64.693)$ & $(-64.359, 52.558)$ \\
    \multirow{2}{*}{Obtained basuco in person} & $6.954$ & $-2.979$ \\
          & $(-38.588, 51.350)$ & $(-51.529, 44.456)$ \\
    \multirow{2}{*}{Obtained basuco through friends} & $3.254$ & $16.780$ \\
          & $(-41.316, 48.748)$ & $(-32.026, 64.190)$ \\
    \multirow{2}{*}{Indirect utility ($y$)} & $-32.921$ & $-19.285$ \\
          & $(-57.618, -9.223)$ & $(-50.865, 8.684)$ \\
    \multirow{2}{*}{Indirect utility$^2$} & $-43.459$ & $-55.035$ \\
          & $(-55.296, -33.796)$ & $(-68.781, -43.371)$ \\
    \multirow{2}{*}{Indirect utility$^3$} & $-4.519$ & $-3.825$ \\
          & $(-7.805, -1.496)$ & $(-8.125, -0.392)$ \\
    \midrule
    \multicolumn{3}{c}{Endogenous} \\
    \midrule
    \multirow{2}{*}{Relative price marijuana} & $17.776$ & $35.868$ \\
          & $(-9.730, 47.792)$ & $(6.314, 69.690)$ \\
    \multirow{2}{*}{Relative price cocaine} & $35.868$ & $14.712$ \\
          & $(6.314, 69.690)$ & $(-26.491, 51.190)$ \\
    \multirow{2}{*}{$\text{Relative price marijuana} \times y$} & $-165.608$ & $-200.421$ \\
          & $(-198.133, -135.220)$ & $(-237.563, -163.980)$ \\
    \multirow{2}{*}{$\text{Relative price cocaine} \times y$} & $-200.421$ & $-189.935$ \\
          & $(-237.563, -163.980)$ & $(-265.410, -122.475)$ \\
    \midrule
    Observations & \multicolumn{2}{c}{167} \\
    \bottomrule
    \caption*{\small \textbf{Notes:} Posterior median coefficients from the Bayesian EASI system structural regressions. Estimates are cluster-specific, where the assignment is obtained automatically through our finite mixture procedure. 95\% posterior quantile credibility intervals are presented underneath the coefficients.}
\end{longtable}

\begin{sidewaystable}[htbp]
    \centering
    \caption{Bayesian estimates of system correlation matrix from EASI model across identified clusters}
    \label{Tab:Correlation}
    \resizebox{1\textwidth}{!}{%
		\begin{threeparttable}
    \begin{tabular}{lcccccc}
    \toprule
    ``Soft'' Cluster & Share of marijuana & Share of cocaine & Relative price marijuana & Relative price cocaine & $\text{Relative price marijuana } \times y$ & $\text{Relative price cocaine } \times y$ \\
    \midrule
    \multirow{2}[1]{*}{Share of marijuana} & \multirow{2}[1]{*}{1.000} &       &       &       &       &  \\
          &       &       &       &       &       &  \\
    \multirow{2}[0]{*}{Share of cocaine} & $-0.665$ & \multirow{2}[0]{*}{1.000} &       &       &       &  \\
          & $(-0.786, -0.148)$ &       &       &       &       &  \\
    \multirow{2}[0]{*}{Relative price marijuana} & $0.023$ & $0.020$ & \multirow{2}[0]{*}{1.000} &       &       &  \\
          & $(-0.075, 0.119)$ & $(-0.117, 0.148)$ &       &       &       &  \\
    \multirow{2}[0]{*}{Relative price cocaine} & $0.032$ & $0.041$ & $0.579$ & \multirow{2}[0]{*}{1.000} &       &  \\
          & $(-0.087, 0.148)$ & $(-0.172, 0.234)$ & $(0.540, 0.615)$ &       &       &  \\
    \multirow{2}[0]{*}{$\text{Relative price marijuana } \times y$} & $-0.133$ & $-0.173$ & $-0.011$ & $-0.082$ & \multirow{2}[0]{*}{1.000} &  \\
          & $(-0.448, 0.230)$ & $(-0.556, 0.293)$ & $(-0.067, 0.048)$ & $(-0.140, -0.024)$ &       &  \\
    \multirow{2}[1]{*}{$\text{Relative price cocaine } \times y$} & $-0.197$ & $-0.153$ & $-0.079$ & $-0.079$ & $0.612$ & \multirow{2}[1]{*}{1.000} \\
          & $(-0.551, 0.214)$ & $(-0.600, 0.385)$ & $(-0.138, -0.021)$ & $(-0.137, -0.019)$ & $(0.576, 0.647)$ &  \\
    \midrule
    ``Hard'' Cluster & Share of marijuana & Share of cocaine & Relative price marijuana & Relative price cocaine & $\text{Relative price marijuana } \times y$ & $\text{Relative price cocaine } \times y$ \\
    \midrule
    \multirow{2}[1]{*}{Share of marijuana} & \multirow{2}[1]{*}{1.000} &       &       &       &       &  \\
          &       &       &       &       &       &  \\
    \multirow{2}[0]{*}{Share of cocaine} & $0.992$ & \multirow{2}[0]{*}{1.000} &       &       &       &  \\
          & $(0.982, 0.996)$ &       &       &       &       &  \\
    \multirow{2}[0]{*}{Relative price marijuana} & $-0.114$ & $-0.128$ & \multirow{2}[0]{*}{1.000} &       &       &  \\
          & $(-0.196, -0.036)$ & $(-0.207, -0.048)$ &       &       &       &  \\
    \multirow{2}[0]{*}{Relative price cocaine} & $-0.156$ & $-0.140$ & $0.579$ & \multirow{2}[0]{*}{1.000} &       &  \\
          & $(-0.240, -0.073)$ & $(-0.223, -0.051)$ & $(0.540, 0.615)$ &       &       &  \\
    \multirow{2}[0]{*}{$\text{Relative price marijuana } \times y$} & $0.844$ & $0.876$ & $-0.011$ & $-0.082$ & \multirow{2}[0]{*}{1.000} &  \\
          & $(0.808, 0.876)$ & $(0.839, 0.910)$ & $(-0.067, 0.048)$ & $(-0.140, -0.024)$ &       &  \\
    \multirow{2}[1]{*}{$\text{Relative price cocaine } \times y$} & $0.932$ & $0.905$ & $-0.079$ & $-0.079$ & $0.612$ & \multirow{2}[1]{*}{1.000} \\
          & $(0.909, 0.951)$ & $(0.863, 0.933)$ & $(-0.138, -0.021)$ & $(-0.137, -0.019)$ & $(0.576, 0.647)$ &  \\
    \bottomrule
    \end{tabular}
    \begin{tablenotes}
    \item \small{\textbf{Notes:}Table provides posterior median of lower-triangular correlation matrix elements from Bayesian estimates. Highest posterior density (HPD) intervals covering 95\% probability are presented underneath the estimated unrestricted coefficients.}
    \end{tablenotes}
    \end{threeparttable}%
    }
\end{sidewaystable}

\begin{table}[htbp]
    \centering
    \caption{Frequentist price elasticities of demand for the full sample of consumers}\label{Tab:Results_FullSample}
    \resizebox{1\textwidth}{!}{%
		\begin{threeparttable}
    \begin{tabular}{cccc}
    \toprule
Good demand & Price Marijuana & Price Cocaine & Price Basuco \\
\midrule
\multicolumn{4}{c}{Frequentist OLS} \\
\midrule
\multirow{2}[1]{*}{Marijuana} & $-0.9744$ & $-0.0132$ & $-0.0118$ \\
      & $(-0.9955, -0.9507)$ & $(-0.0310, 0.0030)$ & $(-0.0280, 0.0039)$ \\
\multirow{2}[0]{*}{Cocaine} & $-0.1277$ & $-0.9728$ & $0.0773$ \\
      & $(-0.2606, 0.0017)$ & $(-1.1258, -0.8093)$ & $(-0.0268, 0.1953)$ \\
\multirow{2}[1]{*}{Basuco} & $-0.3550$ & $0.3581$ & $-0.9194$ \\
      & $(-0.9067, 0.1764)$ & $(-0.1227, 0.8833)$ & $(-1.3157, -0.5058)$ \\
\midrule
\multicolumn{4}{c}{Frequentist 2SLS} \\
\midrule
\multirow{2}[1]{*}{Marijuana} & $-0.9617$ & $-0.0256$ & $-0.0152$ \\
      & $(-0.9912, -0.9338)$ & $(-0.0482, -0.0016)$ & $(-0.0379, 0.0055)$ \\
\multirow{2}[0]{*}{Cocaine} & $-0.2298$ & $-1.5257$ & $0.7263$ \\
      & $(-0.4270, -0.0311)$ & $(-2.1373, -0.9626)$ & $(0.1589, 1.3445)$ \\
\multirow{2}[1]{*}{Basuco} & $-0.3574$ & $3.2823$ & $-3.7030$ \\
      & $(-1.1399, 0.4179)$ & $(0.7246, 6.0796)$ & $(-6.3126, -1.3334)$ \\
\bottomrule
    \end{tabular}
    \begin{tablenotes}
    \item \small{\textbf{Notes:} Full sample includes 1,236 consumers. For frequentist estimators, intervals are 95\% percentile-$t$ intervals from 1,000 bootstrap replications. For Bayesian estimators, 95\% highest posterior density (HPD) intervals are provided.}
    \end{tablenotes}
    \end{threeparttable}%
    }
\end{table}%

\begin{table}[htbp]
    \centering
    \caption{Price elasticities of demand for the sub-sample of consumers from the ``soft" cluster}
    \label{Tab:Results_SubSample}%
	\begin{threeparttable}
        \begin{tabular}{cccc}
            \toprule
            Good demand & Price Marijuana & Price Cocaine & Price Basuco \\
            \midrule
            \multicolumn{4}{c}{Frequentist OLS} \\
            \midrule
            \multirow{2}[1]{*}{Marijuana} & $-0.9374$ & $-0.0205$ & $-0.0309$ \\
                  & $(-0.9660, -0.9093)$ & $(-0.0422, 0.0012)$ & $(-0.0517, -0.0097)$ \\
            \multirow{2}[0]{*}{Cocaine} & $-0.2135$ & $-1.0313$ & $0.2010$ \\
                  & $(-0.3666, -0.0543)$ & $(-1.2392, -0.8226)$ & $(0.0517, 0.3478)$ \\
            \multirow{2}[1]{*}{Basuco} & $-1.3104$ & $0.8832$ & $-0.7802$ \\
                  & $(-1.9986, -0.5760)$ & $(0.2037, 1.5598)$ & $(-1.2045, -0.3269)$ \\
            \midrule
            \multicolumn{4}{c}{Frequentist 2SLS} \\
            \midrule
            \multirow{2}[1]{*}{Marijuana} & $-0.9352$ & $-0.0387$ & $-0.0139$ \\
                  & $(-0.9695, -0.9000)$ & $(-0.0690, -0.0062)$ & $(-0.0440, 0.0163)$ \\
            \multirow{2}[0]{*}{Cocaine} & $-0.3577$ & $-1.4700$ & $0.7884$ \\
                  & $(-0.5955, -0.1172)$ & $(-2.1777, -0.8098)$ & $(0.1574, 1.4261)$ \\
            \multirow{2}[1]{*}{Basuco} & $-0.7452$ & $3.5092$ & $-4.0316$ \\
                  & $(-1.7570, 0.3027)$ & $(0.6723, 6.3652)$ & $(-6.9141, -1.5811)$ \\
            \midrule
            \multicolumn{4}{c}{Bayesian Censored} \\
            \midrule
            \multirow{2}[1]{*}{Marijuana} & $-0.9842$ & $-0.0116$ & $0.0004$ \\
                  & $(-1.0104, -0.9586)$ & $(-0.0339, 0.0114)$ & $(-0.0044, 0.0053)$ \\
            \multirow{2}[0]{*}{Cocaine} & $-0.1330$ & $-0.9080$ & $0.0022$ \\
                  & $(-0.3452, 0.0699)$ & $(-1.0920, -0.7225)$ & $(-0.0335, 0.0386)$ \\
            \multirow{2}[1]{*}{Basuco} & $0.2690$ & $0.1306$ & $-1.3144$ \\
                  & $(-2.1879, 2.6402)$ & $(-1.8721, 2.2195)$ & $(-3.4434, 1.0741)$ \\
            \midrule
            \multicolumn{4}{c}{Bayesian Censored with Endogeneity} \\
            \midrule
            \multirow{2}[1]{*}{Marijuana} & $-0.9617$ & $-0.0372$ & $0.0016$ \\
                  & $(-0.9968, -0.9251)$ & $(-0.0720, -0.0009)$ & $(-0.0053, 0.0089)$ \\
            \multirow{2}[0]{*}{Cocaine} & $-0.3254$ & $-0.7462$ & $0.0482$ \\
                  & $(-0.6292, -0.0462)$ & $(-1.0506, -0.4085)$ & $(-0.1007, 0.1911)$ \\
            \multirow{2}[1]{*}{Basuco} & $0.7953$ & $2.7265$ & $-4.4424$ \\
                  & $(-2.7469, 3.9169)$ & $(-5.5409, 10.9176)$ & $(-13.0818, 3.7845)$ \\
            \bottomrule
        \end{tabular}    
    \begin{tablenotes}
    \item \small{\textbf{Notes:} Identified Safe cluster sample includes 1,202 consumers. For frequentist estimators, intervals are 95\% percentile-$t$ intervals from 1,000 bootstrap replications. For Bayesian estimators, point-estimate presented is the median of chains across 5,000 burn-in and 10,000 kept iterations, with 95\% highest posterior density (HPD) intervals of chains provided in parenthesis.}
    \end{tablenotes}
    \end{threeparttable}
\end{table}%

\begin{figure}
    \centering
    \caption{Inclusion probability to identified ``soft'' cluster of users}
    \label{Fig:Inclusion_Probability}
    \includegraphics[scale = 0.6]{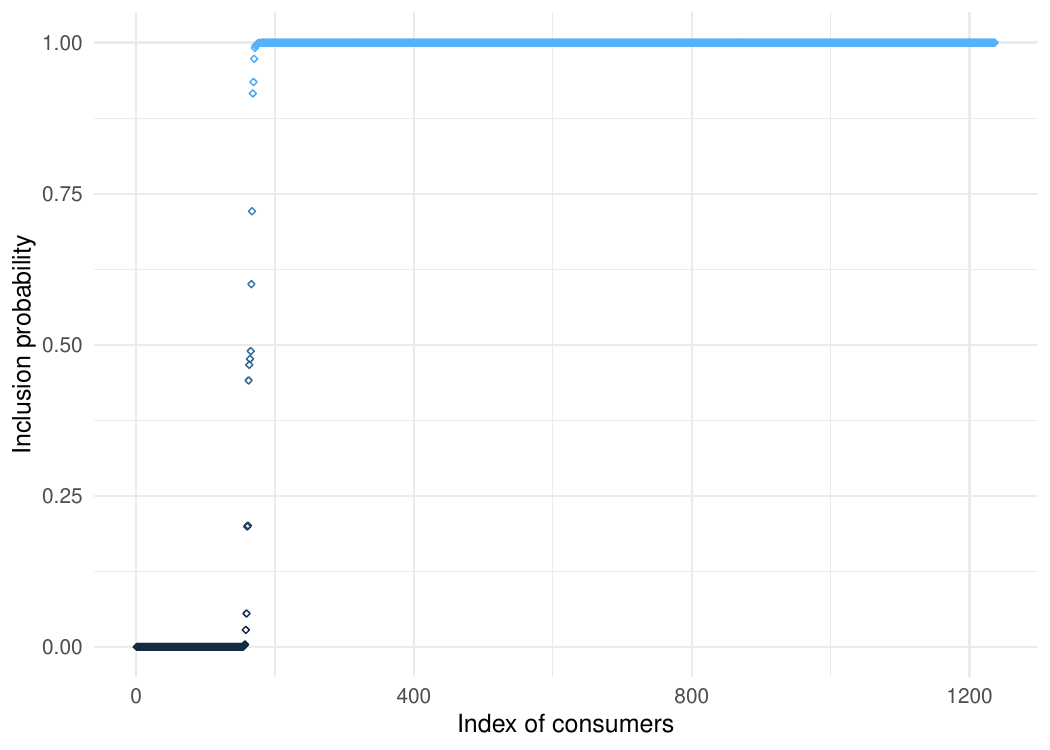}
    \begin{minipage}{\textwidth}
        \begin{tablenotes}
            \item \small{\textbf{Notes:} Sorted by the individual-specific posterior probability of inclusion to the ``soft'' cluster of drug consumers. Posterior average taken over single chain run of 10,000 iterations after a burn-in window of 5,000 and thinning every 10 draws.}
        \end{tablenotes}
    \end{minipage}
\end{figure}

\begin{figure}
    \centering
    \caption{Balance test between ``hard'' and ``soft'' clusters of users}
    \label{Fig:Balance_Tstats}
    \begin{adjustbox}{max totalsize = {\textwidth}{\textheight}, center}    
        \includegraphics[page = 1]{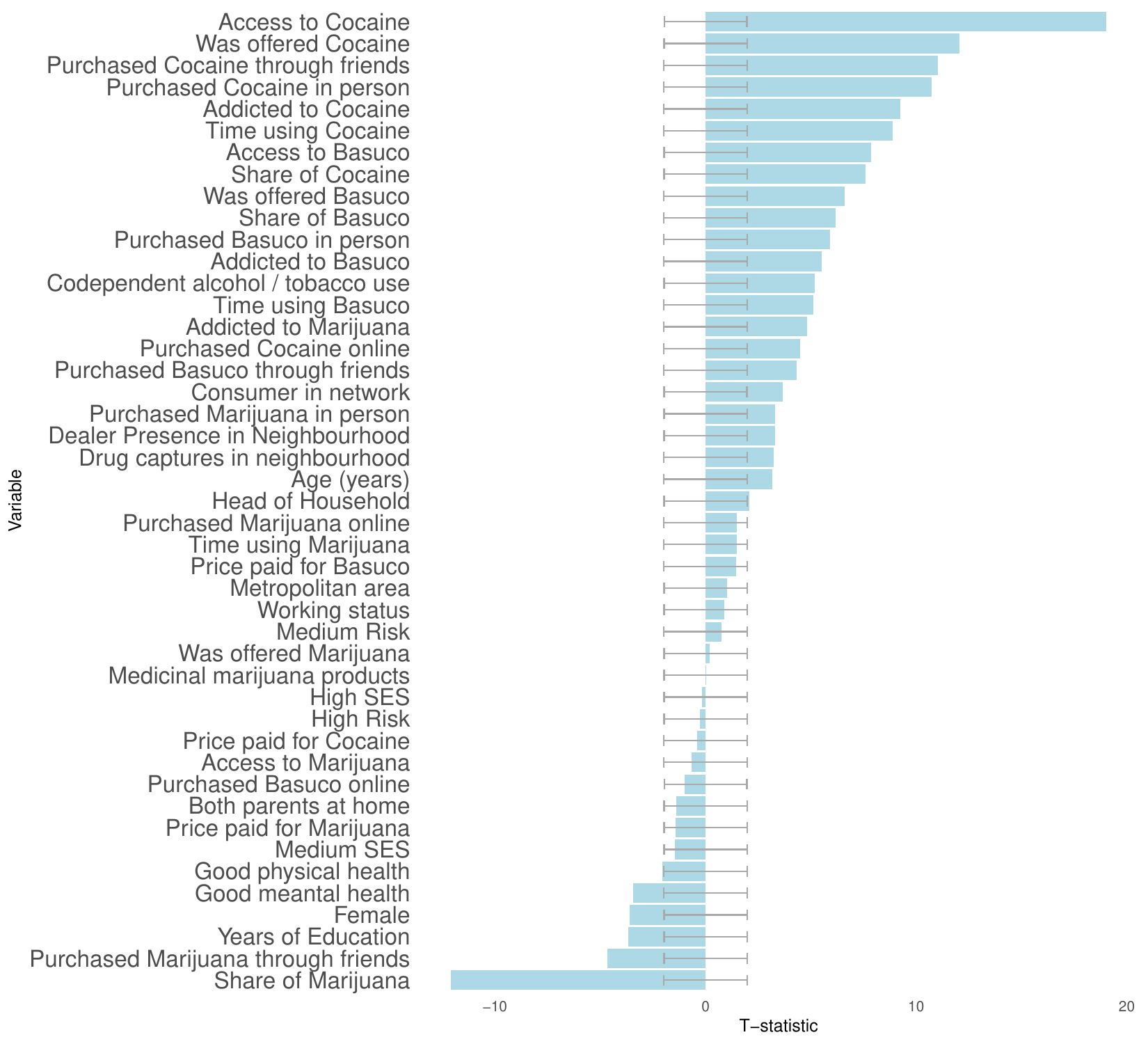}
    \end{adjustbox}
    \begin{minipage}{\textwidth}
        \begin{tablenotes}
            \item \small{\textbf{Notes:} T-statistic of a two-sided mean-difference test between the identified ``hard'' and ``soft'' clusters of consumers. Plotted interval provides critical values for significance of the mean-difference at 5\% level.}
        \end{tablenotes}
    \end{minipage}
\end{figure}

\begin{table}[htbp]
    \centering
    \caption{Balance of covariates comparing ``hard'' and ``soft'' clusters of users}\label{Tab:Balance_Table}
    \begin{threeparttable}
    \begin{tabular}{lccc}
        \toprule
        Covariate & T-statistic & P-value & Confidence Interval \\
        \midrule
        Share of Marijuana & $-18.975$ & 2.695E-20 & $(-0.822, -0.663)$ \\
        Share of Cocaine & $-2.135$ & 0.039 & $(-0.114, -0.003)$ \\
        Share of Basuco & $18.341$ & 7.039E-19 & $(0.712, 0.890)$ \\
        Price paid for Marijuana & $-0.989$ & 0.329 & $(-0.256, 0.088)$ \\
        Price paid for Cocaine & $-0.593$ & 0.557 & $(-0.154, 0.085)$ \\
        Price paid for Basuco & $1.655$ & 0.107 & $(-0.139, 1.353)$ \\
        Drug Expenditure & $0.845$ & 0.404 & $(-0.037, 0.089)$ \\
        Dealer Presence & $2.990$ & 0.005 & $(0.075, 0.392)$ \\
        Medium Risk & $-0.819$ & 0.418 & $(-0.086, 0.036)$ \\
        High Risk & $0.365$ & 0.717 & $(-0.069, 0.100)$ \\
        Consumer in network & $8.812$ & 4.217E-18 & $(0.047, 0.074)$ \\
        Female & $-2.339$ & 0.025 & $(-0.251, -0.018)$ \\
        Both parents at home & $-0.665$ & 0.511 & $(-0.154, 0.078)$ \\
        Medicinal marijuana products & $-1.275$ & 0.211 & $(-0.266, 0.061)$ \\
        Good physical health & $-2.957$ & 0.006 & $(-0.437, -0.081)$ \\
        Good meantal health & $-1.563$ & 0.127 & $(-0.314, 0.041)$ \\
        Years of education & $-6.150$ & 4.884E-07 & $(-5.149, -2.593)$ \\
        Head of household & $-0.043$ & 0.966 & $(-0.178, 0.170)$ \\
        Medium SES & $-4.499$ & 6.484E-05 & $(-0.377, -0.143)$ \\
        High SES & $-0.955$ & 0.346 & $(-0.090, 0.032)$ \\
        Working status & $0.579$ & 0.566 & $(-0.122, 0.220)$ \\
        Age minus 30 & $4.101$ & 2.426E-04 & $(5.109, 15.147)$ \\
        Codependent substance use & $2.192$ & 0.035 & $(0.007, 0.178)$ \\
        Metropolitan area & $0.151$ & 0.881 & $(-0.161, 0.187)$ \\
        Time using marijuana & $0.149$ & 0.882 & $(-4.522, 5.238)$ \\
        Time using cocaine & $0.831$ & 0.412 & $(-2.215, 5.277)$ \\
        Time using basuco & $7.166$ & 3.086E-08 & $(10.984, 19.690)$ \\
        Drug captures & $0.567$ & 0.574 & $(-0.422, 0.750)$ \\
        Access to marijuana & $-0.943$ & 0.353 & $(-0.088, 0.032)$ \\
        Access to cocaine & $18.697$ & 1.138E-68 & $(0.202, 0.249)$ \\
        Access to basuco & $32.912$ & 7.442E-170 & $(0.446, 0.502)$ \\
        Offered marijuana & $-0.651$ & 0.519 & $(-0.167, 0.086)$ \\
        Offered cocaine & $5.385$ & 4.147E-06 & $(0.194, 0.429)$ \\
        Offered basuco & $11.435$ & 1.461E-13 & $(0.540, 0.772)$ \\
        \bottomrule
    \end{tabular}%
    \begin{tablenotes}
        \item \small{\textbf{Notes:} T-statistic and P-values from a mean-difference test across clusters (using different sample sizes). 95\% confidence intervals of the mean estimate are also presented, where the difference is computed as ``hard'' minus ``soft'' cluster.}
    \end{tablenotes}
    \end{threeparttable}
\end{table}

\begin{figure}[htbp]
    \centering
    \caption{Trace plots of price elasticities of demand implied by EASI system estimates}
    \label{Fig:Trace_DemandPrice}
    \resizebox{1\textwidth}{!}{%
		\begin{threeparttable}
    \includegraphics[width = \linewidth, page = 1]{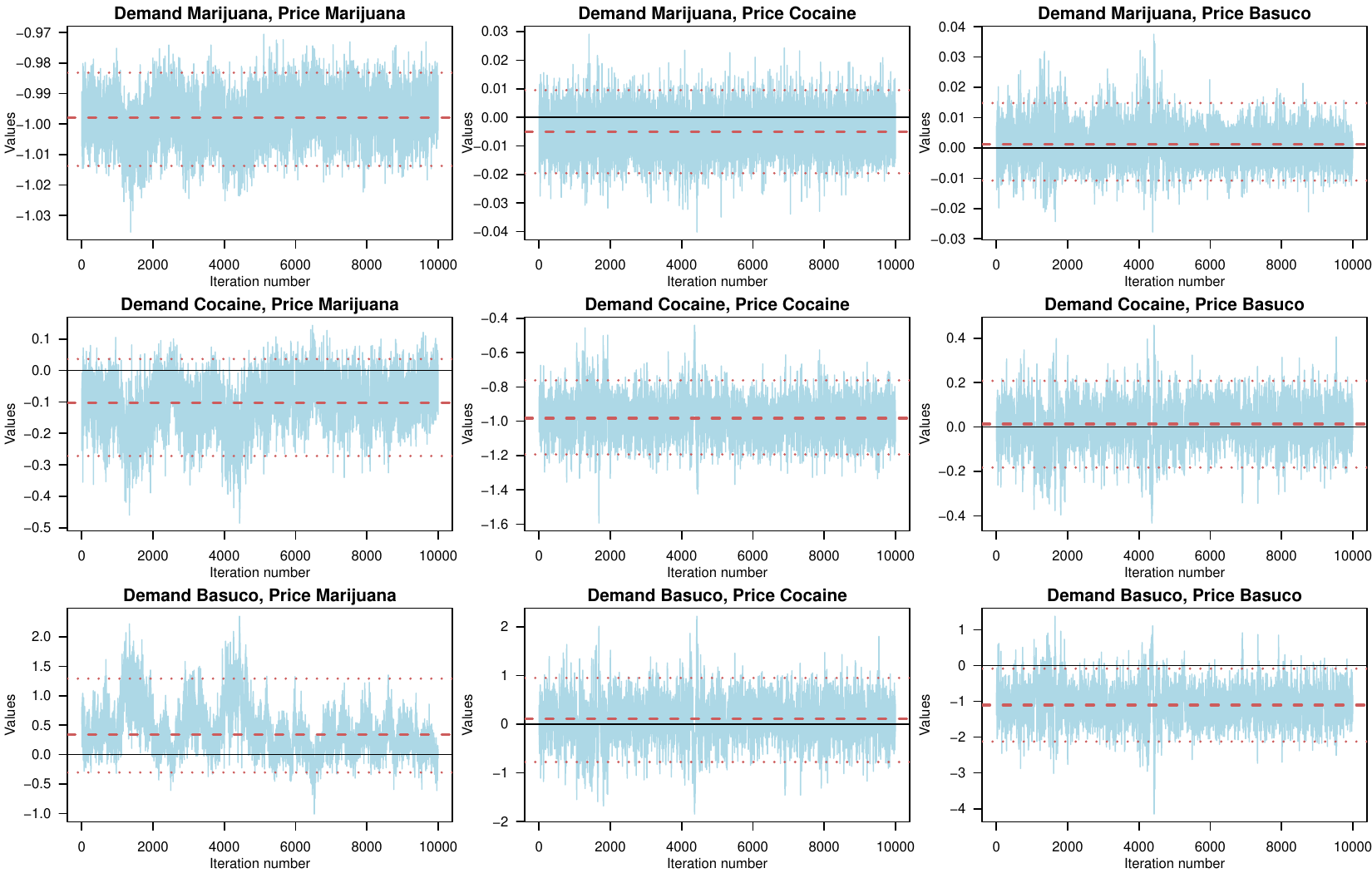}
    \begin{tablenotes}
    \item \small{\textbf{Notes:} Full trace of drawn price elasticity of demand parameter values (own and crossed between illicit drugs). Dashed line represents median of chains, dotted line provides the 95\% HPD for each parameter, and a solid line at 0 is drawn for reference. Single chain run of 10,000 after a burn-in window of 5,000 iterations.}
    \end{tablenotes}
    \end{threeparttable}%
    }
\end{figure}

\begin{figure}[htbp]
    \centering
    \caption{ACF plots of price elasticities of demand implied by EASI system estimates}
    \label{Fig:ACF_DemandPrice}
    \resizebox{1\textwidth}{!}{%
		\begin{threeparttable}
    \includegraphics[width = \linewidth, page = 3]{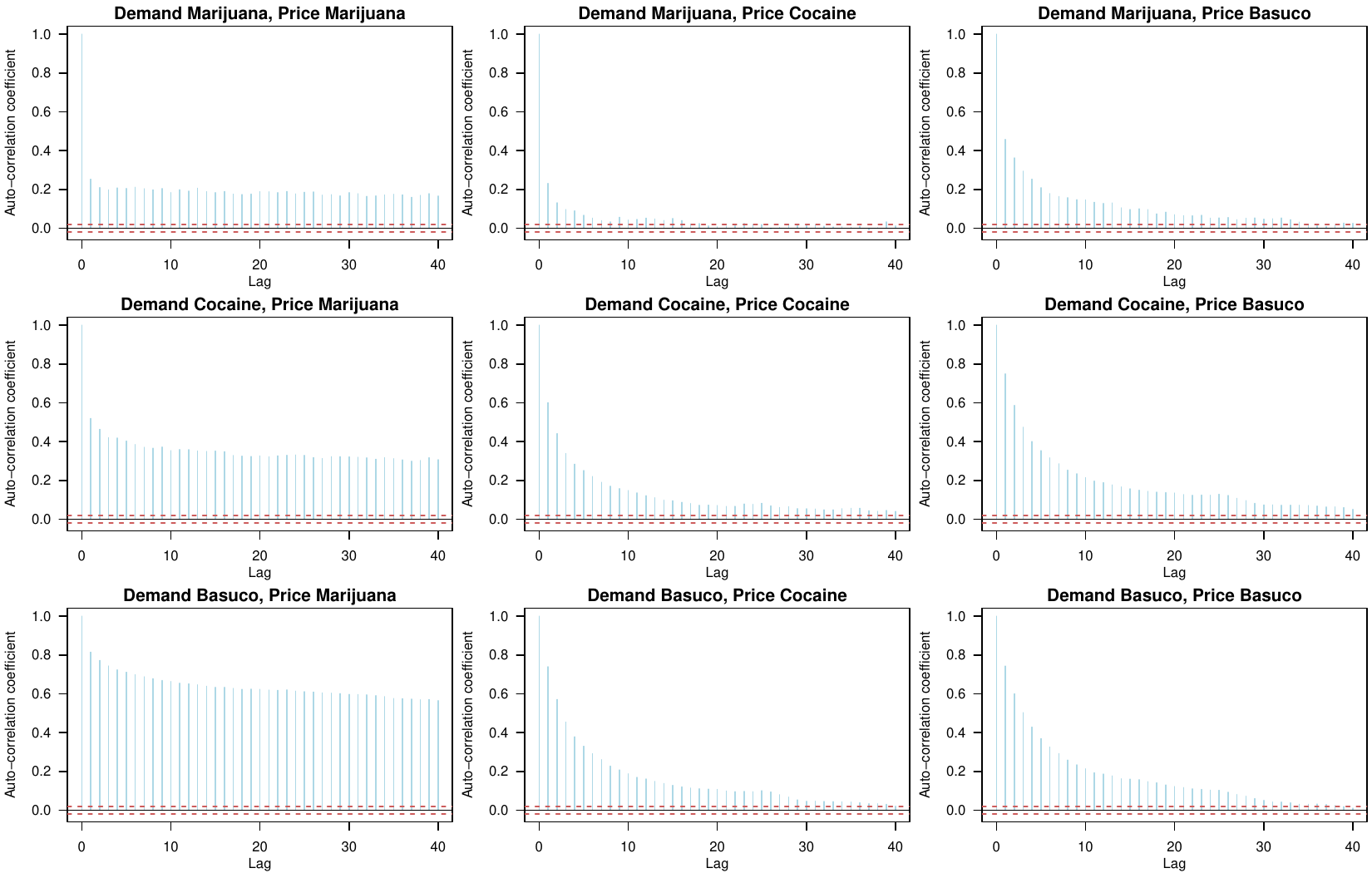}
    \begin{tablenotes}
    \item \small{\textbf{Notes:} Autocorrelation plots for price elasticity of demand parameters (own and crossed between illicit drugs). Dashed line represents critical values for significance of correlation at a given lag. Single chain run of 10,000 after a burn-in window of 5,000 iterations.}
    \end{tablenotes}
    \end{threeparttable}%
    }
\end{figure}

\subsection{Engel Curves and Income Effects}
\label{apx_sec:engel}

While a majority of the changes due to legalization are likely to be presented through changes in drug prices, it is also important to consider how potential relaxations of consumers' budget restrictions influence their demand behavior. In Section \ref{Sec:Data} we presented Engel curves for the full sample to support the use of the EASI model that could recover the flexible non-linear structure in these curves.  We provide evidence that this non-linearity can also be recovered as a mixture of the behavior of the two clusters. 

\begin{figure}
    \centering
    \caption{Non-parametric Engel curves of demand for illicit drugs from identified ``soft'' cluster}
    \label{Fig:Engel_Descriptive_Subsample}
    \begin{adjustbox}{max totalsize={\textwidth}{0.78\textheight}, center}
        \begin{minipage}{\textwidth}
            \begin{subfigure}{\textwidth}
            \centering
            \includegraphics[width = 0.56\textwidth, page = 1]{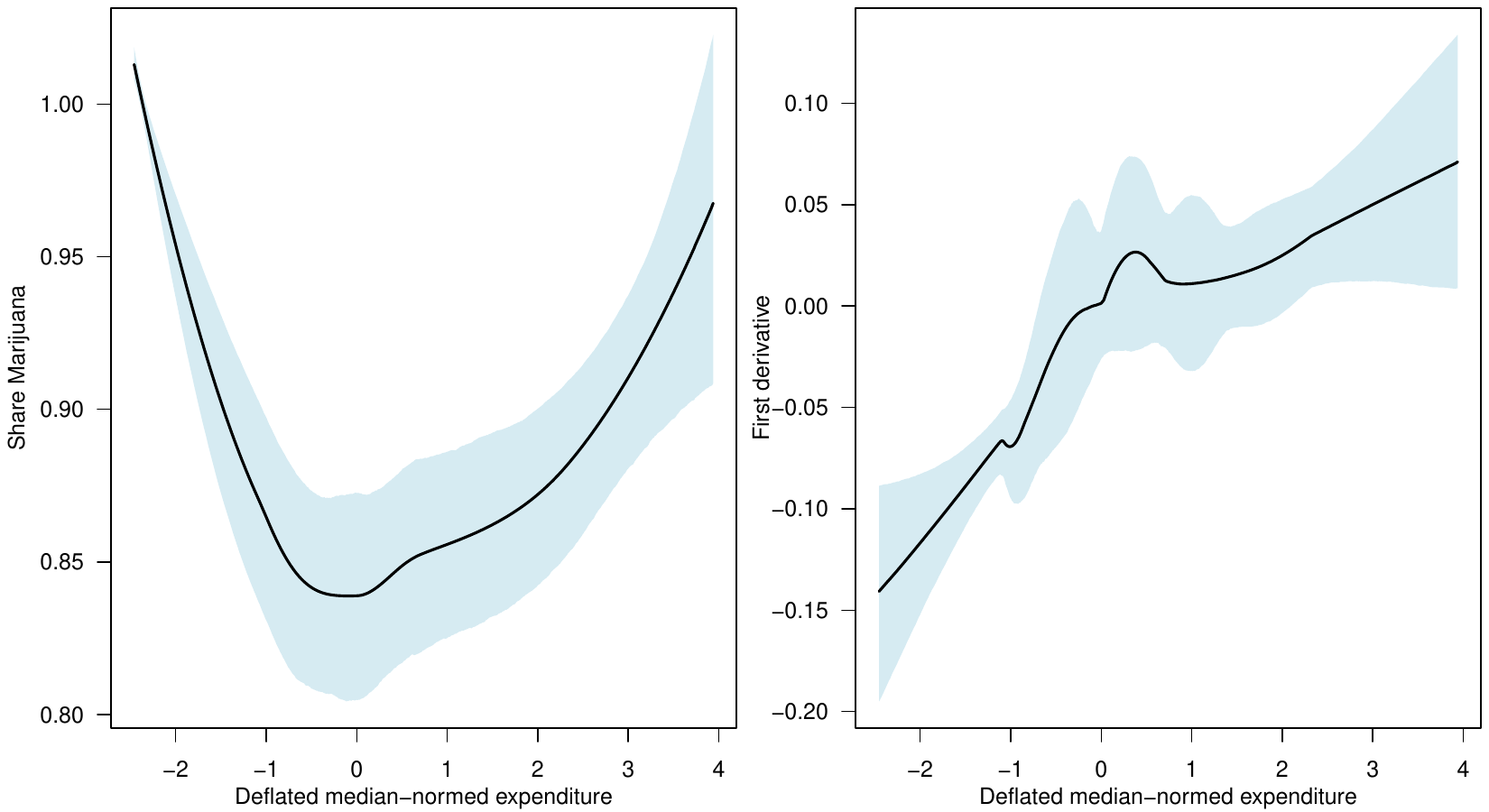}
            \caption{Marijuana}
            \label{Fig:Engel_Descriptive_Subsample_Marijuana}
        \end{subfigure}
        
        \begin{subfigure}{\textwidth}
            \centering
            \includegraphics[width = 0.56\textwidth, page = 2]{Figures/Engel_Descriptive_SubSample.pdf}
            \caption{Cocaine}
            \label{Fig:Engel_Descriptive_Subsample_Cocaine}
        \end{subfigure}
        
        \begin{subfigure}{\textwidth}
            \centering
            \includegraphics[width = 0.56\textwidth, page = 3]{Figures/Engel_Descriptive_SubSample.pdf}
            \caption{Basuco}
            \label{Fig:Engel_Descriptive_Subsample_Basuco}
        \end{subfigure}
        \end{minipage}
    \end{adjustbox}
    \begin{tablenotes}
        \item \small{\textbf{Notes:} Engel curves estimated using a local regression procedure \citep{Cleveland2017}. First derivative computed using simple sorted differences along a grid of expenditure values. Confidence intervals are point-wise 95\% percentile-$t$ intervals from 5,000 bootstrap replications. Engel curves for identified sub-sample exhibits patterns closer to those expected from rational consumption.}
    \end{tablenotes}
\end{figure}

Specifically, we first showcase in Figure \ref{Fig:Engel_Descriptive_Subsample} similar non-parametric estimates of Engel curves as before (see Figure \ref{Fig:Engel_Descriptive}), except that these are constructed for the identified ``soft'' cluster of users. The range of non-linearities in these curves are reduced across all drugs, such that quadratic polynomials are sufficient to capture the relevant curvature. The complexity of the Engel curves observed in Figure \ref{Fig:Engel_Descriptive} arises from a mixture of Engel curves from individuals with differences in unobserved preference heterogeneity. Once we uncover the hidden clusters, we observe less complex Engel curves. 

\begin{figure}
    \centering
    \caption{Pointwise posterior Engel curves from EASI model for demand of illicit drugs across identified ``soft'' and ``hard'' clusters}
    \label{Fig:Engel_Bayes}
    \begin{adjustbox}{max totalsize={\textwidth}{0.77\textheight}, center}
        \begin{minipage}{\textwidth}
            \begin{subfigure}{\textwidth}
                \centering
                \begin{subfigure}{0.49\textwidth}
                    \centering
                    \includegraphics[scale = 0.4, page = 4]{Figures/Engel_Bayes_UnitSum.pdf}
                \end{subfigure}
                \begin{subfigure}{0.49\textwidth}
                    \centering
                    \includegraphics[scale = 0.4, page = 1]{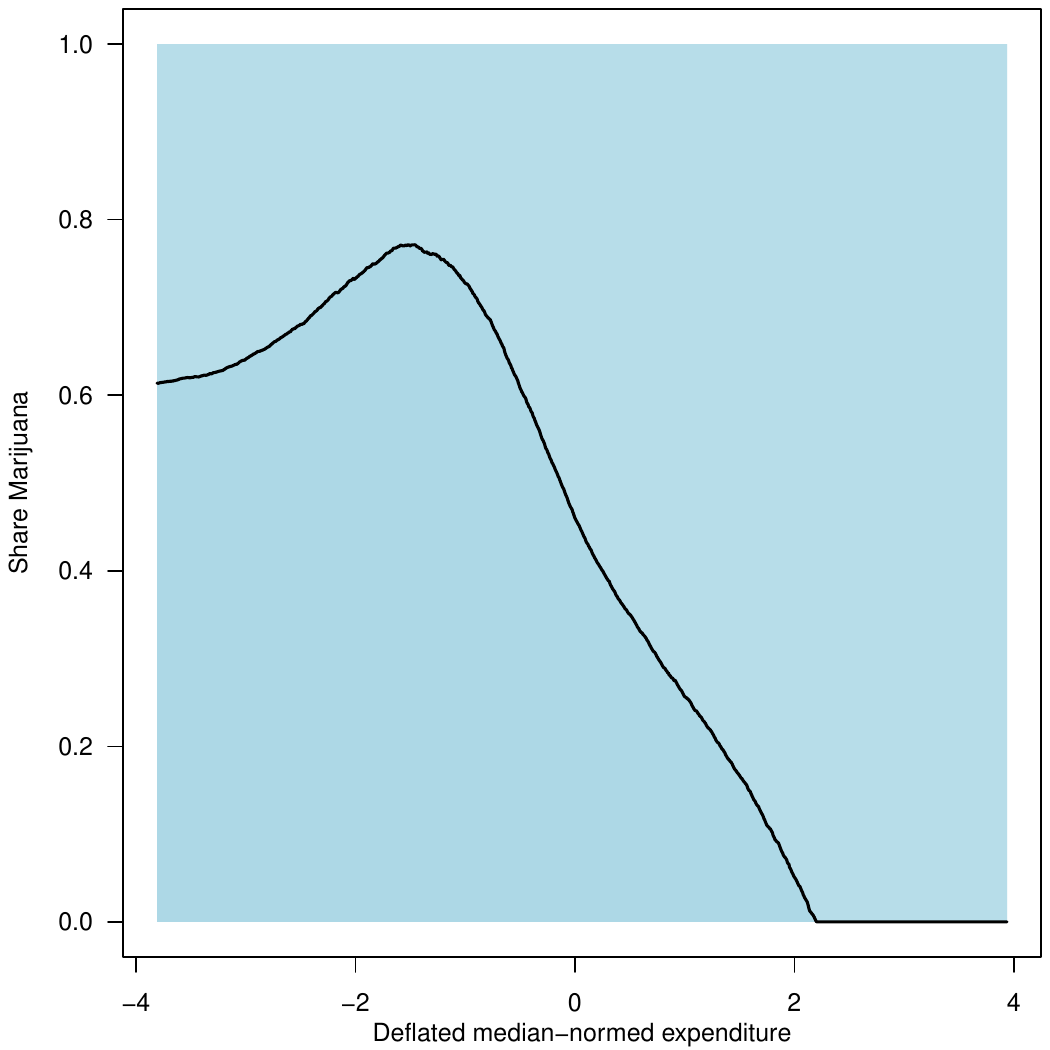}
                \end{subfigure}
                \caption{Marijuana. Left: ``soft'', right: ``hard''.}
                \label{Fig:Engel_Bayes_Marijuana}
            \end{subfigure}
            
            \begin{subfigure}{\textwidth}
                \centering
                \begin{subfigure}{0.49\textwidth}
                    \centering
                    \includegraphics[scale = 0.4, page = 5]{Figures/Engel_Bayes_UnitSum.pdf}
                \end{subfigure}
                \begin{subfigure}{0.49\textwidth}
                    \centering
                    \includegraphics[scale = 0.4, page = 2]{Figures/Engel_Bayes_UnitSum.pdf}
                \end{subfigure}
                \caption{Cocaine. Left: ``soft'', right: ``hard''.}
                \label{Fig:Engel_Bayes_Cocaine}
            \end{subfigure}
            
            \begin{subfigure}{\textwidth}
                \centering
                \begin{subfigure}{0.49\textwidth}
                    \centering
                    \includegraphics[scale = 0.4, page = 6]{Figures/Engel_Bayes_UnitSum.pdf}
                \end{subfigure}
                \begin{subfigure}{0.49\textwidth}
                    \centering
                    \includegraphics[scale = 0.4, page = 3]{Figures/Engel_Bayes_UnitSum.pdf}
                \end{subfigure}
                \caption{Basuco. Left: ``soft'', right: ``hard''.}
                \label{Fig:Engel_Bayes_Basuco}
            \end{subfigure}
        \end{minipage}
    \end{adjustbox}
    \begin{tablenotes}
        \item \footnotesize{\textbf{Notes:} Left panel: individuals assigned into the ``soft" cluster. Right panel: individuals assigned into the ``hard" cluster. Engel curves computed as in Table \ref{Tab:EASI_Summaries} using  solid line plots Bayesian estimate of Engel curves as the point-wise median over a grid of expenditure values. Point-wise highest posterior density (HPD) intervals containing 90\%, 95\% and 99\% posterior probability  provided in blue. Left panel accurately reproduces results for the ``soft'' cluster (compare with the left panel of Figure \ref{Fig:Engel_Descriptive_Subsample}). Engel curves from ``hard'' cluster do not exhibit rational economic behavior.}
    \end{tablenotes}
\end{figure}

We then show the Engel curves implied by our Bayesian estimates, along with their point-wise credibility intervals, in Figure \ref{Fig:Engel_Bayes}. Note how the curves obtained from the ``soft'' cluster estimates closely align with the subsample's non-parametric estimates, particularly for marijuana and cocaine, the two explicitly modeled drugs. Observe that individuals with either larger or smaller expenditure around the median consume more marijuana, with a reverse situation occurring for cocaine and basuco, although the slope change occurs at approximately the median expenditure (slightly less than 0 centered expenditure for marijuana and cocaine, and slightly larger for basuco). Finally, note that the median behavior of the Engel curves for the ``hard" user cluster indicates that most of the expenditure is on basuco as drug expenditure increases; this highlights the addictive nature of the drug. However, the posterior estimates from this cluster are imprecise due to the small sample size in this group.

\subsection{Further Scenario Analyses}
\label{apx_sec:additional_legalization_results}

Figure \ref{Fig:Predicted_Drug_Expenditure} shows the predicted total expenditure on marijuana, cocaine, and basuco for current ``soft'' consumers under a 50\% reduction in marijuana prices, classified by age group. To aggregate into a nationally representative quantity we use our survey expansion factors, approximating the new quantity and expenditures for each individual after modifying their reported price by 50\% (as done for the representative agent). The shares and total expenditure closely resemble those in Figure \ref{Fig:Drug_Expenditure}. This is due to the unitary own-price elasticity of marijuana, unchanged prices for cocaine and basuco, and the relatively low shares of expenditure on these latter drugs.

\begin{figure}[!htbp]
    \centering
    \caption{Distribution of individual-specific prices and average prices under different legalization scenarios}
    \label{Fig:Legalization_Price_Distribution}
    \includegraphics[width = 0.6\textwidth]{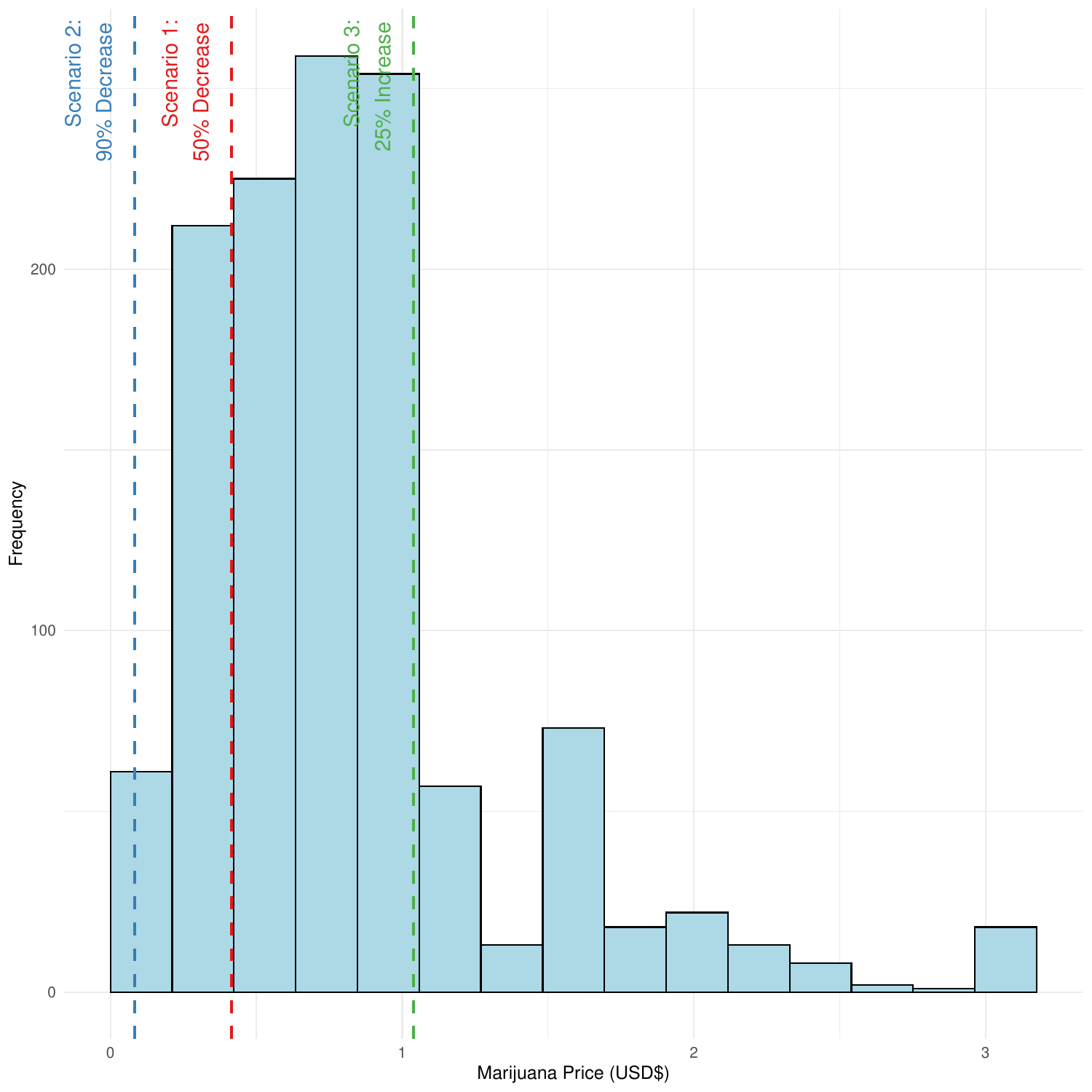}
    \begin{tablenotes}
        \item \small{\textbf{Notes:} Histogram with 15 equal-density bins of the prices paid for marijuana by individuals in the \textit{ENCSPA} survey. Price variation reflects location, proximity to hubs, consumer search, and potential supply effects. Lines showcase the average price paid after each of the counterfactual legalization scenarios. Post-legalization price scenarios all naturally occur within the price distribution.}
    \end{tablenotes}
\end{figure}

\begin{figure}[htbp]
    \centering
    \caption{Shares of predicted total expenditure on drugs in Colombia by age groups under a legalization policy}
    \label{Fig:Predicted_Drug_Expenditure}
    \includegraphics[width = 0.8\textwidth, trim={18pt 18pt 18pt 18pt}, clip = true]{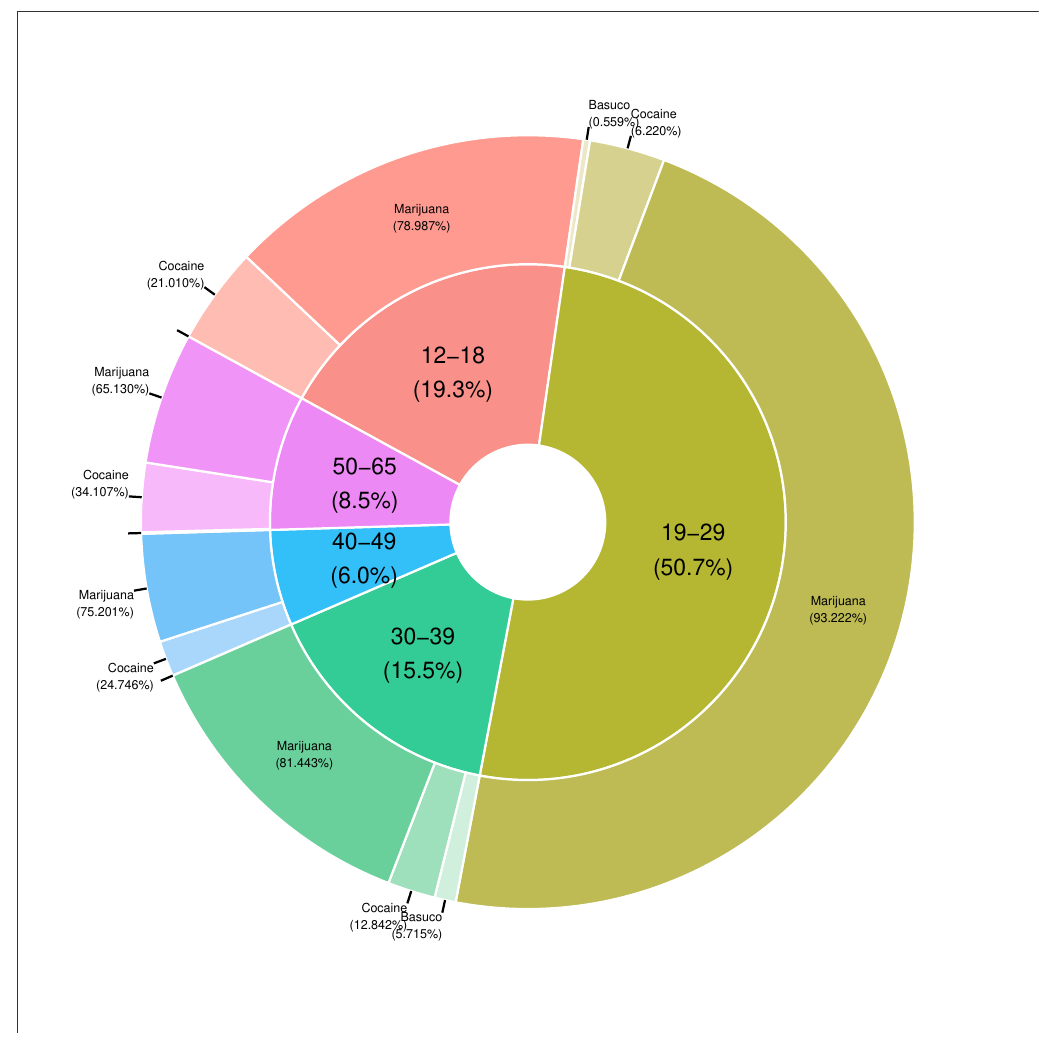}
    \begin{tablenotes}
    \item \small{\textbf{Notes:} Total drug market expenditure in Colombia as predicted by the counterfactual quantities from a 50\% decrease in the price of marijuana, and presuming full access to a dealer post-legalization. We use the average exchange rate in 2019 (COP/USD 3,274) to transform to US dollars. Total size of the market after implementing the policy is estimated to be 228 million USD.}
    \end{tablenotes}
\end{figure}

\begin{figure}
    \centering
    \caption{Density plots of equivalent variation following a price change in marijuana implied by EASI system estimates from ``soft'' cluster (annual USD)}
    \label{Fig:Density_EV_OtherScenarios}
    \resizebox{1\textwidth}{!}{%
    \begin{threeparttable}
    \begin{subfigure}[b]{0.48\textwidth}
        \centering
        \includegraphics[width = \textwidth, page = 2]{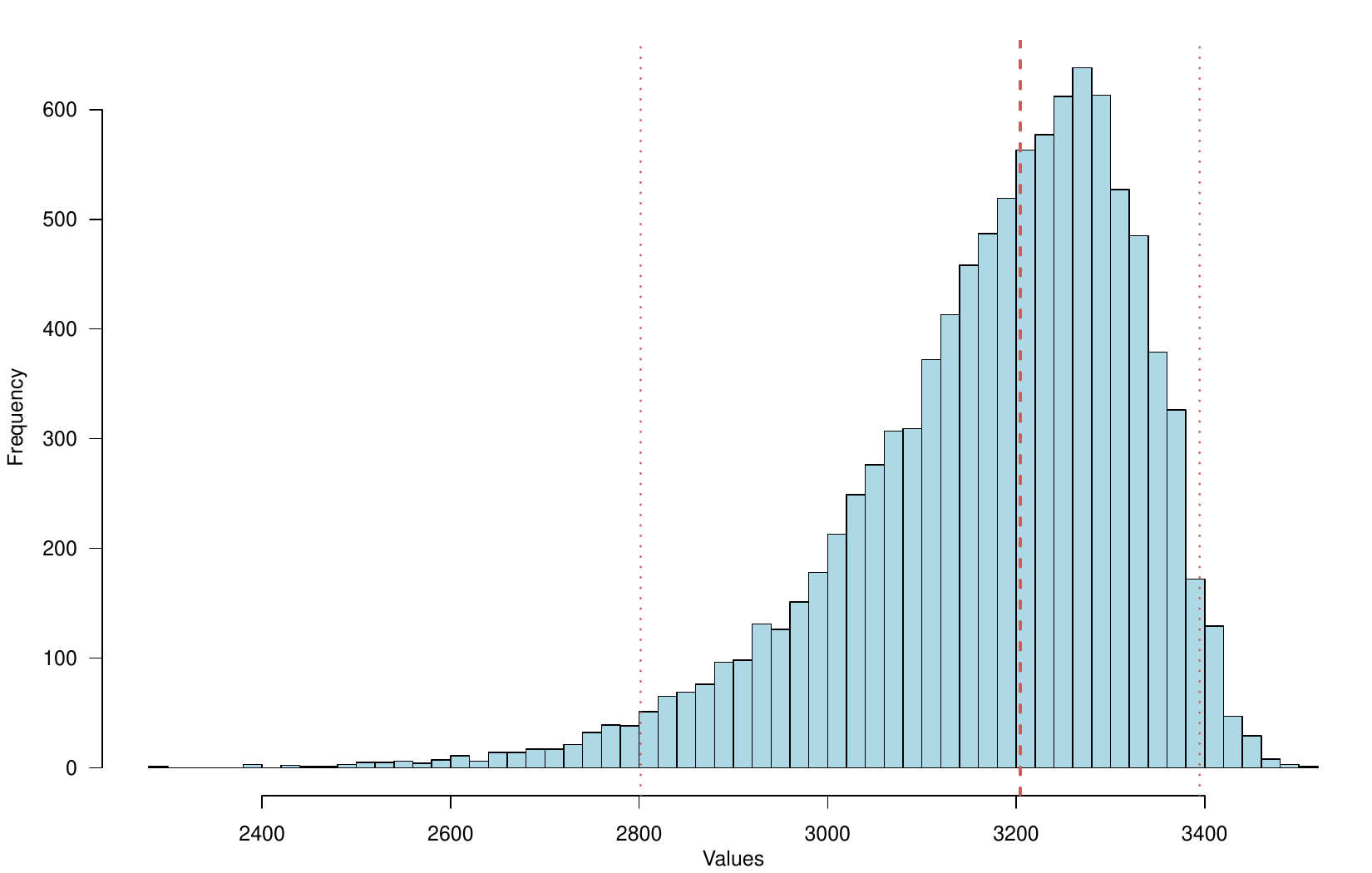}
        \caption{90\% decrease}
    \end{subfigure}
    \begin{subfigure}[b]{0.48\textwidth}
        \centering
        \includegraphics[width = \textwidth, page = 2]{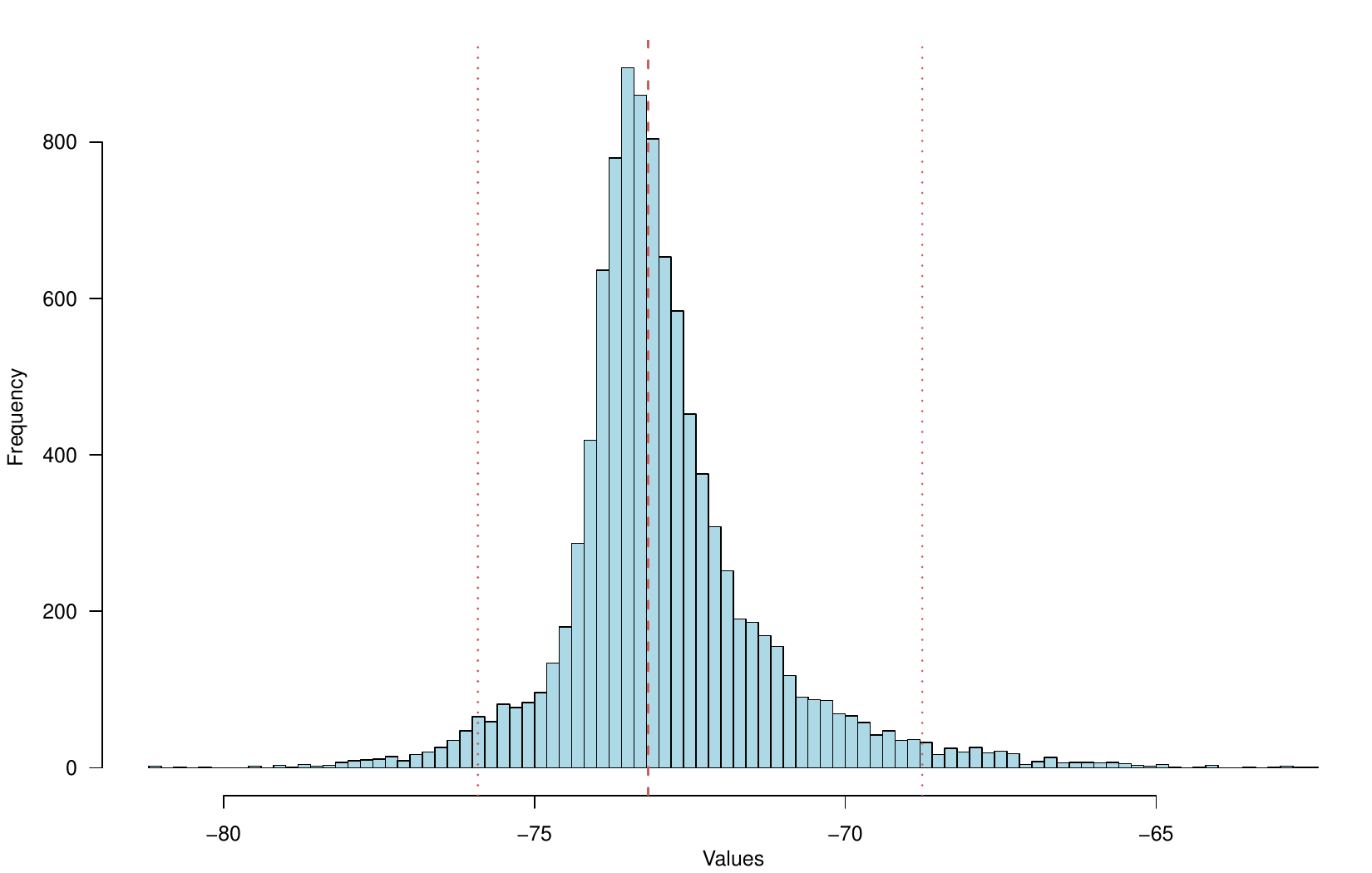}
        \caption{25\% increase}
    \end{subfigure}
    \begin{tablenotes}
    \item \small{\textbf{Notes:} Full posterior density over equivalent variation in 2019 USD\$ after changes in the price of marijuana and complete access to dealers following legalization of marijuana. Dashed line represents median of chains, dotted line provides the 95\% HPD for each parameter, and a solid line at 0 is drawn for reference. Single chain run keeping 10,000 draws after a burn-in window of 5,000 iterations, discarding every 10 draws.}
    \end{tablenotes}
    \end{threeparttable}
    }
\end{figure}

\addtocounter{figure}{-1}

\begin{figure}
    \centering
    \caption{Density plots of government revenue following a price change in marijuana implied by EASI system estimates from ``soft'' cluster (annual USD million)}
    \label{Fig:Density_GovernmentRevenue_OtherScenarios}
    \resizebox{1\textwidth}{!}{%
    \begin{threeparttable}
    \begin{subfigure}[b]{0.48\textwidth}
        \centering
        \includegraphics[width = \textwidth, page = 2]{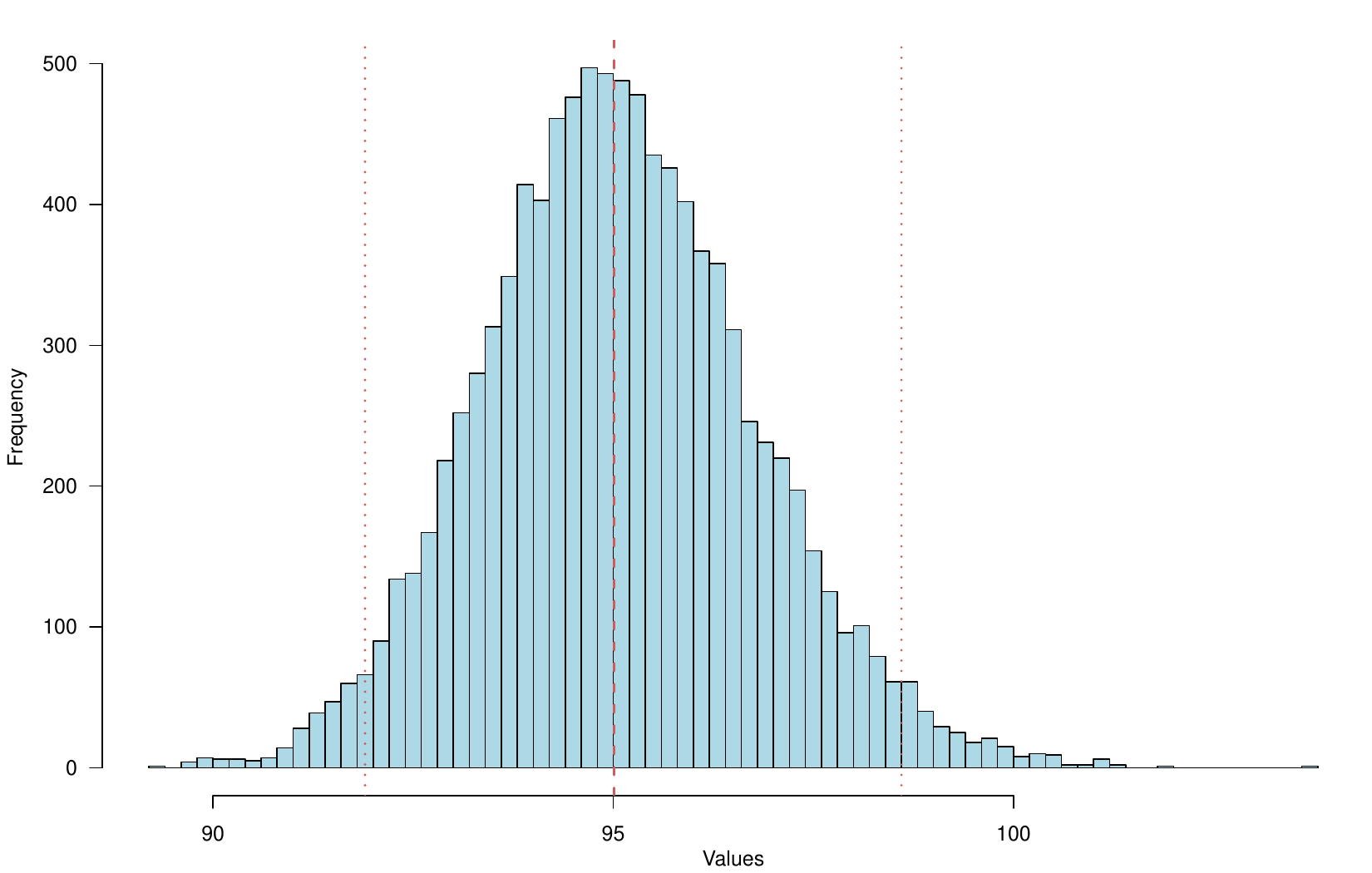}
        \caption{90\% decrease}
    \end{subfigure}
    \begin{subfigure}[b]{0.48\textwidth}
        \centering
        \includegraphics[width = \textwidth, page = 2]{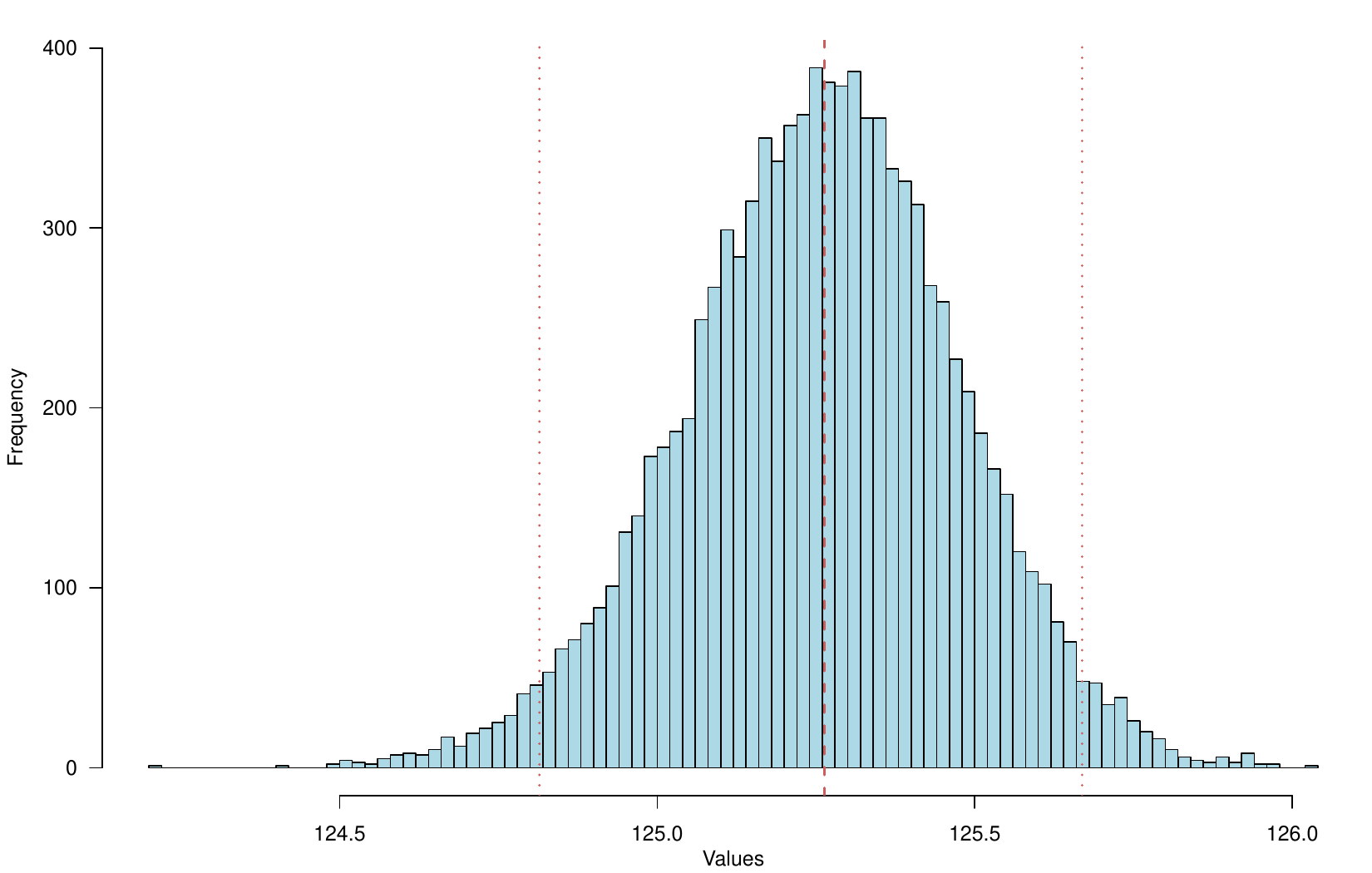}
        \caption{25\% increase}
    \end{subfigure}
    \begin{tablenotes}
    \item \small{\textbf{Notes:} Full posterior density over total annual government revenue given in millions of 2019 USD\$ after changes in the price of marijuana and complete access to dealers following legalization of marijuana. Dashed line represents median of chains, dotted line provides the 95\% HPD for each parameter, and a solid line at 0 is drawn for reference. Single chain run keeping 10,000 draws after a burn-in window of 5,000 iterations, discarding every 10 draws.}
    \end{tablenotes}
    \end{threeparttable}
    }
\end{figure}

\addtocounter{figure}{-1}

\begin{figure}
    \centering
    \caption{Density plot of change in dealer revenue following a price change in marijuana implied by EASI system estimates from ``soft'' cluster (annual USD million)}
    \label{Fig:Density_RevenueChange_OtherScenarios}
    \resizebox{1\textwidth}{!}{%
    \begin{threeparttable}
    \begin{subfigure}[b]{0.48\textwidth}
        \centering
        \includegraphics[width = \textwidth, page = 2]{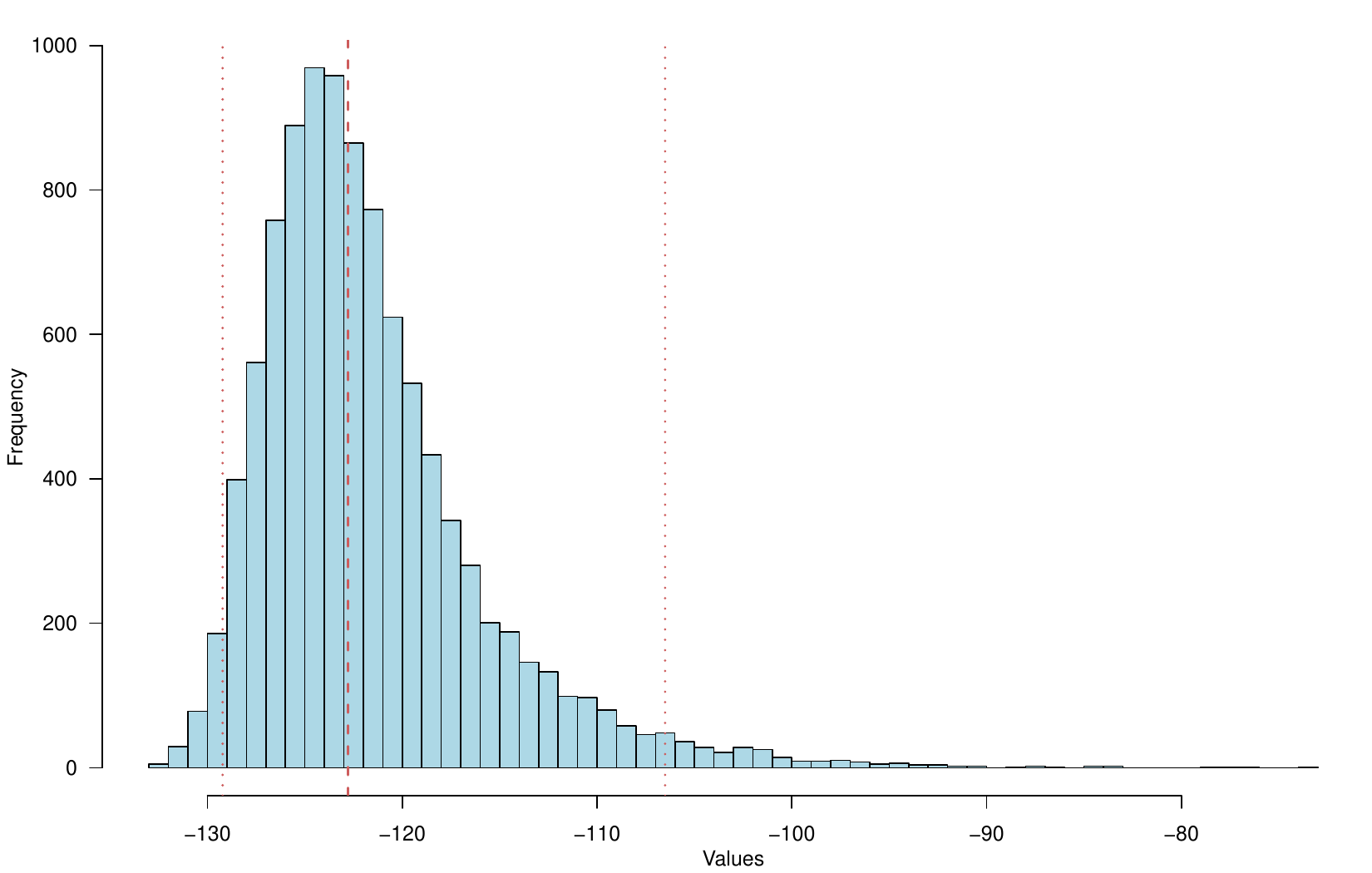}
        \caption{90\% decrease}
    \end{subfigure}
    \begin{subfigure}[b]{0.48\textwidth}
        \centering
        \includegraphics[width = \textwidth, page = 2]{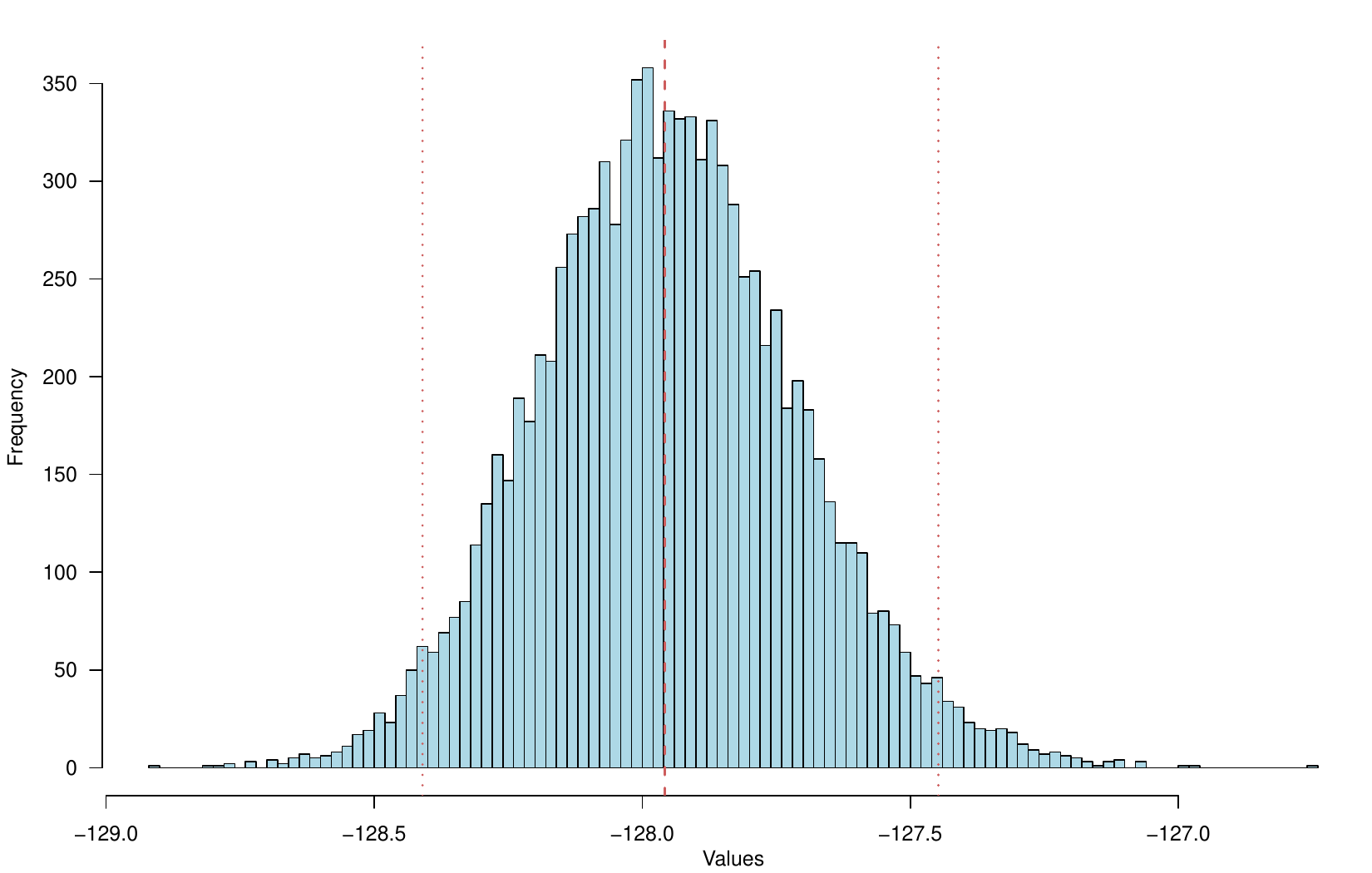}
        \caption{25\% increase}
    \end{subfigure}
    \begin{tablenotes}
    \item \small{\textbf{Notes:} Full posterior density over change in total annual dealer revenue given in millions of 2019 USD\$ after changes in the price of marijuana and complete access to dealers following legalization of marijuana. Dashed line represents median of chains, dotted line provides the 95\% HPD for each parameter, and a solid line at 0 is drawn for reference. Single chain run keeping 10,000 draws after a burn-in window of 5,000 iterations, discarding every 10 draws.}
    \end{tablenotes}
    \end{threeparttable}
    }
\end{figure}

\addtocounter{figure}{-1}

\begin{figure}
    \centering
    \caption{Density plot of percentage change in number of users required to offset dealer revenue change following a price change in marijuana as implied by EASI system estimates from ``soft'' cluster}
    \label{Fig:Density_UserChange_OtherScenarios}
    \resizebox{1\textwidth}{!}{%
    \begin{threeparttable}
    \begin{subfigure}[b]{0.48\textwidth}
        \centering
        \includegraphics[width = \textwidth, page = 2]{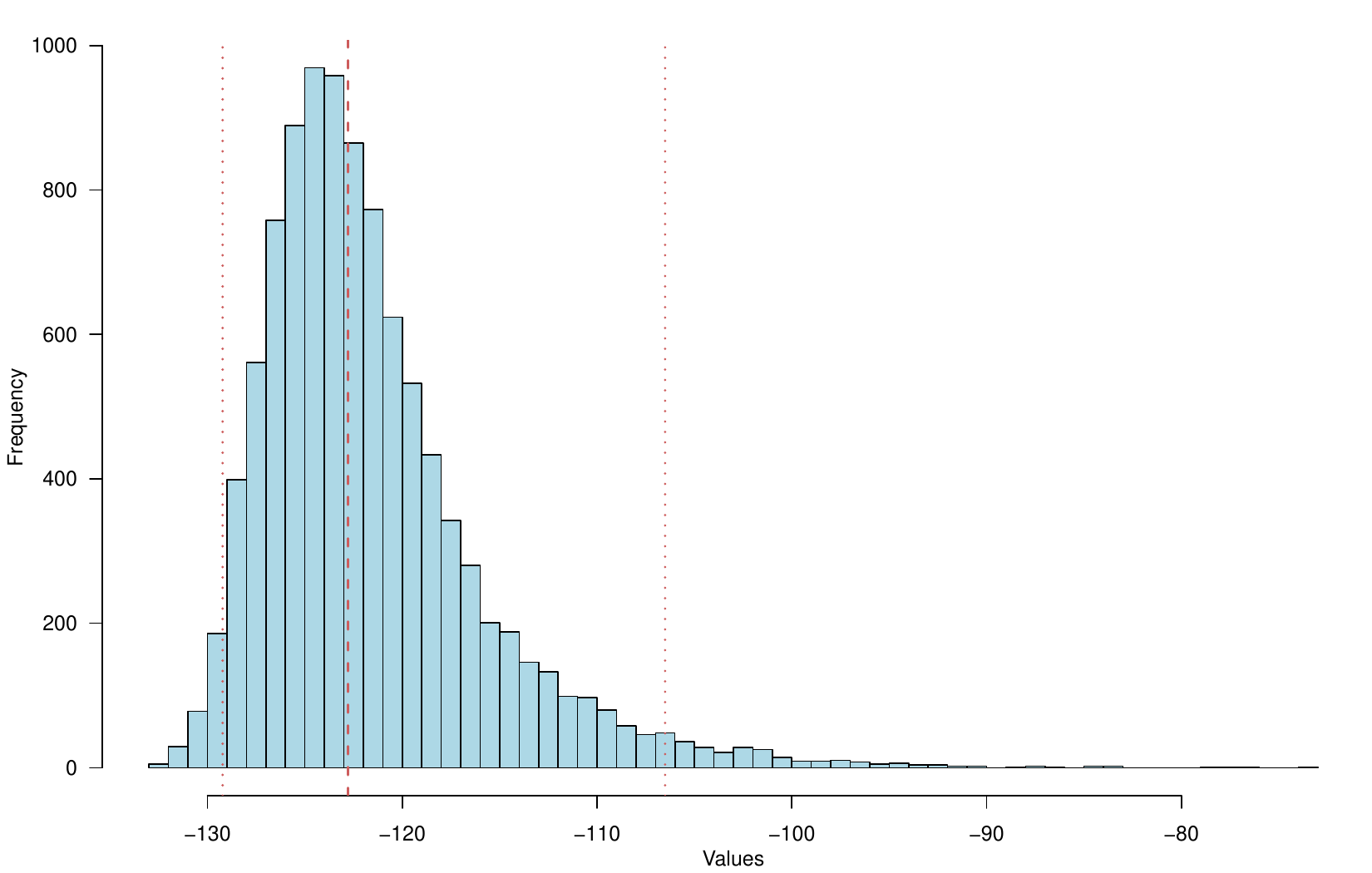}
        \caption{90\% decrease}
    \end{subfigure}
    \begin{subfigure}[b]{0.48\textwidth}
        \centering
        \includegraphics[width = \textwidth, page = 2]{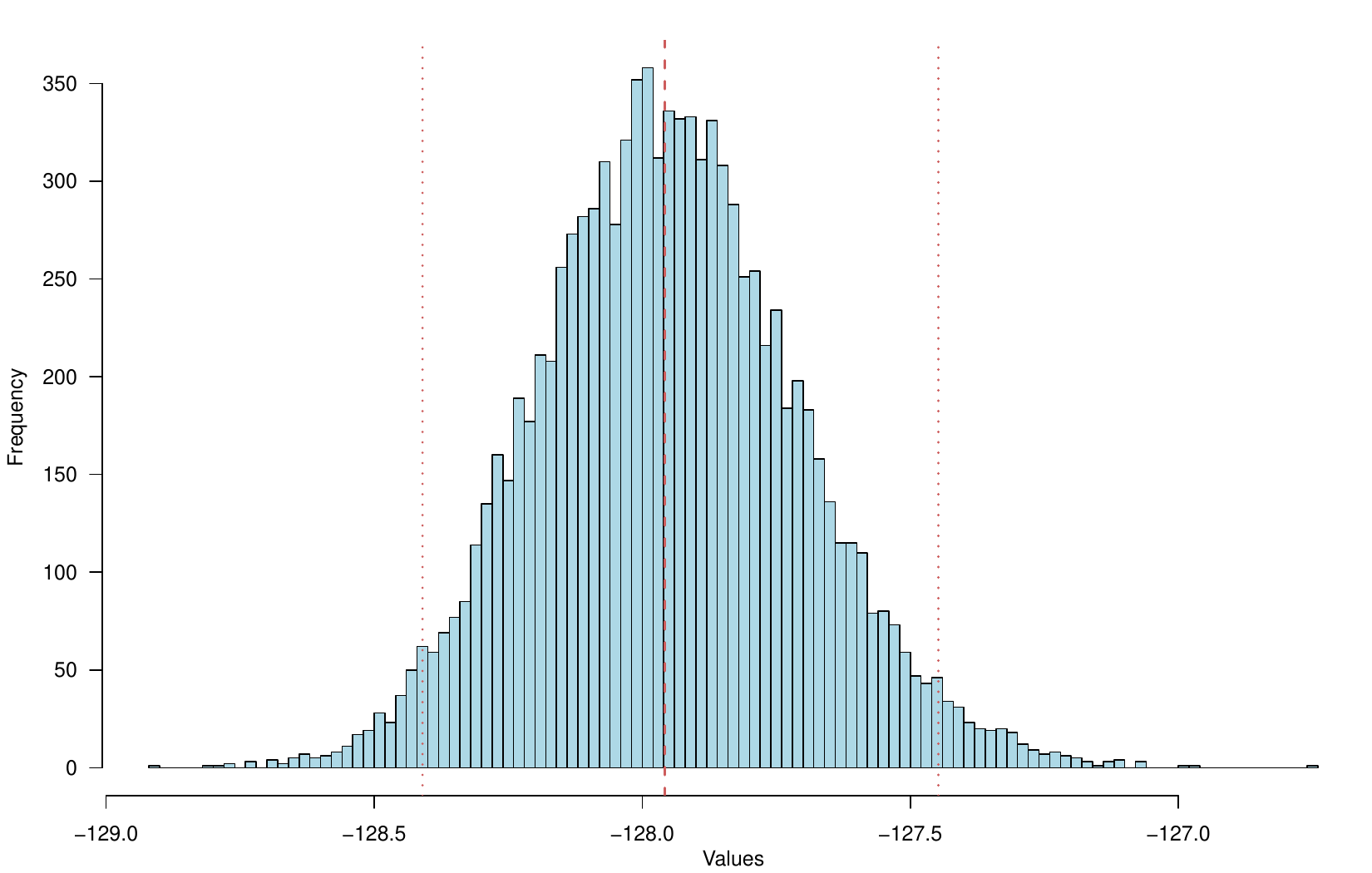}
        \caption{25\% increase}
    \end{subfigure}
    \begin{tablenotes}
    \item \small{\textbf{Notes:} Full posterior density over the percentage change in the number of users required to offset the total change in annual dealer revenue after changes in the price of marijuana and complete access to dealers following legalization of marijuana. Dashed line represents median of chains, dotted line provides the 95\% HPD for each parameter, and a solid line at 0 is drawn for reference. Single chain run keeping 10,000 draws after a burn-in window of 5,000 iterations, discarding every 10 draws.}
    \end{tablenotes}
    \end{threeparttable}
    }
\end{figure}

\addtocounter{figure}{-1}

\begin{figure}
    \centering
    \caption{Density plot of change in drug dealer marijuana profits following a price change in marijuana implied by EASI system estimates from ``soft'' cluster}
    \label{Fig:Density_ProfitChange_OtherScenarios}
    \resizebox{1\textwidth}{!}{%
    \begin{threeparttable}
    \begin{subfigure}[b]{0.32\textwidth}
        \centering
        \includegraphics[width = \textwidth, page = 2]{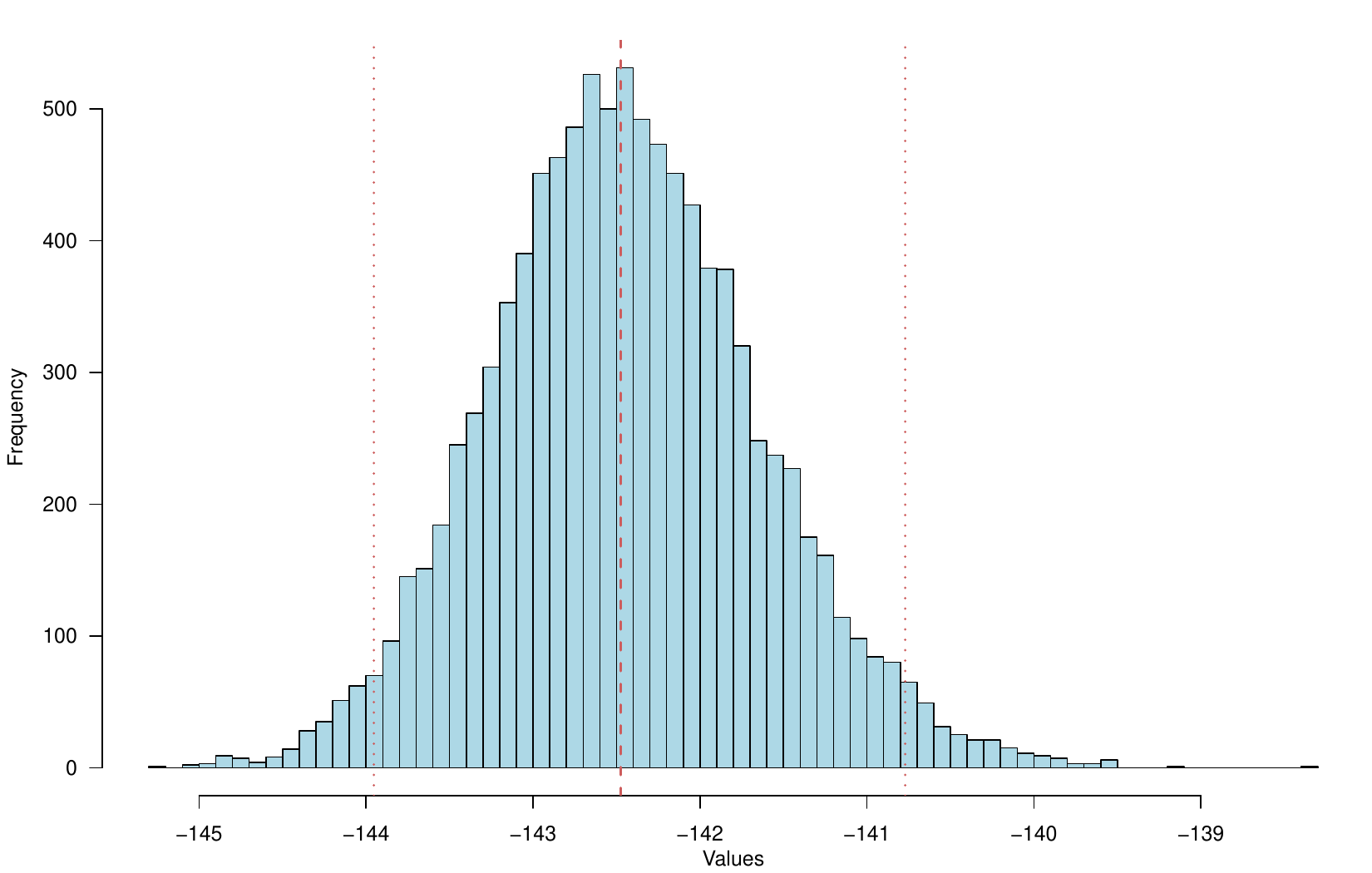}
        \caption{90\% decrease}
    \end{subfigure}
    \begin{subfigure}[b]{0.32\textwidth}
        \centering
        \includegraphics[width = \textwidth, page = 2]{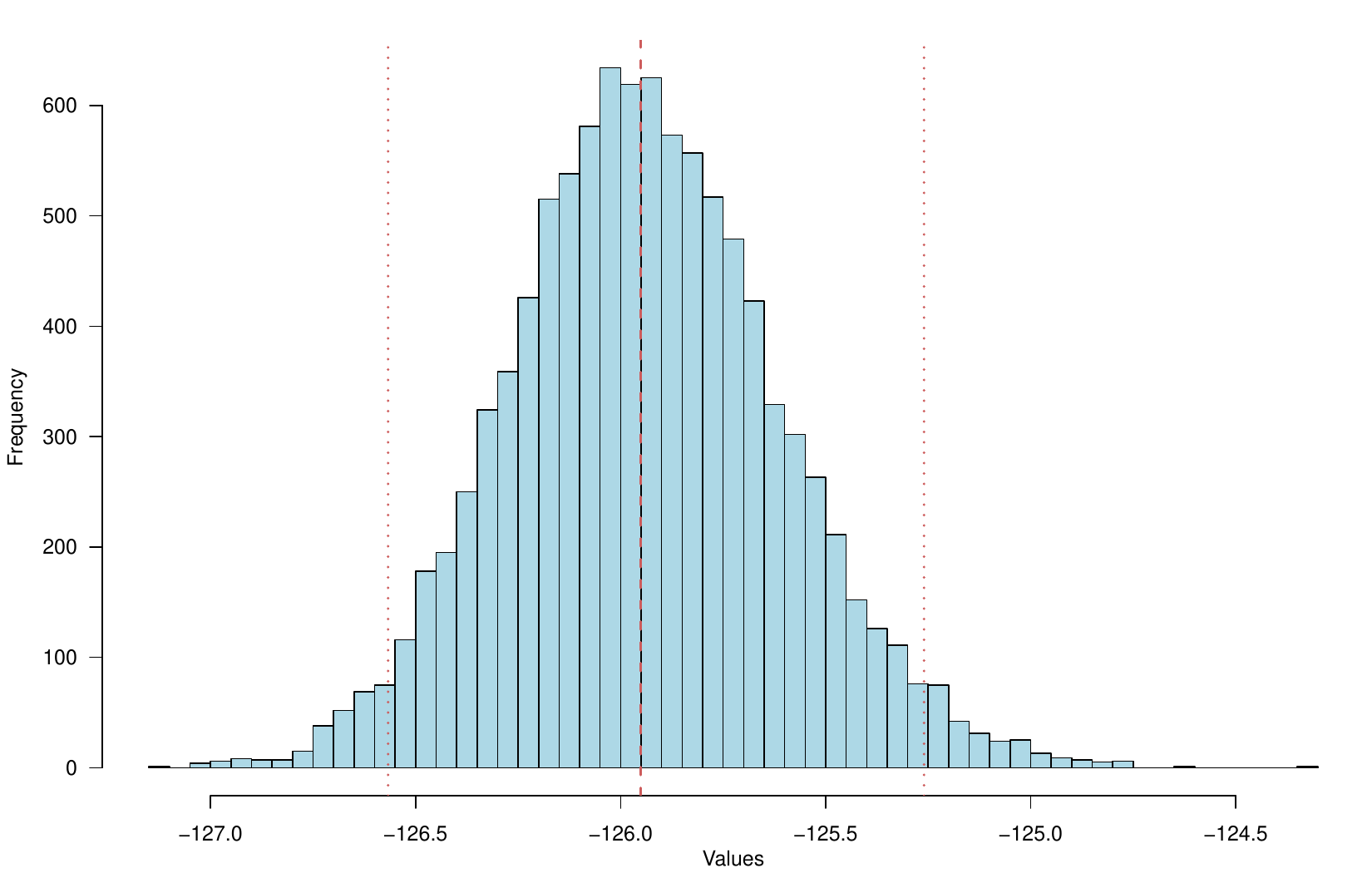}
        \caption{50\% decrease}
    \end{subfigure}
    \begin{subfigure}[b]{0.32\textwidth}
        \centering
        \includegraphics[width = \textwidth, page = 2]{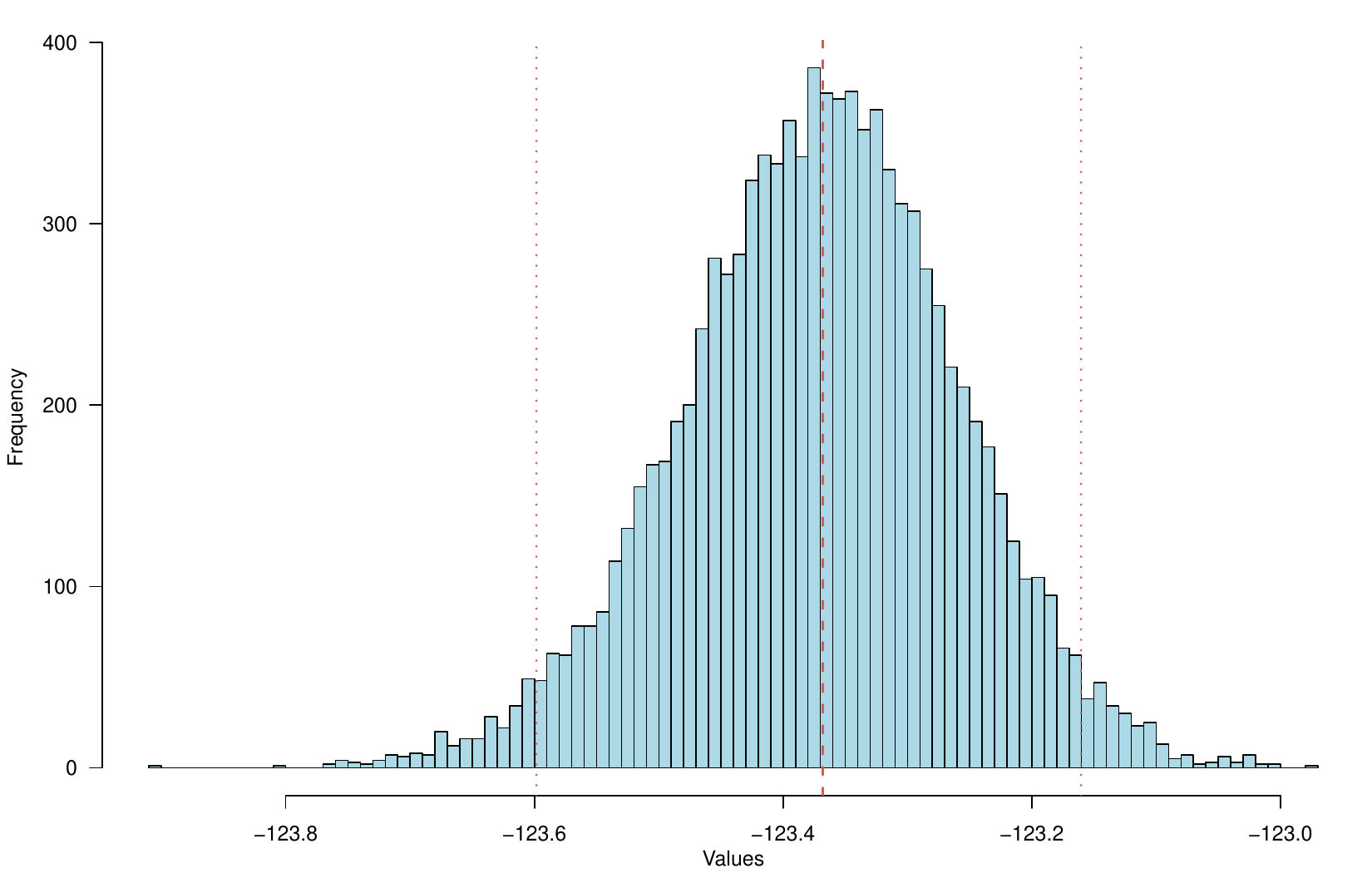}
        \caption{25\% increase}
    \end{subfigure}
    \begin{tablenotes}
    \item \small{\textbf{Notes:} Full posterior density over the change in total marijuana profits given in millions of 2019 USD\$ after changes in the price of marijuana and complete access to dealers following legalization of marijuana. Dashed line represents median of chains, dotted line provides the 95\% HPD for each parameter, and a solid line at 0 is drawn for reference. Single chain run keeping 10,000 draws after a burn-in window of 5,000 iterations, discarding every 10 draws.}
    \end{tablenotes}
    \end{threeparttable}
    }
\end{figure}

\end{appendices}

\end{document}